\documentclass[12pt,preprint]{aastex}
\begin{document}

\shorttitle{Debris Disks in NGC 2232}
\shortauthors{Currie, T., Plavchan, P., Kenyon, S. J.}

\title{A Spitzer Study of Debris Disks In The Young Nearby Cluster NGC 2232: 
Icy Planets Are Common Around $\sim$ 1.5--3 M$_{\odot}$ Stars}
%\title{Possible Evidence for Terrestrial Planet Formation from Strong IRAC-excesses in h \& $\chi$ Persei }
%\title{Strong IRAC-excess sources in h \& $\chi$ Persei: Possible Evidence for Terrestrial Planet Formation}
\author{Thayne Currie\altaffilmark{1}, Peter Plavchan\altaffilmark{2}, and
Scott J. Kenyon\altaffilmark{1}}
\altaffiltext{1}{Harvard-Smithsonian Center for Astrophysics, 60 Garden St. Cambridge, MA 02140}
%\altaffiltext{2}{Department of Physics \& Astronomy, University of California-Los Angeles, Los Angeles, CA 90095}
\altaffiltext{2}{Michelson Science Center, M/C 100-22, California Institute of Technology, 770 S. Wilson Ave., 
Pasadena, CA 91125}
%\altaffiltext{4}{Department of Physics, University of Utah}
\email{tcurrie@cfa.harvard.edu, plavchan@ipac.caltech.edu, skenyon@cfa.harvard.edu}
\begin{abstract}
We describe Spitzer IRAC and MIPS observations of the
nearby 25 Myr-old open cluster NGC 2232. Combining
these data with ROSAT All-Sky Survey observations,
proper motions, and optical photometry/spectroscopy,
we construct a list of highly probable cluster members.
We identify 1 A-type star, HD 45435, with definite excess emission at
4.5--24 $\mu$m indicative of debris from 
terrestrial planet formation.   We also identify 2--4 late-type 
stars with possible 8 $\mu$m
excesses, and 8 early-type stars with definite 24 $\mu$m
excesses. Constraints on the dust luminosity and temperature
suggest that the detected excesses are produced by debris disks.
From our sample of B and A stars, stellar rotation appears correlated with 
24 $\mu m$ excess, a result expected if massive 
primordial disks evolve into massive debris disks.
 To explore the evolution
of the frequency and magnitude of debris around A-type
stars, we combine our results with data for other young
clusters.  The frequency of debris disks 
around A-type stars appears to increase from $\sim$ 25\%
at 5 Myr to $\sim$ 50--60\% at 20--25 Myr. Older A-type stars
have smaller debris disk frequencies: $\sim$ 20\% at
50--100 Myr. For these ages, the typical level of debris
emission rises from 5--20 Myr and then declines.  
 Because 24 $\mu$m dust emission probes icy planet formation
around A-type stars, our results suggest that the
frequency of icy planet formation is $\eta_{i}$ $\gtrsim$
0.5--0.6. Thus, most A-type stars ($\approx$ 1.5--3 M$_{\odot}$) produce icy planets.
%Although NGC 2232 has been virtually ignored 
%by the star/planet formation community for the past 35 years, this work shows it to 
%be a potentially important laboratory for understanding debris disk evolution and planet formation.
\end{abstract}
\section{Introduction}
Recent studies of young stellar clusters from the \textit{Spitzer Space Telescope} \citep{We04}
 provide powerful insights into the emergence of mature planetary systems from  
primordial disks surrounding young stars.  Spitzer observations 
indicate that the initial stages of primordial disk evolution proceed rapidly.  Within $\sim$ 1-3 Myr, many 
optically-thick primordial disks undergo dust 
settling, form inner regions cleared of dust, and undergo grain growth to sizes larger than the 
sub-micron sizes characteristic of the interstellar medium \citep[e.g.][]{Cal05, La06, Sa06, Bou08}.  
As these processes occur, the level of inner disk emission drops and the frequency of primordial disks 
declines (e.g. \citealt{He07}, see also \citealt{Hi05}).  Primordial disks around early type stars disappear
faster than around later-type stars \citep{Ca06, He07, Ck08}.  

By $\sim$ 10--15 Myr, primordial disks 
are exceedingly rare. The disk population is then dominated by optically thin, gas-poor \textit{debris disks} 
\citep{He06, Cu07a, Ck08}.  Because debris dust is removed on timescales much shorter 
than the age of the star, the dust must be replenished to sustain disk emission.  
Collisions between $\gtrsim$ km-sized planetesimals excited by large-scale 
planet formation \citep[e.g.][]{Bp93, Kb04} can provide this replenishment.
Though debris disks around early-type stars can appear as soon as $\sim$ 3 Myr \citep{Ck08}, the mid-IR debris emission 
 does not peak until $\approx$ 10--20 Myr \citep{Cu08a}.  This rise in debris emission 
is consistent with planetesimal collisions during the growth of icy protoplanets in the outer disk \citep{Kb08}.
  After protoplanets reach their maximum size at $\approx$ 20--30 Myr, the level of debris emission decays as $\sim$ t$^{-1}$ \citep{Ri05, Su06}.
This decay is consistent with steady-state collision models \citep[e.g.][]{Wy07}.
 
The epoch of maximum debris emission ($\sim$ 10--30 Myr) is also important 
for terrestrial planet formation.  Planet formation models and radiometric 
dating suggest that terrestrial planets reach their final mass by $\sim$ 10--30 Myr \citep{Ws93, Yi02, 
Kb06}.  Terrestrial planet formation produces debris emission observable at $\approx$ 5--10 $\mu m$.  
Mid-IR Spitzer observations reveal terrestrial zone dust emission around many 
13 Myr old stars in the massive Double Cluster, h and $\chi$ Persei, and several stars in 
 other $\lesssim$ 40 Myr old clusters/associations including Sco-Cen, NGC 2547, and the $\beta$ Pic Moving Group 
\citep{Zs04, Ch05, Ch06, Cu07b, Rh07a, Go07, Cu08a, Li08}\footnote{At least two other, much older stars harbor warm debris disks 
that may be indicative of stochastic collisions in the terrestrial zone: BD +20$^{o}$ 307 and HD 23514 \citep{Sz05, Rh07b}}.
 Observations of many warm, IRAC-excess stars in h and $\chi$ Persei point to a spectral-type/stellar-mass dependent 
evolution of terrestrial zone dust, suggesting that the terrestrial planet formation process runs to completion faster
for high-mass stars than for intermediate-mass stars \citep{Cu07a, Cu08b}.  

Reconstructing the time history of terrestrial and icy planet formation 
 at the 10--40 Myr epoch requires Spitzer IRAC and MIPS observations of many clusters 
besides h and $\chi$ Persei.
While observations of stars in 10--20 Myr old Sco-Cen and the $\beta$ Pic moving group provide 
important studies of cold dust probed by MIPS \citep{Ch05, Re08}, strong constraints on warm dust
probed by $\le$ 10--15 $\mu m$ broadband photometry of cluster members 
are limited to IRAS, MSX, and ground-based campaigns with 
 much poorer sensitivity.  The FEPS (\textit{Formation and Evolution of Planetary Systems}; 
\citealt{Me06}) Legacy Program observed $\sim$ 35 10--20 Myr-old Sco-Cen sources in IRAC with a range of 
spectral types ($\sim$ F8/G0 to K3).  However, this sample size is 
too small to provide statistically robust constraints given the low overall frequency of warm dust emission 
\citep{Ma04, Cu07a, Go07}.  

Though h and $\chi$ Persei is massive enough to investigate the frequency of warm dust 
from terrestrial planet formation in one environment, 
observations of other clusters are needed to constrain the time evolution of warm dust 
in many environments.  The Spitzer survey of NGC 2547 \citep{Go07} potentially fill this void in sampling, 
yielding a large population of cluster stars ($\sim$ 400-500) and revealing warm dust around 
at least $\sim$ 6 FGKM stars.  However, recent work has 
revised the age of NGC 2547 upward from 25 Myr to $\sim$ 35--40 Myr \citep{Nj06, Mn08}.  Spitzer studies 
of other $\sim$ 15--35 Myr-old clusters would then clearly aid in constraining the 
evolution of the observable signatures of terrestrial planet formation.

Understanding the evolution of cold dust emission from debris disks
 also requires additional observations of 15--35 Myr-old clusters.  While 24 $\mu m$ debris disk emission 
around early-type stars peaks at $\approx$ 10--20 Myr, the evolution 
of debris emission at $\approx$ 20--30 Myr is not well constrained.  
Few clusters in this age range have been observed with Spitzer.  The 
most populous cluster previously assigned to this age range, NGC 2547, 
is likely older.  Observing 20--30 Myr old clusters may also reveal 
a delay in the peak debris emission for later type, lower-mass stars
relative to early-type stars as implied by planet formation models \citep[e.g.][]{Kb08}.  
Observations of young stars at 70 $\mu m$ may provide tentative evidence for this 
 delay \citep{Hi08}.  Furthermore, the disk environment around late type, 
low-mass stars may be qualitatively different due to the importance of corpuscular wind drag 
in removing debris dust \citep{Pl05}.
Recent Spitzer studies argue that detectable debris disks around M stars may be rarer 
 than around intermediate/high-mass stars \citep{Ga07, Pl08}.
To further investigate this possibility, MIPS observations of M dwarf stars provide 
new constraints on planet formation processes.

In this paper, we investigate debris disk evolution and planet formation 
in the young open cluster NGC 2232 located in Gould's Belt close to the 
Orion Nebula Cluster.  Until very recently, the most complete studies of this cluster 
date from over 30 years ago \citep{Cl72, Lm74}, which identify NGC 2232 as a group 
of proper motion stars with B-F spectral types and an age of $\sim$ 20-25 Myr.  Based on 
pre-main sequence evolutionary models,  \citet{Ly06} estimate an age of 25--30 Myr 
 and find a 25 Myr nuclear age.  Both \citet{Cl72} and \citet{Ly06} show that 
the cluster is nearby ($\sim$ 320-360 pc).  At this distance, IRAC observations  
can probe well into the M dwarf regime; MIPS observations can 
 detect the photospheres of early and intermediate-mass stars.  The cluster has an extremely low mean 
reddening (E(B-V) $\sim$ 0.07; \citealt{Ly06}).
Thus, NGC 2232's distance, age, and reddening combine to make investigating debris disk 
evolution feasible in a way that complements previous studies of h and $\chi$ Persei and 
 NGC 2547.

Our investigation is organized as follows.
After describing the observations, image processing, and photometry of Spitzer IRAC/MIPS 
sources, we match sources with 2MASS/optical VI data in \S 2.  Analysis of these sources 
shows that many may have excess emission from circumstellar dust.  
To identify cluster members, we rely on ROSAT x-ray observations, proper 
motion data, optical/IR color-magnitude diagrams and spectroscopy in \S 3.
In \S 4 and 5, we analyze Spitzer data of x-ray selected cluster members to identify stars with 
evidence for circumstellar dust, identify stars with active terrestrial planet formation,
 and explore any statistical trends in the disk population 
with the stellar properties and in the level of debris emission with time.  

NGC 2232 harbors a substantial population of stars with 24 $\mu m$ excess with a 
range of spectral types from late B to late K spectral type.  At least one star (HD 45435) shows evidence for 
warm 8 $\mu m$ excess.  Constraining the disk luminosity and dust temperature of this source and comparing 
its SED to debris disk models shows that its warm dust emission is plausibly 
produced by terrestrial planet formation.  

%The frequency of 24 $\mu m$ 
%excess from A stars is high ($\ge$ 50\%).  Because this emission likely comes 
%from disk regions beyond the ice line ($\approx$ 170 K), these results imply that 
%icy planets around A stars are common.  
This survey yields robust constraints on the evolution of 
cold debris disks.
% and important implications for the 
%frequency of icy planets around high-mass, early-type stars.
We identify a correlation between stellar rotation and 24 $\mu m$ 
excess for high-mass stars that may reveal an evolutionary link between massive primordial disks 
and massive debris disks.  When combined with Spitzer data of other young 
clusters, our survey shows that the level of debris emission around 
A stars in the cluster is consistent with a peak in the 24 $\mu m$ debris emission 
at $\sim$ 10--20 Myr, confirming the results of recent surveys \citep{Cu08a}.  
The frequency of 24 $\mu m$ debris emission 
\textit{increases} from $\sim$ 5 Myr and peaks between 10 Myr and 30 Myr.

Finally, this work has important implications for the frequency of icy planets around 
high-mass, early-type stars.
The fraction of A stars ($\approx$ 1.5--3 M$_{\odot}$) 
with 24 $\mu m$ excess emission is high ($\ge$ 50\%).  Because this emission likely comes
from disk regions beyond the ice line ($\approx$ 170 K), these results imply that
icy planets around A stars are common. 

%ecitet{Cu07a} 
\section{Data}
\subsection{Spitzer IRAC and MIPS photometry}
\subsubsection{Observations and Image Processing}
Observations of NGC 2232 were taken with the Infrared Array Camera (IRAC; \citealt{Faz04}) and 
Multiband Imaging Photometer for Spitzer (MIPS; \citealt{Rie04}) on March 29, 2005 and April 4, 2005 as a part of the 
\textit{Guaranteed Time Observations} program (Program ID 37).  
Both sets of data for NGC 2232 (l= 215$^{o}$, b = -7.4$^{o}$) cover $\sim$ 1.4 square degrees on the 
sky with boundaries of $\alpha_{2000}$ $\sim$ 6$^{h}$25$^{m}$50$^{s}$ -- 
6$^{h}$30$^{m}$00$^{s}$ and $\delta_{2000}$ $\sim$ -5$^{o}$30$^{m}$00$^{s}$ --
-4$^{o}$05$^{m}$00$^{s}$.  
The IRAC observations consist of high-dynamic range exposures of 0.6s and 10.4s at
3.6$\mu m$, 4.5$\mu m$, 5.8$\mu m$, and 8$\mu m$.  Sources 
were typically observed at four dither points, which yielded an average
integration time/pixel of $\sim$ 2.4s and 41.6s for the short and long exposures. 
The integration time/pixel was less for the edges of the region ($\sim$ 20.8 s) and more for central 
 regions where dithered images overlap.  The MIPS observations at 24 $\mu m$ and 70 $\mu m$ 
were taken in scan mode with a typical integration time/pixel of 80s.
Background cirrus levels are low ($\sim$ 28 MJy/sr) and vary little 
($\sim$ 3\%) across the MIPS field.

While the MIPS Basic Calibration Data (BCD) are largely free of artifacts, the IRAC 
BCD data require additional post-BCD processing before mosaicing. 
Using a custom IDL script written by T. C., 
we applied the appropriate array-dependent correction for point sources 
on each IRAC BCD frame \citep{Qu04}.  
Inspection of the 10.4s BCD frames showed that many bright stars produce
 severe 'striping' and column pulldown effects, especially in the [3.6] and [4.5] channels.  
  For columns with bright stars, column pulldown shifts the background flux  
levels by up to $\sim$ 2 MJy/sr.  We removed these artifacts  
 by applying a modified version of the \textit{muxstriping} and \textit{column pulldown} 
algorithms developed by Jason Surace and Leonidas Moustakas available on the \textit{Spitzer Science Center} 
website.  These steps drastically reduce striping and bias 
level artifacts.  Flux levels for stars change by up to $\sim$ 5-10\% 
in all channels due to the array-dependent correction and up to 30\% for faint stars 
in channels 1 and 2 in columns affected by bright stars.

We then processed both the IRAC and MIPS data with the MOPEX/APEX pipeline, interpolating 
over bad pixels.  Using a bicubic interpolation with outlier rejection, the individual processed BCD 
frames were combined together into a single image in each filter 
for each exposure time.  The final mosaics
were inspected and cleared of any remaining image artifacts or pattern noise that could 
seriously compromise the photometry.  
\subsubsection{Photometry}
Because the IRAC data are not well sampled (except for the brightest stars) 
and the point-spread function core is poorly characterized, we performed 
source finding and aperture photometry with a combination of 
SExtractor \citep{Be96} and IDLPHOT routines in lieu of PRF fitting with APEX.
For source identification with SExtractor, we constructed a background map for 
each IRAC mosaic in a 24x24 pixel area smoothed by an 8x8 pixel box.  Sources were 
identified by their fluxes relative to the standard deviation in the background rms 
($\sigma_{bkgd, rms}$).  Groups of pixels with 
 counts greater than 5$\sigma_{bkgd, rms}$ (2$\sigma_{bkgd, rms}$) were identified as sources 
for the [3.6] and [4.5] ([5.8] and [8]) filters.  The resulting source lists 
were used as input for aperture photometry with the \textit{aper.pro} routine.
%same as the output of the \textit{find.pro} algorithm in IDLPHOT and were used as input for aperture 
%photometry with the \textit{aper.pro} routine.  
The source flux was computed in a 2 ($\sim$ 2.44") and 
3 pixel ($\sim$ 3.66") aperture radius, using a local background calculated from a 4 pixel-wide annulus surrounding 
each source extending from 2-6 and 3-7 pixels.  We multiplied the source fluxes by the appropriate aperture 
corrections from the IRAC data handbook, version 
3.0\footnote{http://ssc.spitzer.caltech.edu/irac/dh/iracdatahandbook3.0.pdf}.  To fine-tune the aperture 
correction, we compared the photometry for the brightest unsaturated sources 
derived above with that using a 10 pixel aperture, which should require no aperture correction.  

Photometry from the 2 and 3 pixel apertures showed 
excellent agreement (dispersion of $\approx$ 0.02-0.05 magnitudes) through $\sim$ 13.5-14th (15th) magnitude for the [5.8] and [8] ([3.6] and [4.5]) 
channels.  Beyond these limits, the dispersion in magnitudes increased to $\sim$ 0.15 mag.   
The measured pixel area of $\gtrsim$ 13th magnitude stars is completely enclosed by the 2-pixel aperture radius. 
Therefore, we chose photometry from the 2-pixel aperture for stars fainter than 13th magnitude in all filters
and the 3-pixel aperture for brighter stars.  Catalogs were constructed for both the 0.6s and 10.4s exposures 
and trimmed of sources lying within 5 pixels of the mosaic edges.

To check the photometry, we compared aperture magnitudes with results from SExtractor.
  The differences between SExtractor and IDLPHOT magnitudes 
 are typically small, $\lesssim$ 0.025 mags, through 13th-14th magnitude in 
all filters.  For the long IRAC exposures, stars brighter than $\sim$ 7th-9th magnitude are
 saturated in all bands, with negative pixel counts in the centers of the brightest stars.  Therefore, 
we identified stars with m[IRAC] $\le$ 9 in any band from the 0.6s catalog and replaced their counterparts 
in the 10.4s catalog.  To merge the final IRAC catalog with the 2MASS JHK$_{s}$ catalog, 
we employed a 2" matching radius between 2MASS and IRAC, resulting in 16204, 16477, 9417, and 
6087 matches in the [3.6], [4.5], [5.8], and [8] channels.  

\begin{figure}
\epsscale{0.75}
\centering
%\plotone{ch4dist.ps}
%\plotone{mipsdist.ps}
\plotone{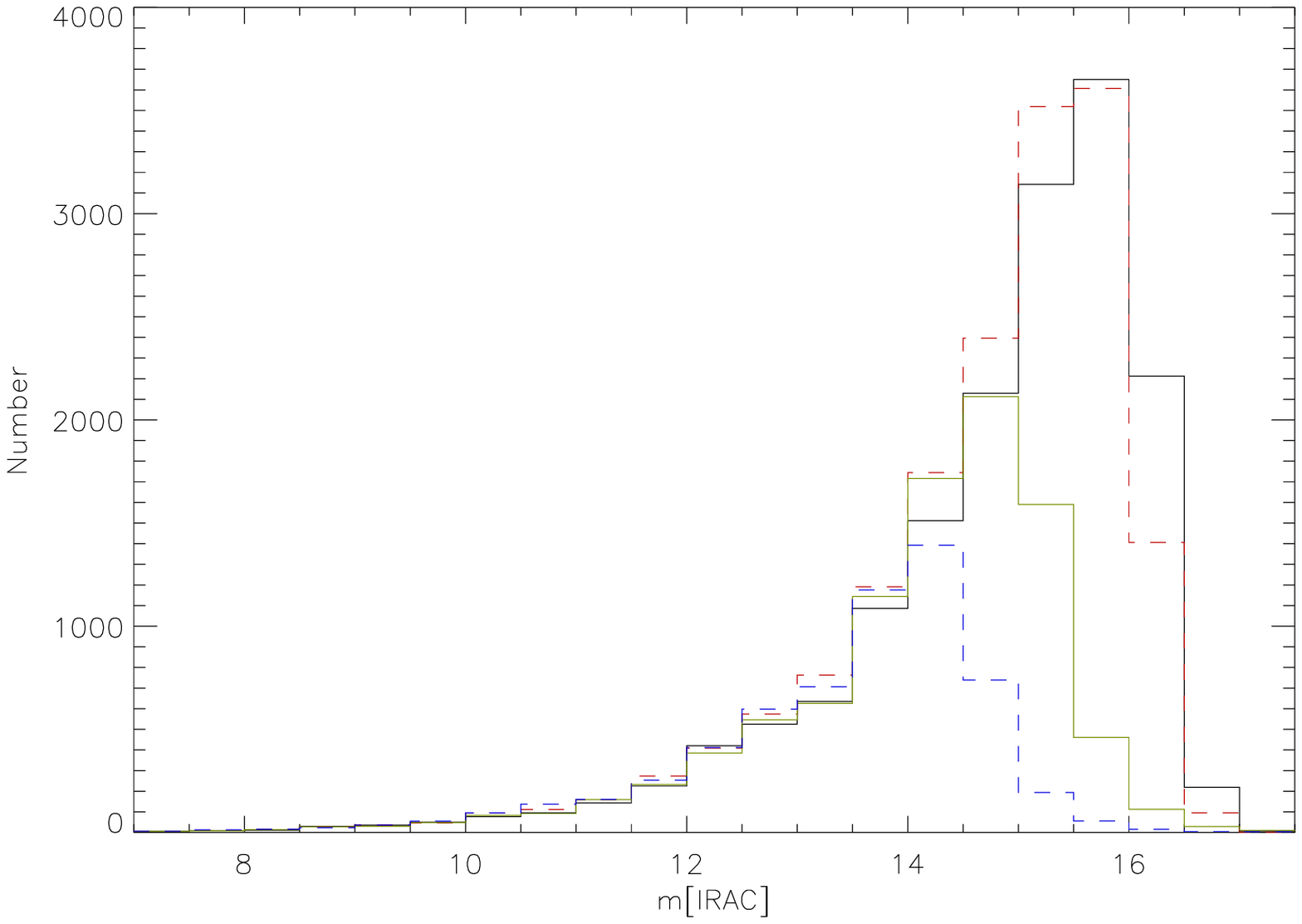}
\plotone{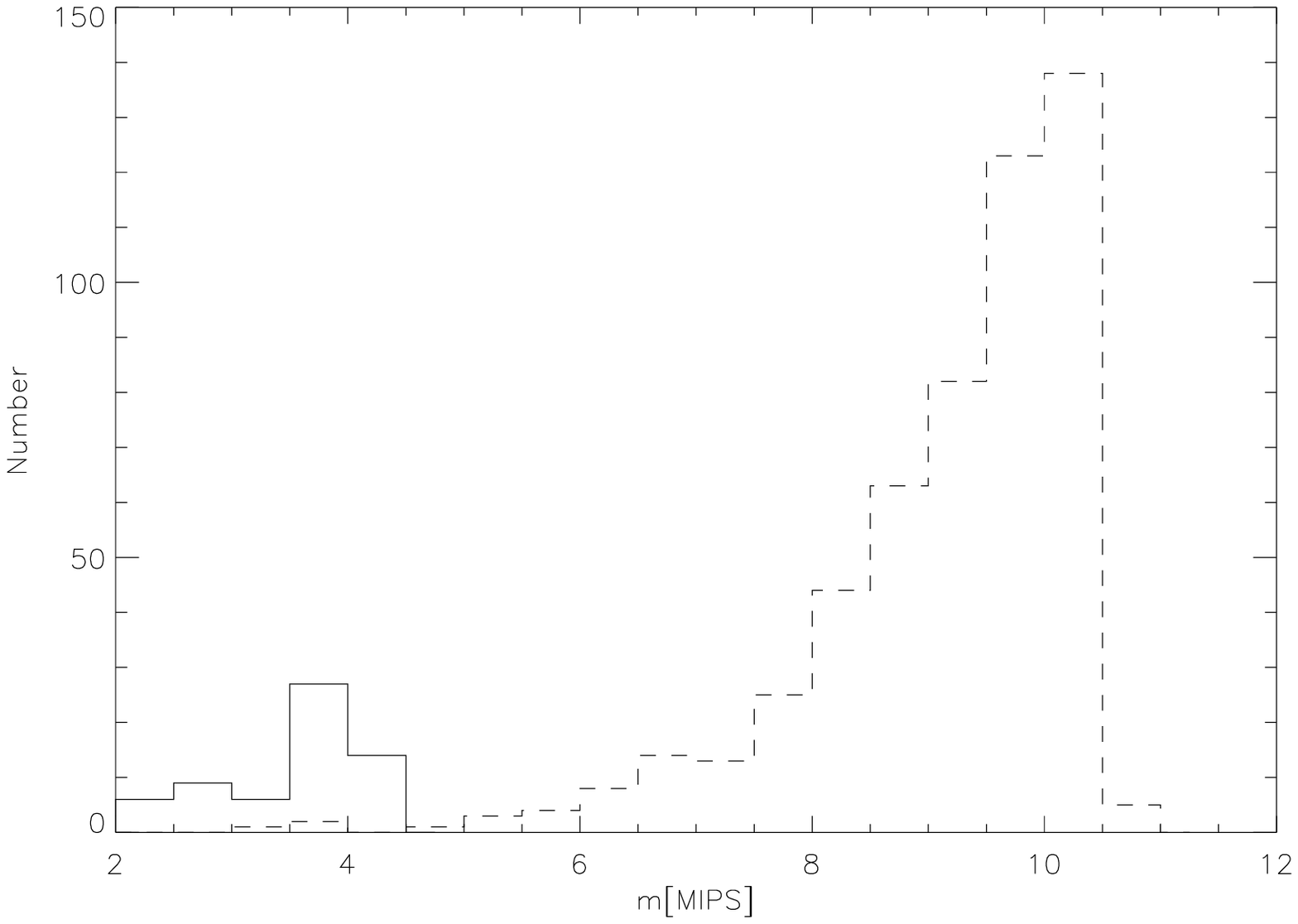}
\caption{(Top) Distribution of IRAC detections with 2MASS counterparts.  From right to left, the distributions 
are for the [3.6], [4.5], [5.8], and [8] channels.  The source counts peak at a magnitude of 16, 16, 15, and 14.5 
for the [3.6], [4.5], [5.8], and [8] channels, respectively.  
(Bottom) Distribution of MIPS 24 $\mu m$ and 70 $\mu m$ detections with 2MASS counterparts.  The number counts in the
24 $\mu m$ filter peak at [24] $\sim$ 10.5 while those in the 70 $\mu m$ filter peak at [70] $\sim$ 4.}
\label{iracdist}
\end{figure}

The photometric errors are calculated from the Poisson error in the source counts, the read noise, 
the Poisson error in the background level, and the uncertainty in the background ($\sigma_{bkgd,rms}$).
Figure \ref{iracdist} (top panel) shows that the source counts in the IRAC bands peak at [3.6] $\sim$ 16, 
[4.5] $\sim$ 16, [5.8] $\sim$ 15, and [8] $\sim$ 14.25-14.5.  
%According to Figure \ref{errordist}, 
The IRAC photometry approaches the 10$\sigma$ limit at [3.6] $\sim$ 16.5 (70.56 $\mu$Jy), 
[4.5] $\sim$ 16.5 (45.13 $\mu$Jy), [5.8] $\sim$ 14.25 (229.46 $\mu$Jy), and [8] $\sim$ 14 (161.09 $\mu$Jy).  

Because the MIPS point-spread function is well characterized, we performed photometry on the 24$\mu m$ and 70$\mu m$ MIPS 
data with pixel response function (PRF) fitting with APEX.  We detect 2,702 sources at [24] and 
438 sources at [70].  To identify candidate stellar sources, we merged the MIPS catalogs with 2MASS  
using matching radii comparable to the median position errors in MIPS: 2" for the 24 $\mu m$ channel and 4" for 
70 $\mu m$\footnote{Setting the matching radii to the median position errors guards against false detections with 
the risk of not identifying true MIPS counterparts whose positional offsets are greater than 2".}.
There are 526 MIPS 24$\mu m$ sources with 
2MASS counterparts and 66 70$\mu m$ sources with 2MASS counterparts.
The bottom panel of Figure \ref{iracdist} shows the distribution of the 24 $\mu m$ and 70 $\mu m$ MIPS magnitudes for these 
sources.  The source counts peak at [24] $\sim$ 10.5 ($\sim$ 460 $\mu$Jy) and [70] ($\sim$ 20 mJy).  
The corresponding peak in the 2MASS J filter for sources detected at [24] is 
J $\sim$ 11.  The (5) 10$\sigma$ limits are [24] $\sim$ (10.5) 9.75 and 
[70] $\sim$ (4) 3.5\footnote{The uncertainties quoted here  
do not include the zero-point uncertainty, which is $\approx$ 4\%.}.  
  Within the IRAC coverage, most MIPS 24 $\mu m$ detections with 2MASS counterparts are 
brighter than J=11 but the 70$\mu m$ data yield no 2MASS matches brighter than J=14.5.
  Most of the 70 $\mu m$ sources are faint in 2MASS (J $\sim$ 16),  
have very red near-IR colors, and are probably extragalactic sources.  
Therefore, we restrict our analysis to the IRAC and MIPS 24 $\mu m$ bands.
Table \ref{fullphot} lists the full photometric catalog for sources detected in at least 
one IRAC channel or the MIPS 24 $\mu m$ band.

%Of the $\sim$ 2,702 MIPS 24 $\mu m$ sources, 526 have 2MASS counterparts.  
Many of the MIPS sources without 2MASS magnitudes 
are galaxies or highly-reddened background stars.
Using the number density of galaxies in the 24 $\mu m$ filter from 
\citet{Pa04}, we expect $\sim$ 1400-1700 extragalactic sources in 
1.4 square degrees.   Typical fluctuations in the galaxy counts 
are $\sim$ 50\% \citep{Pa04}; thus all of the unmatched MIPS sources 
could be galaxies.  Alternatively, highly-reddened background MIPS sources with 
2MASS JHK$_{s}$ fainter than the 10$\sigma$ completion limit may lack near-IR 
detections.  This region of the sky contains clusters with regions of high column 
density interstellar gas (e.g. 
Orion Nebula Cluster), which may have a very large distributed population as well as 
background embedded clusters (e.g. Rosette Nebula).  
Some sources may also have positions determined from the MIPS mosaic 
that lie $>$ 2" away from their 2MASS positions.   

\subsection{ROSAT X-Ray Observations of NGC 2232}
To supplement the Spitzer data, we queried the High 
Energy Astrophysics Science Archive Research Center database (HEASAR; NASA-Goddard 
Space Flight Center\footnote{http://heasarc.gsfc.nasa.gov/W3Browse/}) 
for ROSAT detections from the High-Resolution Imager (HRI) within 2 degrees of NGC 2232.  
Our x-ray source list is drawn from the ROSHRITOTAL, Bright and Faint Source, and Brera Multi-scale Wavelet 
catalogs \citep{Vog99, Pa03}.   Observations consist of two separate pointings 
which cover most of the Spitzer field.  Exposures were taken in 20ks, 48ks or 30ks, 36ks
 pairs.  Two additional bright x-ray sources were observed with 0.148 ks integrations.
The HRI instrument has a resolution of $\sim$ 5" in the center of the field.
%, the pixel scale is much smaller which 
%results in oversampled data.  

The ROSHRITOTAL catalog lists all sources detected by the Standard Analysis  Software System 
in processed public HRI datasets.  This catalog 
lists 299 detections, many of which are multiple detections of the same source.
The Brera catalog consists of sources identified from a wavelet detection algorithm, which  
accurately identifies point sources and extended sources \citep{Pa03}. 
The Brera catalog contains 68 sources within the NGC 2232 
field detected at a signal-to-noise (SNR) $>$ 4.2.  The faint and bright source catalogs add 
35 and 5 detections.

To find 2MASS sources with x-ray counterparts, we combined the ROSAT and 2MASS/Spitzer catalogs 
using a 5" matching radius comparable to the ROSAT positional accuracy.
%Even though the field is more sparsely populated than other 
%10-30 Myr-old clusters (e.g. h and $\chi$ Persei), it is possible that more than one 2MASS source will lie 
%within 5" of an x-ray source.  For now, we keep all possible 2MASS/ROSAT matches and remove unlikely 
%matches later.
Merging the catalogs yields 79 2MASS sources with x-ray counterparts.  The distribution of positional offsets 
between 2MASS and ROSAT coordinates show that most fall within $\sim$ 2"-3" of the 2MASS positions.  
In one case, a single ROSAT source falls within 5" of multiple 2MASS sources.  We chose the source with 
the smaller positional offset (2.8" vs. 3.2"). The source 3.2" from the ROSAT coordinate also has 2MASS colors 
indicating it is a background early-type star.
%In these cases, we 
%selected the 2MASS source with the smallest positional offset\footnote{In two of three cases, selecting the 
%more likely 2MASS counterpart was clear.  The 2MASS/ROSAT offsets were 5.3" vs. 2.4", 5.3" vs. 2.3" and 3.2" vs. 2.8"}.
\subsection{Optical Photometry from \citet{Cl72} and \citet{Ly06}}
We also include optical UBV and BVI photometry from \citet{Cl72} and \citet{Ly06}.  
These data 1) identify a cluster sequence for bright stars and 2) provide a 
longer wavelength baseline.  
  The \citet{Cl72} photometry catalog was downloaded from the WEBDA open 
cluster database.  The catalog contains $\sim$ 43 sources with V $\lesssim$ 12.  
Initial positions in B1950 coordinates were inspected for accuracy. 
Fortunately, all stars are bright enough to have HD numbers and precise positions 
from the Hipparcos and Tycho catalogs.  Based on these measurements, we adjusted the 
coordinates from \citet{Cl72} and precessed them into J2000 coordinates. 

The \citet{Ly06} survey focused on a much smaller area ($\sim$ 15'x15' centered on the 
cluster) but went deeper (to V $\sim$ 20).   The survey identified 1,407 
stars within this region.  We merged the \citet{Cl72} and 
\citet{Ly06} catalogs with the 2MASS and Spitzer catalogs, using a 2" matching radius.

\citet{Cl72} used multiple optical color-magnitude and color-color diagrams to 
identify stars defining a locus consistent with a cluster, to derive 
the interstellar reddening to each star, and to estimate the age of the 
cluster.  From these methods, \citet{Cl72} identifies at least 19 early-type 
stars consistent with cluster membership, computes 
a mean reddening of $\bar{E(B-V)}$ $\sim$ 0.01, and estimates a nuclear 
age of $\sim$ 20 Myr.  \citet{Ly06} show that by 'subtracting' 
(in V/V-I space) a field population from that containing the cluster, 
potential cluster members remain which define a clear \textit{empirical} isochrone for the cluster.
They compute an age of $\sim$ 25-30 Myr and a slightly larger typical reddening of 
E(B-V) $\sim$ 0.07.  Our motivation for including these data is to 
compare the colors of candidate cluster members in the ROSAT and 
Spitzer samples with known cluster 
members from \citet{Cl72} and the locus of cluster members from 
\citet{Ly06}.
\subsection{Optical Spectra}
To explore the NGC 2232 stellar population in more detail, we downloaded 
spectra of stars in the NGC 2232 field from the FAST data archive housed 
at the Telescope Data Center at the Smithsonian Astrophysical Observatory.  Spectra were 
taken by the FAST instrument on the 1.5m Tillinghast telescope on January 31, 2001 as a 
part of the FAST combo queue observing program (PI: Brian Patten).  We also 
took FAST spectra of ID 18012 on April 14, 2008 (PI: Thayne Currie).  The 
data were taken with the 270 g mm$^{-1}$ grating, yielding a wavelength 
coverage of 4000-9000 \AA\ with 3 \AA\ resolution \citep{Fa98}.  We identify 38 separate 
sources in the 2001 data archive, 36 of which have high enough signal-to-noise to analyze. 

To derive quantitative spectral types, we first compared equivalent widths of the H I Balmer lines, 
CaII H and K, G band, Na 5890, and TiO bands with spectroscopic standards from \citet{Ja84}.  In a 
second iteration, we computed the spectral types from the semi-automatic spectral typing 
code SPTCLASS \citep{He04} and from manually measuring the equivalent width of indices with 
high correlation coefficients from smoothed spectra.  
Spectral types derived manually and with SPTCLASS agree within 0.5-1 subclasses. Table \ref{fullspectra} lists 
the spectral types of these stars.  Many stars originally observed by \citet{Cl72} were 
reobserved here.  Agreement between spectral types derived here and those from the literature 
are good to within $\sim$ 2-3 subclasses.

\begin{figure}
%\plotone{ngc2232_id18566.ps}
%\plotone{ngc2232_id18622.ps}
\plotone{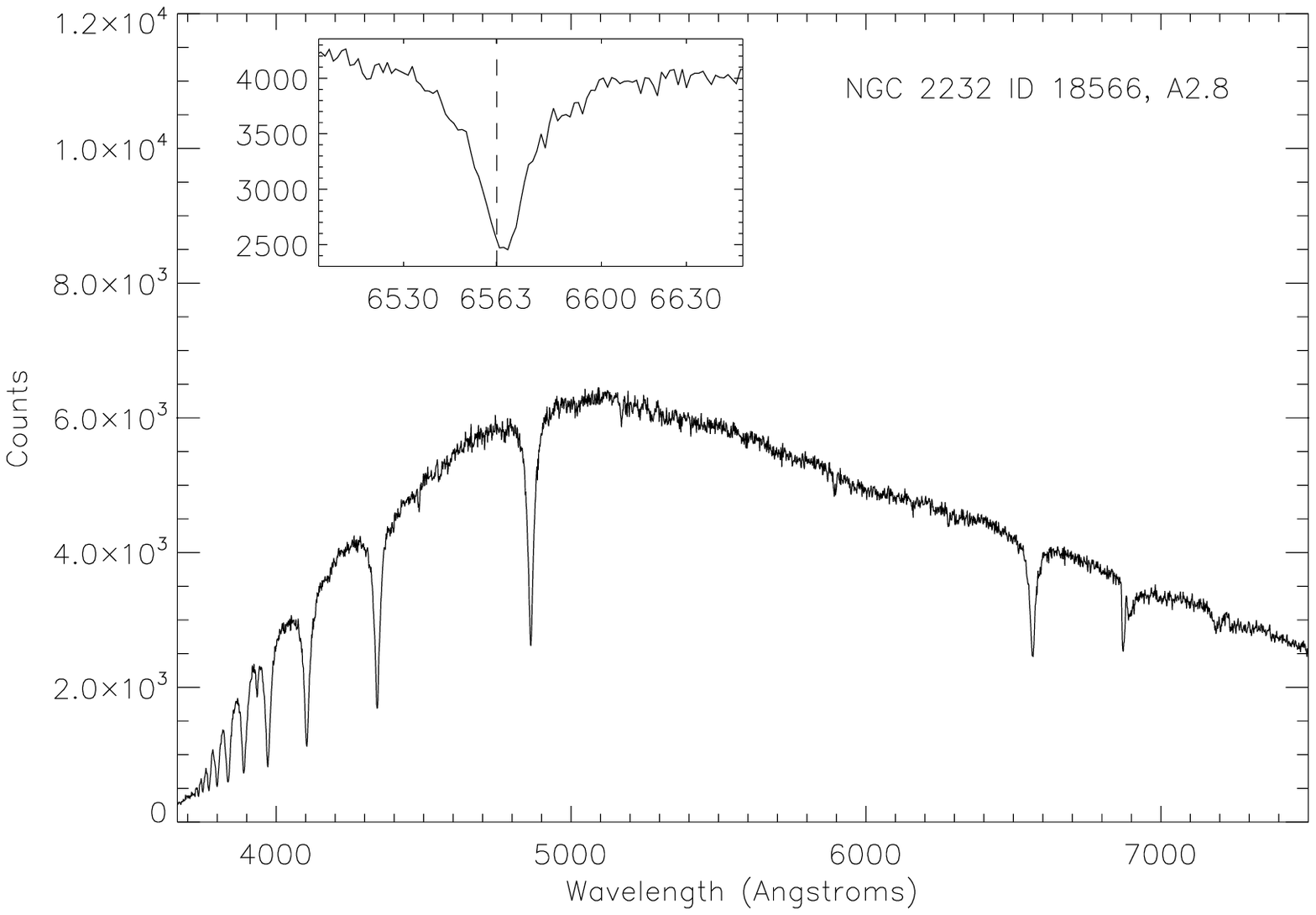}
\plotone{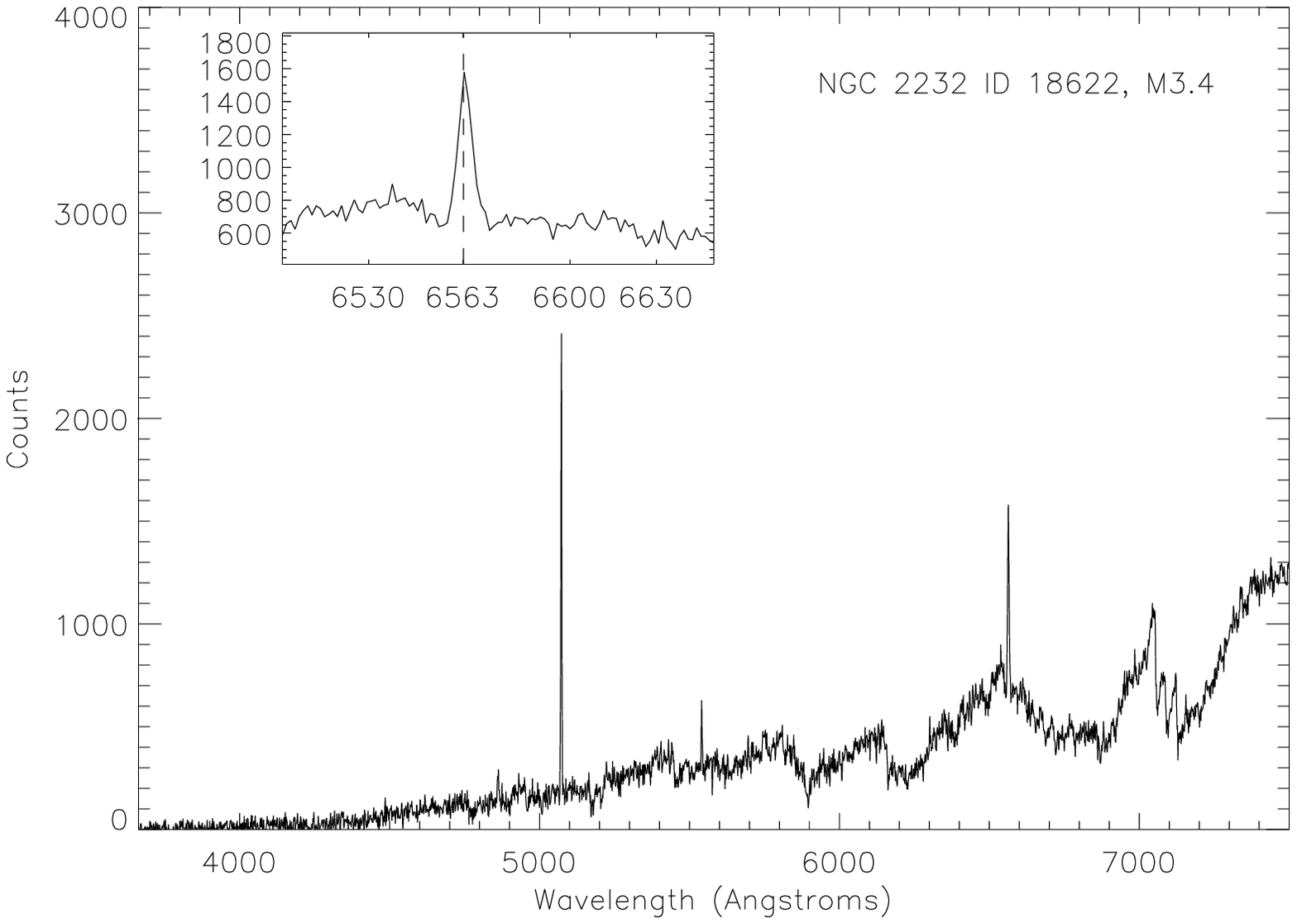}
\caption{Spectra of NGC 2232 cluster members ID 18566 and ID 18662.  Both sources have 
very red K$_{s}$-[5.8] and K$_{s}$ - [8] colors.  The insert show the H$_{\alpha}$ line profile.}
\label{fastspec}
\end{figure}

We show two spectra in Figure \ref{fastspec}.  Both have red K$_{s}$-[8] 
colors but very different spectral types.  The top panel shows ID 18566,  
 listed as a B9.5 star in \citet{Lm74} and classified as A2.8 here.  The individual 
indices used to derive the spectral type (e.g. CaII) yield estimates ranging 
between A0 and A3.  Spectral types estimated from the Balmer lines yield A2-A3.  
ID 18622, shown in the bottom panel, is a Li-poor M3.4 star with 
H$_{\alpha}$ emission (EW(H$_{\alpha}$) $\sim$ 8 \AA).  Because 
this source has a ROSAT detection and the line is single-peaked and 
rather narrow (FWHM $<$ 6 km s$^{-1}$), the emission is likely due to chromospheric activity.  
Several other stars, all with ROSAT detections, show similar levels of H$_{\alpha}$ emission.
  Thus, we find no strong evidence for a population of late-type stars with 
strong H$_{\alpha}$ emission indicative of accretion (EW(H$_{\alpha}$) $\gtrsim$ 10\AA; \citealt{Wb03}).
%The properties of all stars with spectra are listed in 
%Table \ref{fullspectra}.

\subsection{IRAC and MIPS Colors of Sources with 2MASS counterparts}
To make a first assessment of a possible disk population among the
 Spitzer sources, we consider several color-color and color-magnitude
diagrams constructed from 2MASS, IRAC, and MIPS data. Our goal is to
isolate stars with excess IR emission from stars with photospheric
colors.  For the derived age ($\sim$ 25 Myr), distance (320--360 pc),
and reddening ($A_V \lesssim$ 0.2 mag), early to mid-M type stars in NGC 2232 have
J $\sim$ 14--14.5 and intrinsic 2MASS/IRAC/MIPS colors $\le$ 0.5. For
the derived sensitivities of the IRAC data, we expect all cluster stars
to lie above the 10$\sigma$ limits for all IRAC bands.  Thus, we focus our analysis on
sources with 2MASS J $\le$ 14.5.

NGC 2232 is a sparse cluster \citep{Cl72, Ly06}. For a normal mass function,
the observed population of massive cluster stars implies a total cluster
population of $\sim$ 500 stars. Because stars with IRAC excess are
rare in young clusters with ages $\sim$ 10--30 Myr \citep[$\lesssim$ 5--10\%;][]{Cu07a, Go07},
 we expect $\lesssim$ 25--50 stars with IR excess.
There are $\sim$ 18,000 2MASS sources with IRAC counterparts in this
region. Compared to massive clusters in relatively low background regions
\citep[e.g. h and $\chi$ Per;][]{Cu07a}, identifying an obvious main sequence
for cluster stars is challenging. Therefore, we concentrate on color-color and
color-magnitude diagrams that allow us to isolate (i) background/cluster
stars with a well-defined dispersion around the sequence expected for
stellar photospheres and (ii) a handful of stars with IR excesses at least
3$\sigma$ larger than the measured dispersion in photospheric colors.

\begin{figure}
\epsscale{1}
\centering
%\plottwo{jkk41.ps}{k3k4.ps}
\plottwo{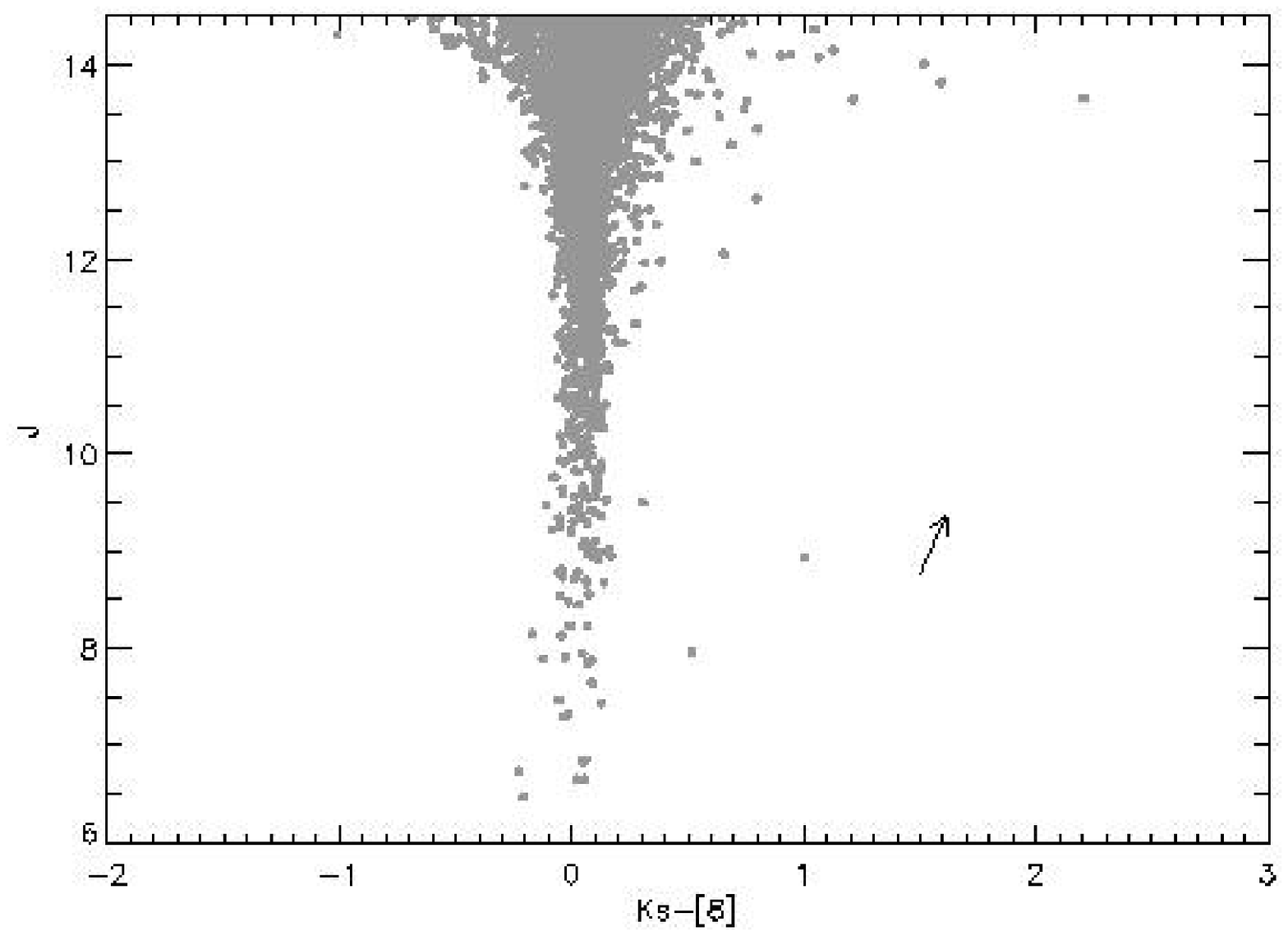}{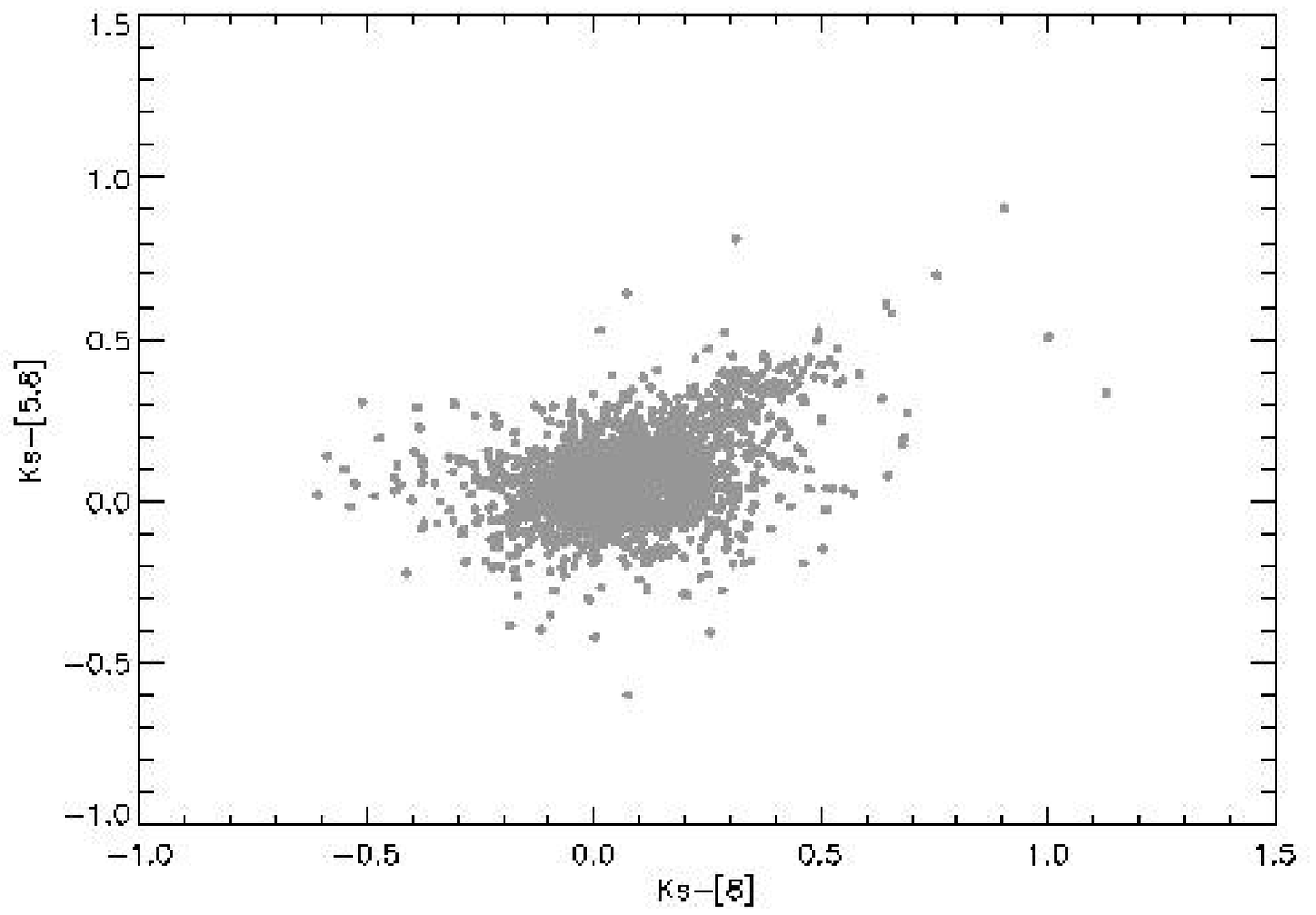}
\centering
%\plottwo{jkk24.ps}{k4k24.ps}
\plottwo{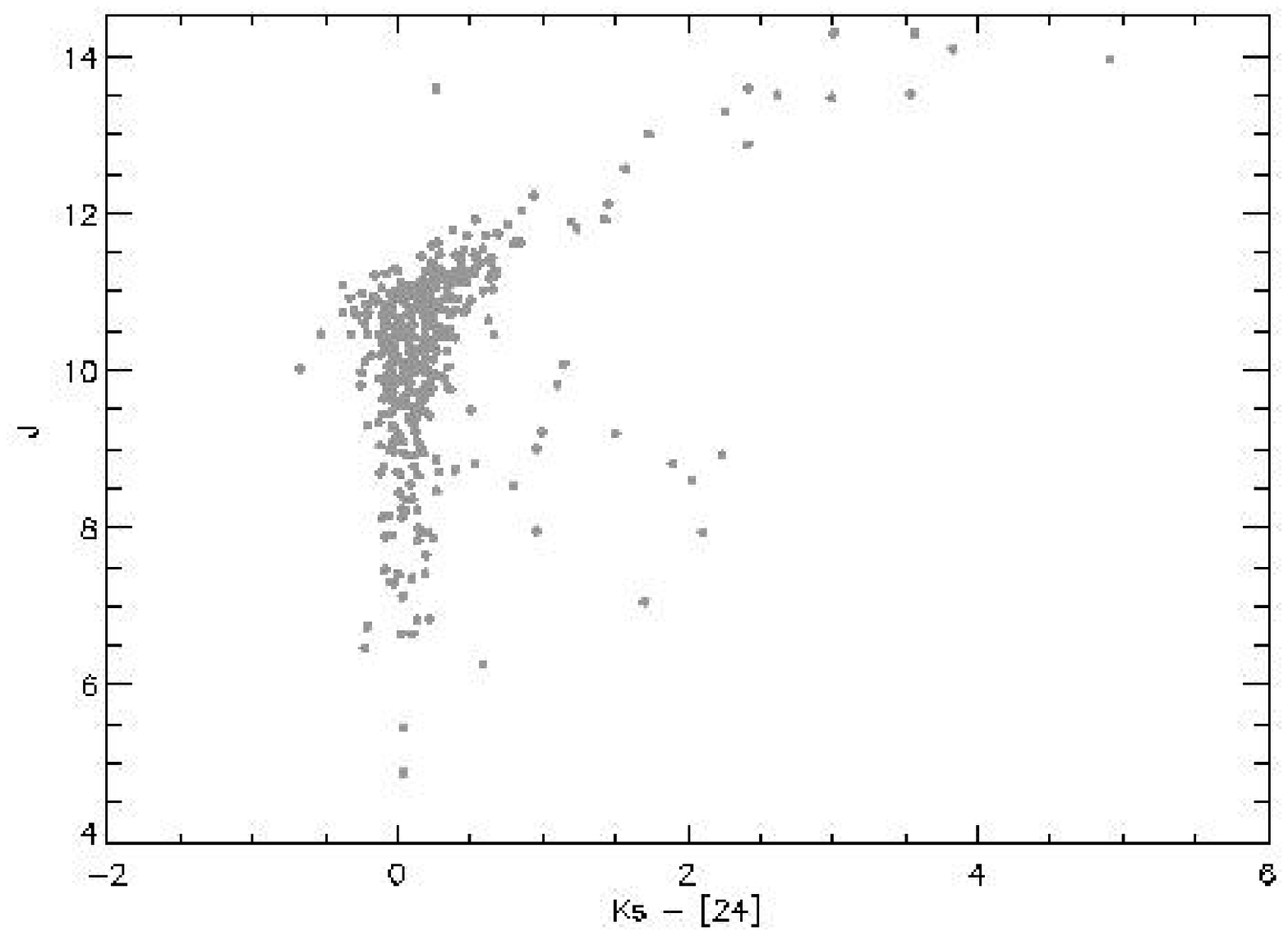}{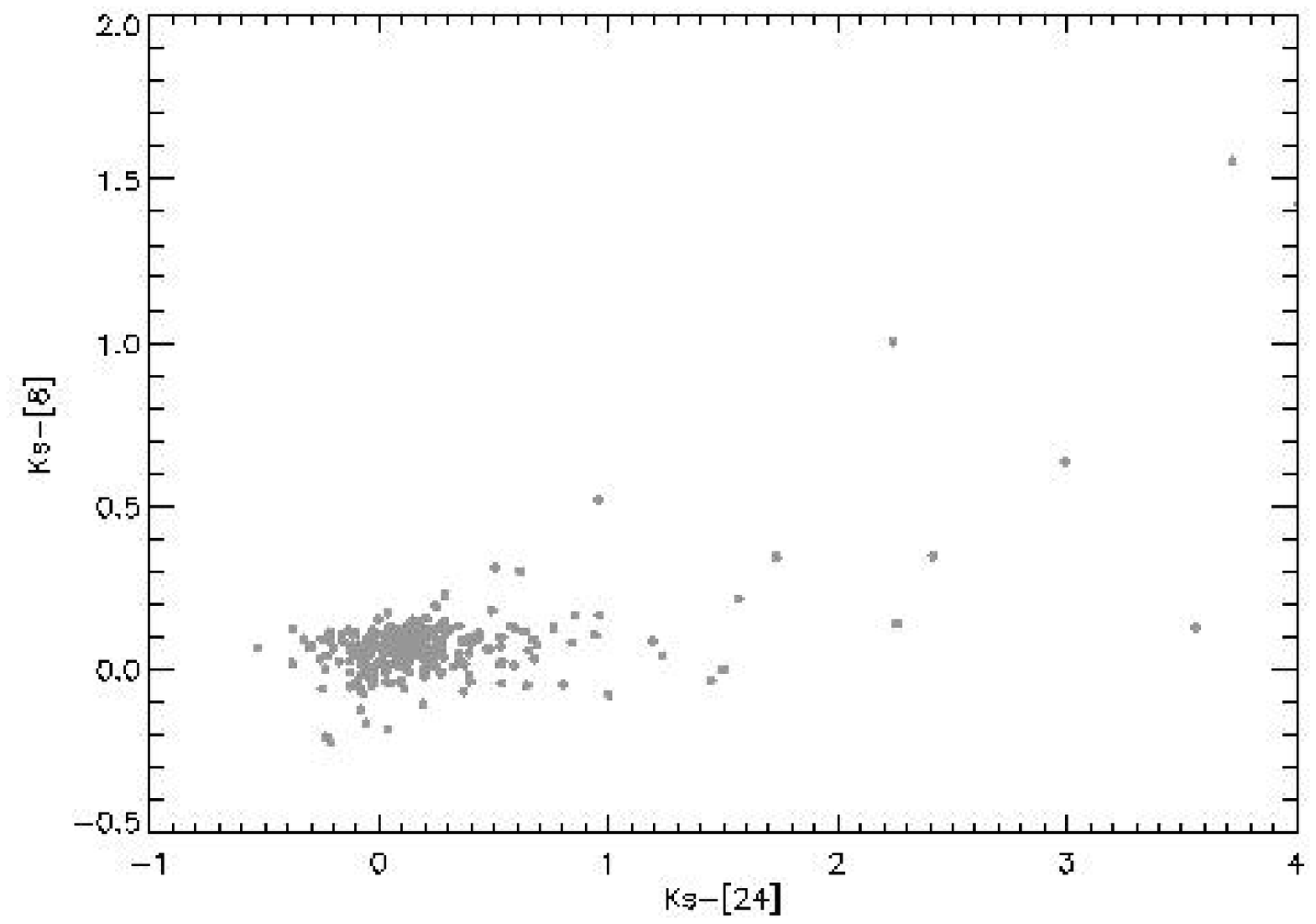}
\caption{Color-magnitude and color-color diagrams for NGC 2232.  
Top left: J vs. K$_{s}$ -[8] color-magnitude 
diagram. Top right: K$_{s}$-[5.8]/K$_{s}$-[8] color-color diagram.  Bottom left: 
J vs. K$_{s}$-[24] color-magnitude diagram.  Bottom right: 
K$_{s}$-[8]/K$_{s}$-[24] color-color diagram.  The reddening vector is 
derived from the reddening laws derived by \citet{In05}} 
\label{what24}
\end{figure}

Figure \ref{what24} shows the J vs. K$_{s}$-[8] color-magnitude diagram and 
K$_{s}$-[5.8] vs. K$_{s}$-[8] color-color diagram (top left and top-right panels, respectively).  
The main distribution of 
K$_{s}$-[8] colors defines a locus that widens from 0.1 mag at J=6 to $\sim$ 0.4 mag 
at J=14.5.  The flaring of this locus to $\sim$ 0.4 mag near the faint 
limit results from a combination of a) larger photometric errors, b) a much wider range 
of intrinsic colors for both cluster and background stars, and c) a wider range of 
reddening for background stars.  Several sources have K$_{s}$-[8] colors far 
redder than the main distribution.  There are no stars with 
similarly blue K$_{s}$-[8] colors.  The main distribution of K$_{s}$-[5.8]/K$_{s}$-[8] colors 
extends from K$_{s}$-[5.8] = -0.1 to 0.2 and K$_{s}$-[8] = -0.2 to 0.3 (Figure \ref{what24}, top-right 
panel).  Few sources with very red K$_{s}$-[5.8] colors have blue K$_{s}$-[8] colors.  Most sources 
with red K$_{s}$-[5.8] colors also have red K$_{s}$-[8] colors.  The sources with 
very red K$_{s}$-[8] colors could plausibly be cluster members with warm 
circumstellar dust.

Many sources in Figure \ref{what24} (bottom-left panel) 
have K$_{s}$-[24] colors greater than $\sim$ 0.5 and as large 
as 5.  These excess sources lie in two groups.  The first group 
has J $\sim$ 6-10 with K$_{s}$-[24] $\sim$ 0.5 to 2.25.  If these stars are 25 Myr-old 
cluster members, they are massive stars with B0--F5 spectral types.  The second group has 
J $\sim$ 12-14 with K$_{s}$-[24] $\sim$ 2-5.  If these sources are 
cluster members, they are low-mass stars with K5-M6 spectral types.   

In addition to stars with IR excess, the NGC 2232 field has 
a clear sequence of stars with 
colors of normal main sequence stars (Figure \ref{what24}, bottom left panel).  
This sequence begins at J, K$_{s}$-[24] $\sim$ 5,0 and extends to J,K$_{s}$ $\sim$ 
10,-0.25 to 0.25.  
At J $\sim$ 10-11, the sequence broadens considerably and then develops 
a red tail of sources with very red K$_{s}$-[24] colors.  With measured 
5--10$\sigma$ sensitivity limits of 9.75--10.5 (\S 2.1), the MIPS images 
cannot detect stellar photospheres with J $\gtrsim$ 11.  Thus, many K and M stars 
detected by MIPS are probably 24 $\mu m$ excess sources.

Figure \ref{what24} (bottom-right panel) shows that many 
of the stars with photospheric K$_{s}$-[8] colors have 
excess emission at 24 $\mu m$.  The main distribution of colors ranges from K$_{s}$-[8] 
$\sim$ -0.1 to 0.1 and K$_{s}$-[24] $\sim$ -0.2 to 0.3.  We find no evidence for sources 
with red K$_{s}$-[8] colors but blue K$_{s}$-[24] colors.  About six sources have 
K$_{s}$-[8] $\gtrsim$ 0.3 and K$_{s}$-[24] $\gtrsim$ 1.  These stars have 
 substantial warm circumstellar dust, have intrinisically red photospheres (see \S 4.1), 
or are heavily reddened.

Figure \ref{what24} demonstrates that some stars in the direction of NGC 2232 have
robust detections of IR excess emission. The number of IR excess stars,
$\sim$ 10--20, is comparable to our predicted number of $\sim$ 25--50.
Thus, some of these excess stars are plausibly associated with the cluster. To
identify {\it bone fide} cluster stars and candidate cluster stars among this group, we consider
 several methods using ROSAT detections, color-magnitude
diagrams, and proper motions.  Once we have a catalog of confirmed/candidate members, 
we analyze the disk population in \S4.

\section{Identifying Cluster Members}
We now use ROSAT x-ray detections, optical/IR colors, and proper motion to 
identify candidate cluster members.  The positions of ROSAT sources on
  optical/IR color-magnitude diagrams 
follow a locus consistent with previously identified cluster members 
from \citet{Cl72} and \citet{Ly06}.  This 
locus allows us to identify additional candidate members from optical/IR 
color-magnitude diagrams and from proper motion data.
\subsection{Catalog of Cluster Members based on X-Ray Active Stars and \citet{Cl72}}
To isolate cluster members from background stars, we first consider the
ROSAT detections. Young late-type stars are X-ray active \citep[e.g.][]{Pr05}. For
stars with ages $\sim$ 10--15 Myr, the typical X-ray flux is $\sim$ 100-1000 times
larger than fluxes for stars of similar mass with ages $\gtrsim$ 100 Myr
\citep[Chandra Observations of Orion, the Pleiades, and h Persei;][]{Pr05, St05, Ces08}.
%  Along the line-of-sight to NGC 2232, background
%stars have typical distances of $\sim$ 500--1000 pc. 
%Thus, X-ray active
%stars in the cluster are $\sim$ 2-4nn times brighter than the background
%population.

Figure \ref{jjmhxray} shows the ROSAT detections and the Claria (1972) cluster members
on a J/J-Ks color-magnitude diagram. These stars define a clear locus
consistent with the isochrone expected for a 25 Myr-old cluster. Bright
cluster members from Claria (1972) trace the isochrone well for J = 6--10.
For J $\ge$ 10, the ROSAT detections clearly indicate the location 
of the cluster, isolating it from brighter (fainter) foreground red dwarfs (background 
red giants) with J $\lesssim$ 10 and J-K$_{s}$ $\gtrsim$ 0.6 and the large 
background population with J $\gtrsim$ 12 and J-K$_{s}$ $\sim$ 0--0.6.
%For J = 10--12 and J-Ks = 0.2--0.6, $\sim$ nn\% of
%stars bounded by the isochrone are ROSAT sources. For fainter stars with
%J = 12--14 and J-Ks = 0.7--0.9, $\sim$ nn\% of stars within the isochrone
%are ROSAT soruces.

ROSAT sources outside the 25 Myr isochrone are mostly faint background stars.
Stars with J = 14--14.5 and J-K$_{s}$ = 0.4--0.8 have bluer J-K$_{s}$ colors than
cluster stars at the same J. The apparent magnitudes of these stars are
consistent with young stars at distances $\sim$ 500--1000 pc and thus might
be members of more distant clusters or star-forming regions.

\subsection{Candidate Members Identified from \cite{Ly06} and 2MASS photometry}
The ROSAT detections suggest that we can use 
optical or optical/IR color-magnitude diagrams to identify candidate members.  
To test this possibility, we 
compare the V/V-J colors for cluster members identified by ROSAT and \citet{Cl72} with other stars in 
the field.  We use V-band photometry from \citet{Ly06} and photometry from \citet{Cl72} for 
sources outside Lyra's coverage.

Figure \ref{vvmjxray} shows that the ROSAT and Claria stars  
 define a tight locus in V/V-J space.  
%Noticeably absent are bright stars with 
%very red colors, as seen in Figure \ref{jjmhxray}.  This absence is likely due to 
%the smaller spatial coverage of \citet{Ly06}, who focus on regions where most of the 
%known cluster members are located.  
From these stars, we define an empirical isochrone for the 
cluster (solid line) with an upper limit to include binaries and a lower limit to allow for
 slight differences in reddening/distance/age.  This empirical isochrone is identical 
to the \citet{Si00} theoretical isochrone for 25 Myr-old $\gtrsim$ 0.5 M$_{\odot}$ (M0-M1) stars, 
 brighter (bluer) than V $\sim$ 16 (V-J $\sim$ 3.2).  To match the empirical 
locus of x-ray active stars\footnote{This shift allows us to match   
 the luminosity and colors of stars with masses $\lesssim$ 0.5-0.6 M$_{\odot}$ 
and to account for uncertainties in low-temperature opacities that 
affect stellar atmosphere models \citep{Ba98}.}, 
we shift the isochrone redward by $\sim$ 0.5 magnitudes for V $\ge$ 16.

\begin{figure}
%\plotone{jjmkxray.ps}
\plotone{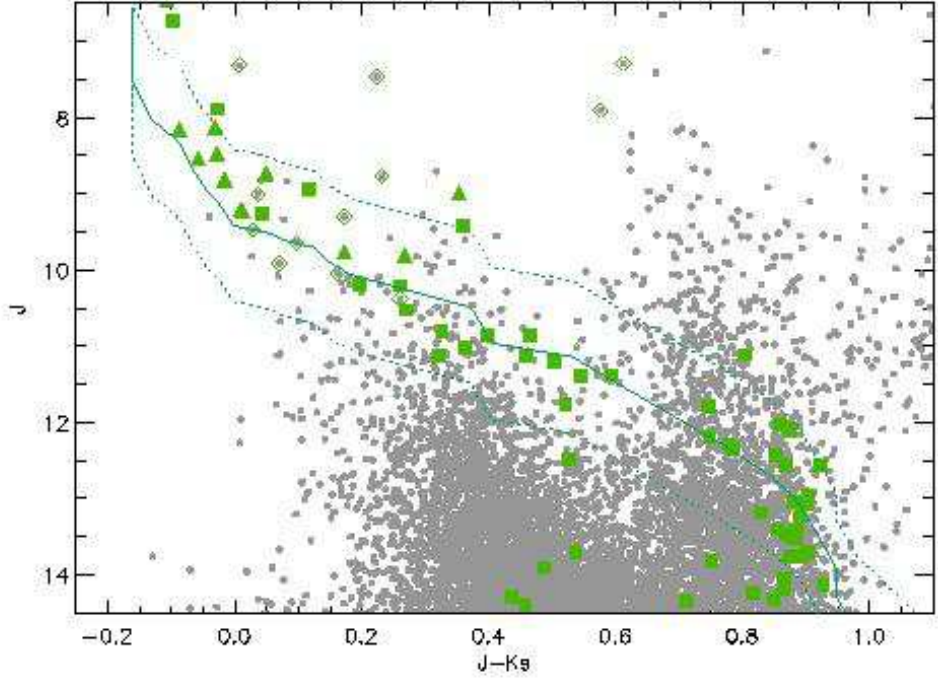}
%\label{jjmhxray}
\caption{J/J-K$_{s}$ color-magnitude diagram identifying probable cluster members in NGC 2232.  
Green triangles denote x-ray quiet cluster members identified by \citet{Cl72}; 
green squares identify x-ray active sources.  Stars rejected as non members 
by \citet{Cl72} are shown as grey dots surrounded by open diamonds.  
  The sequence of x-ray quiet cluster members and x-ray active sources
tracks a clear locus in J/J-K$_{s}$ consistent with a $\approx$ 25 Myr-old cluster.   
The solid and dotted lines show the 25 Myr \citet{Si00} isochrone with upper and 
lower bounds described in the text. 
X-ray active sources below this locus are probably young background stars.}
\label{jjmhxray}
\end{figure}

\begin{figure}
%\plotone{vvmjxray2.ps}
\plotone{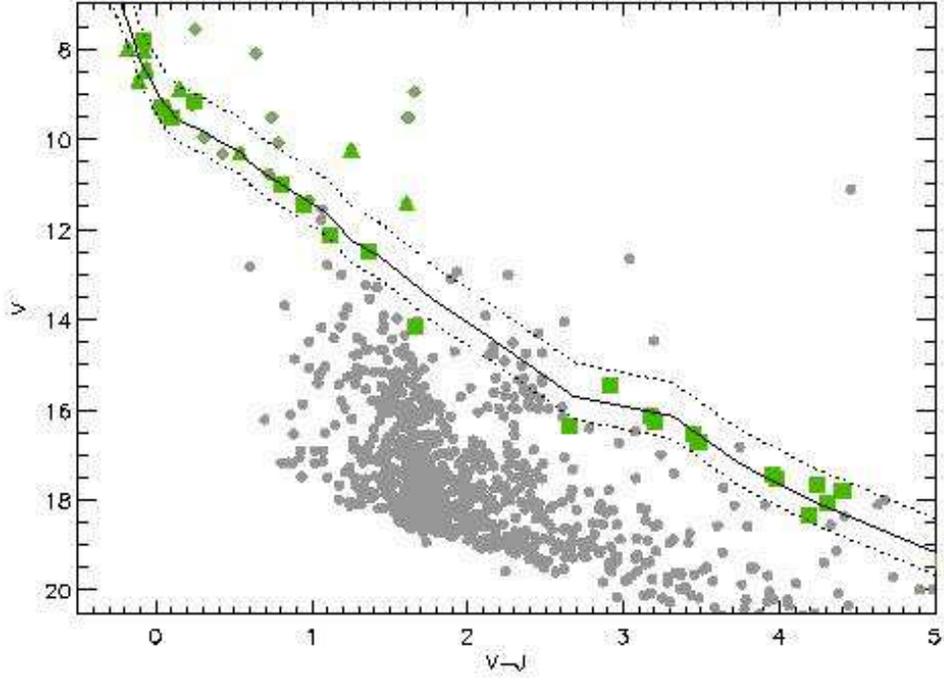}
\caption{V/V-J color-magnitude diagram showing the positions of x-ray active cluster members 
and bright members compared with other stars.  Symbols and lines as in 
Figure \ref{jjmhxray}.  The cluster members define a clear locus 
in V/V-J.}
\label{vvmjxray}
\end{figure}

Compared to Figure \ref{jjmhxray}, Figure \ref{vvmjxray} has many fewer bright red stars lying above
the isochrone, which are likely foreground late-type stars or 
background giants.  This difference probably results from the smaller spatial
coverage of \citet{Ly06}, who focus on the cluster center. To test
this possibility, we compare the relative areas of the \citet{Ly06}
and the IRAC/MIPS surveys. The IRACS/MIPS coverage is $\sim$ 22.5 times
that of the \citet{Ly06} data. Thus, we expect 22.5 times as many
bright, red stars in Figure \ref{jjmhxray} as in Figure \ref{vvmjxray}.  
 The observed ratio of $\sim$ 125/4 $\approx$ 31 is reasonably close to our expectation.

\begin{figure}
%\plottwo{jjmkt.ps}{jjmht.ps}
\plottwo{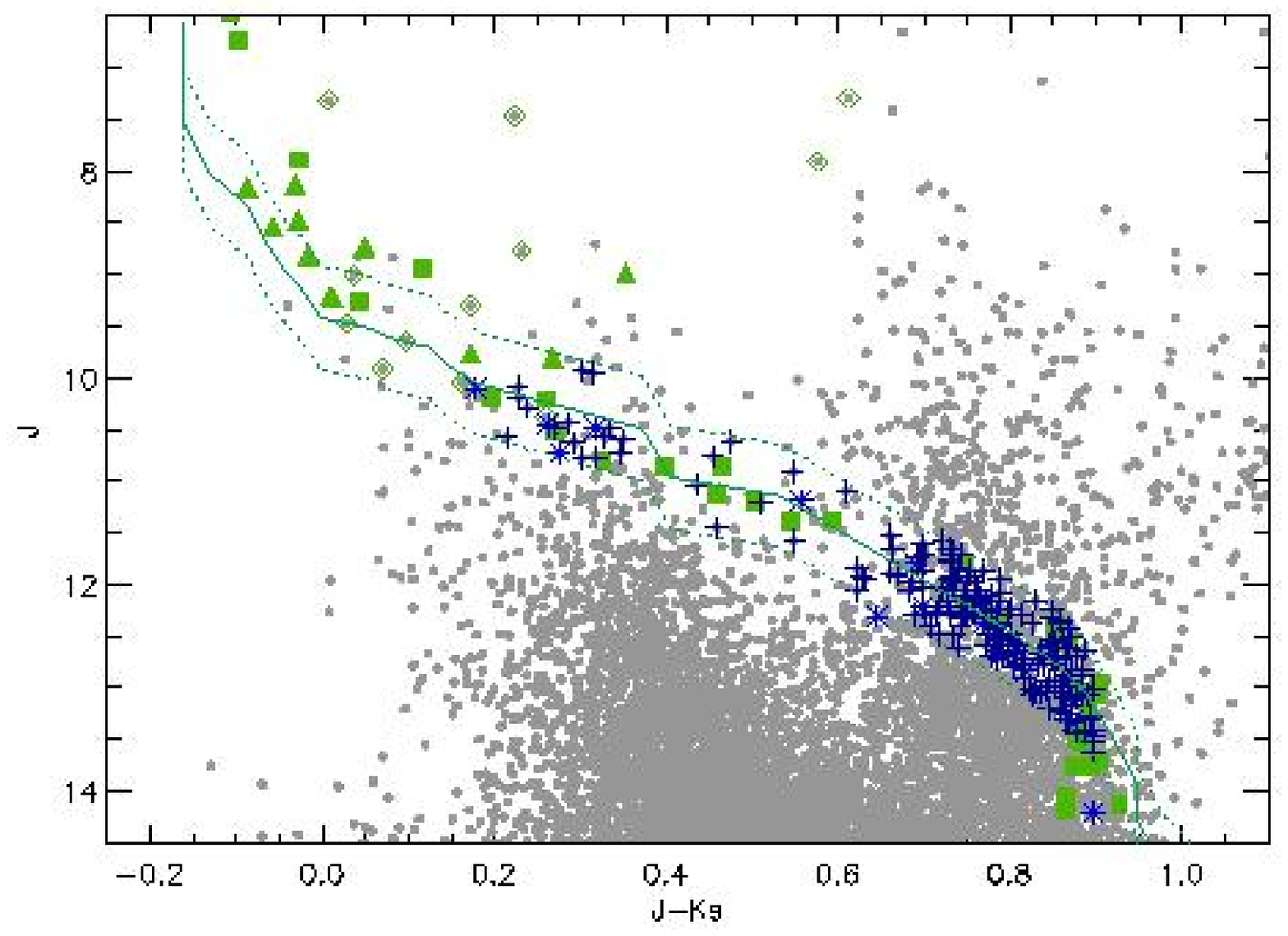}{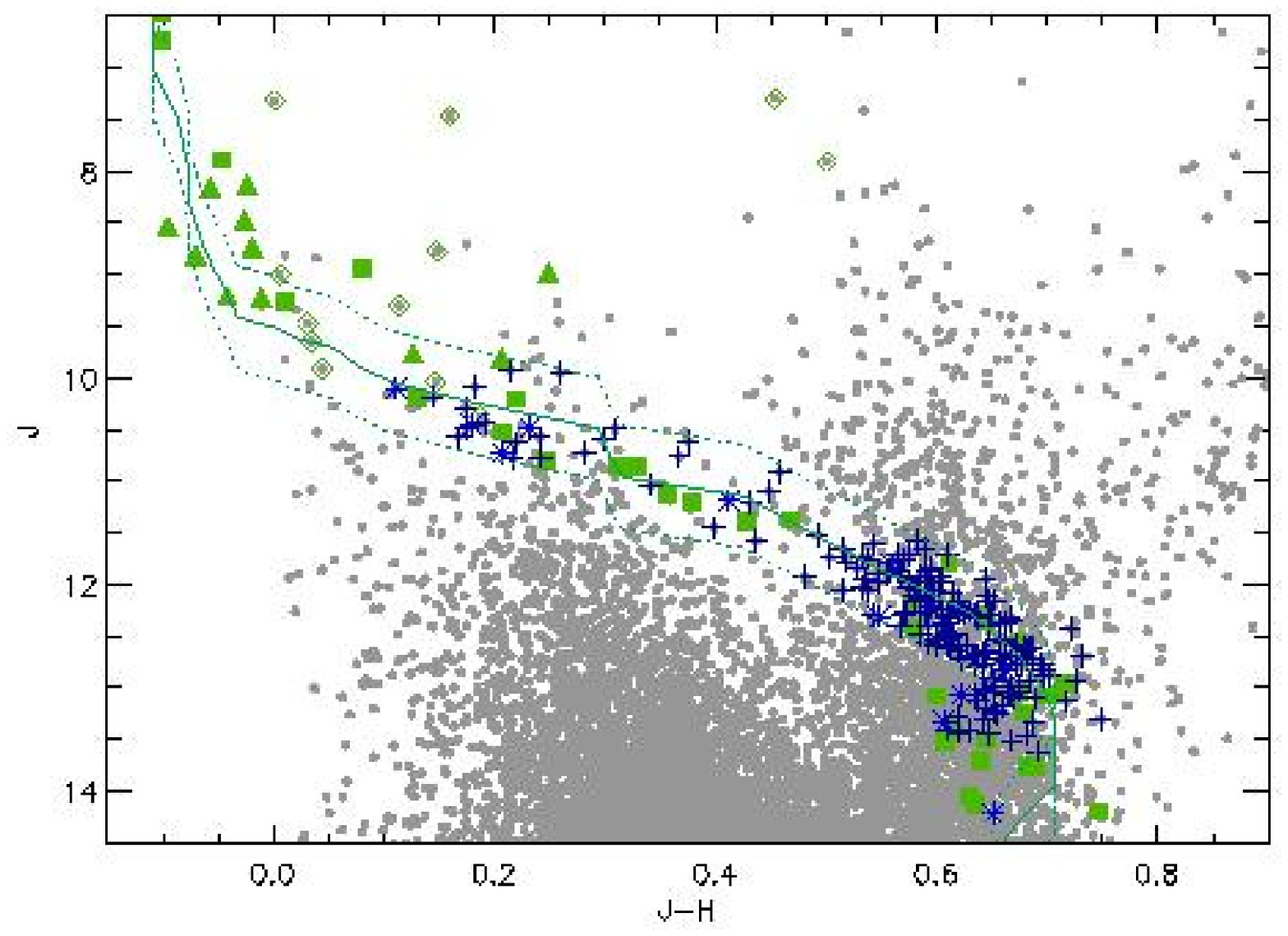}
\caption{J/J-H and J/J-K$_{s}$ color-magnitude diagrams showing positions of confirmed and 
candidate cluster members.  Sources from the \citet{Cl72} catalog (triangles) and 
 ROSAT detections (squares) are 'confirmed' members.  Candidate members were selected 
based on V/V-J (asterisks) and J/J-H and J/J-K$_{s}$ (crosses) color-magnitude diagrams.}
\label{jjmhk}
\end{figure}

Because cluster members define a narrow locus in V/V-J space, we identify 
candidate cluster members lying along this locus between the 
upper and lower bounds (+0.5, -0.75 mags; dotted lines) in Figure \ref{vvmjxray}.
We also identify stars outside the \citet{Ly06} coverage but overlapping the area enclosed by 
confirmed members with positions on the J/J-H and J/J-K$_{s}$ 
color-magnitude diagrams consistent with cluster membership (Figure \ref{jjmhk}).  
Candidate members are drawn from a more narrowly defined locus ($\pm$ 0.5 mags).  
We make additional passes through J/J-K$_{s}$ and J/J-H color space and reject all 
candidate members that clearly lie outside the plausible color-magnitude range in either diagram.
%brighter than J=13.5 that lie outside $\ge$ 0.5 mags outside the 
%isochrone.  The isochrone in J/J-H 
% This method yields an additional 151 'candidate' members.
\subsection{Additional Selection Criteria and the Final Membership List}
To provide a final check on our membership catalog, we compare the proper motions of previously identified 
members and other stars on the field.  The median proper motion of the cluster is derived from 
Tycho observations of the early-type members in the \citet{Cl72} catalog.  These stars have a mean 
proper motion of $\Delta_{ra, dec}$ $\sim$ -5.33 $\pm$ 1.27, -2.68 $\pm$ 1.34 mas yr$^{-1}$.  We identify 
all stars within 2$\sigma$ (of their measurement errors) as having a proper motion consistent 
with cluster membership.  All members in the \citet{Cl72} catalog meet this criteria.  Using 
the Tycho catalog, Guide Star Catalog, and Henry Draper catalog, we identify field stars 
and potential cluster members.  We reject three bright stars (IDs 2162, 15449, and 18442) and add two 
stars (IDs 5494 and 5625) whose proper motions are within \textit{1$\sigma$} of the mean cluster 
motion.  We also add one source from our spectroscopic catalog, ID 10258, 
which has J/J-K$_{s}$ and J/J-H positions barely outside our photometric membership criteria. 
This star has a J magnitude and colors very similar to other K5-M0 cluster members.

 In total, we identify 11 members from the \citet{Cl72} catalog, 37 members from 
x-ray activity, and two members from proper motion.  To this sample we add, 15 candidate members from V/V-J colors and 
176 candidate members from J/J-K$_{s}$ and J/J-H diagrams, 
bringing the total to 239 confirmed/candidate members.  Of these candidates, 209 have IRAC and/or MIPS photometry.
Thus, the cleaned list of x-ray active and previously identified members  
are photometrically (in two color-magnitude diagrams) and, in some cases, astrometrically confirmed 
as members.  The list of candidate members are all photometrically consistent with membership.  
Of confirmed/candidate cluster members, 32 have optical spectra listed in Table \ref{fullspectra}.
Table \ref{fullmember} lists all the confirmed/candidate members with Spitzer photometry.

The candidate members identified from color-magnitude diagrams 
 contain a potentially large population of field stars, 
especially for stars with near-IR colors  
consistent with foreground M stars and background M giants (J-H $\sim$ 0.6-0.7, J-K$_{s}$ $\sim$ 0.75-0.85).
  To better constrain their 
status, these sources need additional photometric and spectroscopic data.  In the analysis 
section, we retain these sources with the qualifier that some are probably not cluster members.
Thus, though candidate members identified from optical and near-IR color-magnitude diagrams 
may include NGC 2232 sources with IR excess, including all candidate members identified 
from these diagrams necessarily 
introduces uncertainties in determining the frequency of disks for stars later than $\sim$ F0-G0.
The samples of confirmed and candidate cluster members are, in order from most to least 
robust, the x-ray active sources and the \citet{Cl72} catalog, sources selected based on V/V-J 
and proper motion, and sources selected based on J/J-K$_{s}$.
\section{Analysis of the NGC 2232 Disk Population}
\subsection{IRAC/MIPS colors of Cluster Members}
\begin{figure}
%\plottwo{jmk3.ps}{jmk4.ps}
\plottwo{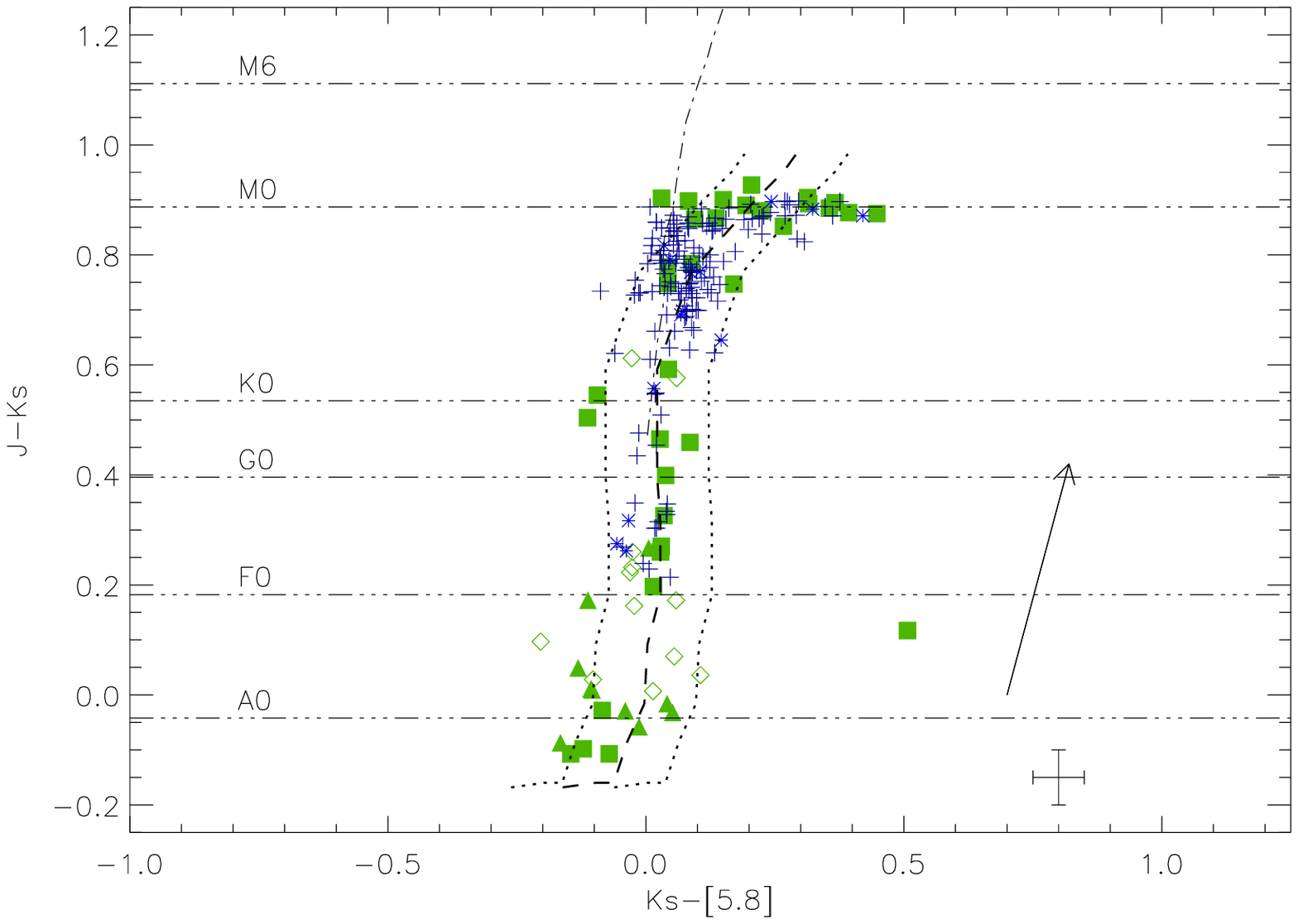}{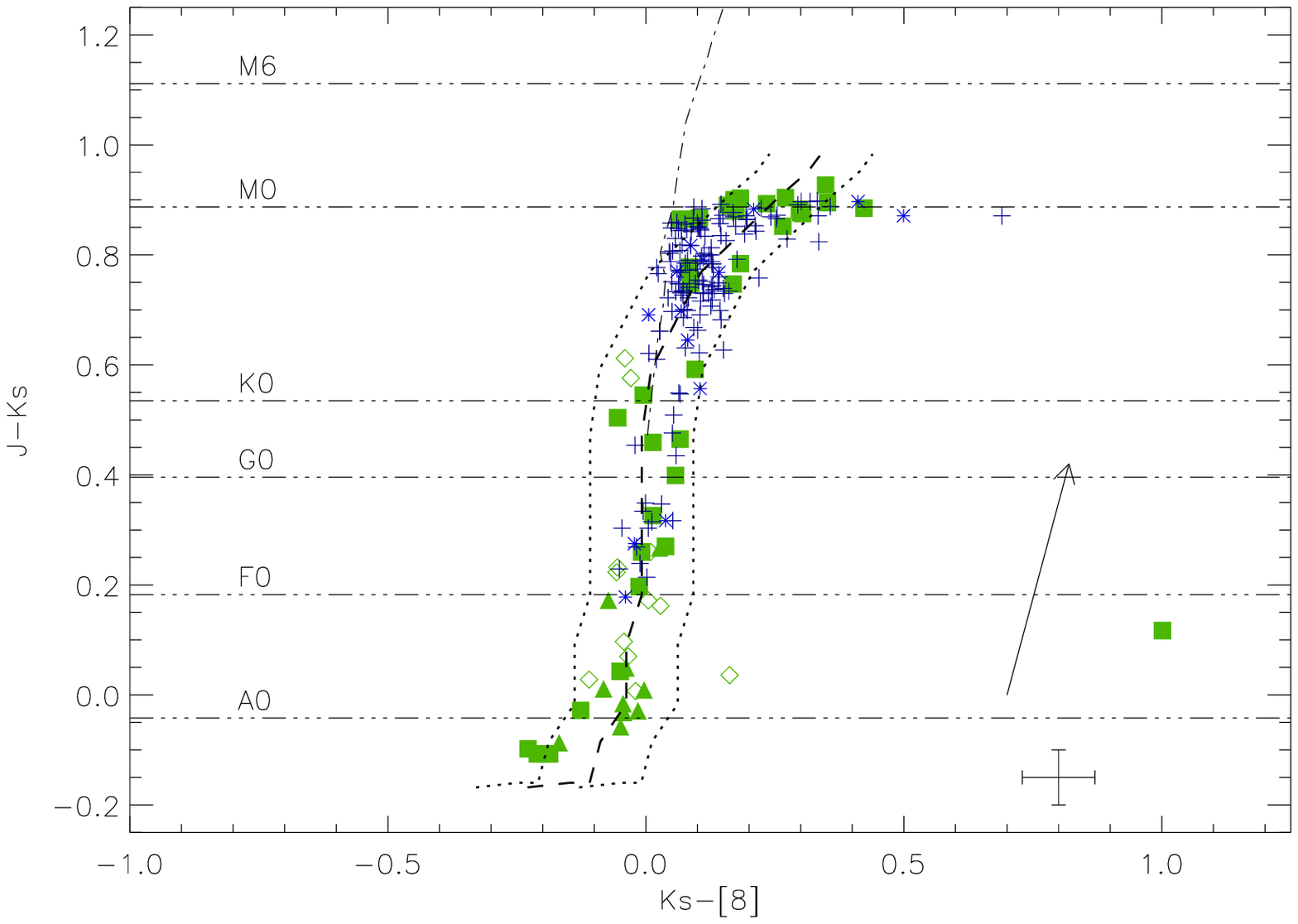}
\caption{J-K$_{s}$ vs. K$_{s}$-[5.8] and K$_{s}$-[8] for NGC 2232 cluster members and 
field stars in the \citet{Cl72} catalog.  The symbols are the same as in the previous 
figures.  Most members shown are 'candidate' members 
selected from their positions on the J/J-K$_{s}$ and J/J-H color-magnitude diagrams.  
We also show the J-K$_{s}$ colors expected for star with a 
range of spectral types (dash-three dots).  The maximum photometric errors are shown in the 
lower right-hand corner.  Most faint sources have $\sigma$[5.8,8] $\sim$ 0.02-0.04.  
%In both 
%plots, the vast majority of sources define a narrow locus centered on photospheric colors.  
There are 1-2 sources with blue J-K$_{s}$ colors and [5.8] and [8] excess; a second potential 
population of excess sources is centered on J-K$_{s}$ $\sim$ 0.85 and has 
K$_{s}$-[5.8, 8] $\lesssim$ 0.3-0.4.}
%\end{document}
\label{jkiracmem}
\end{figure}

\begin{figure}
\centering
%\plottwo{jmk24.ps}{k4k24t.ps}
\plottwo{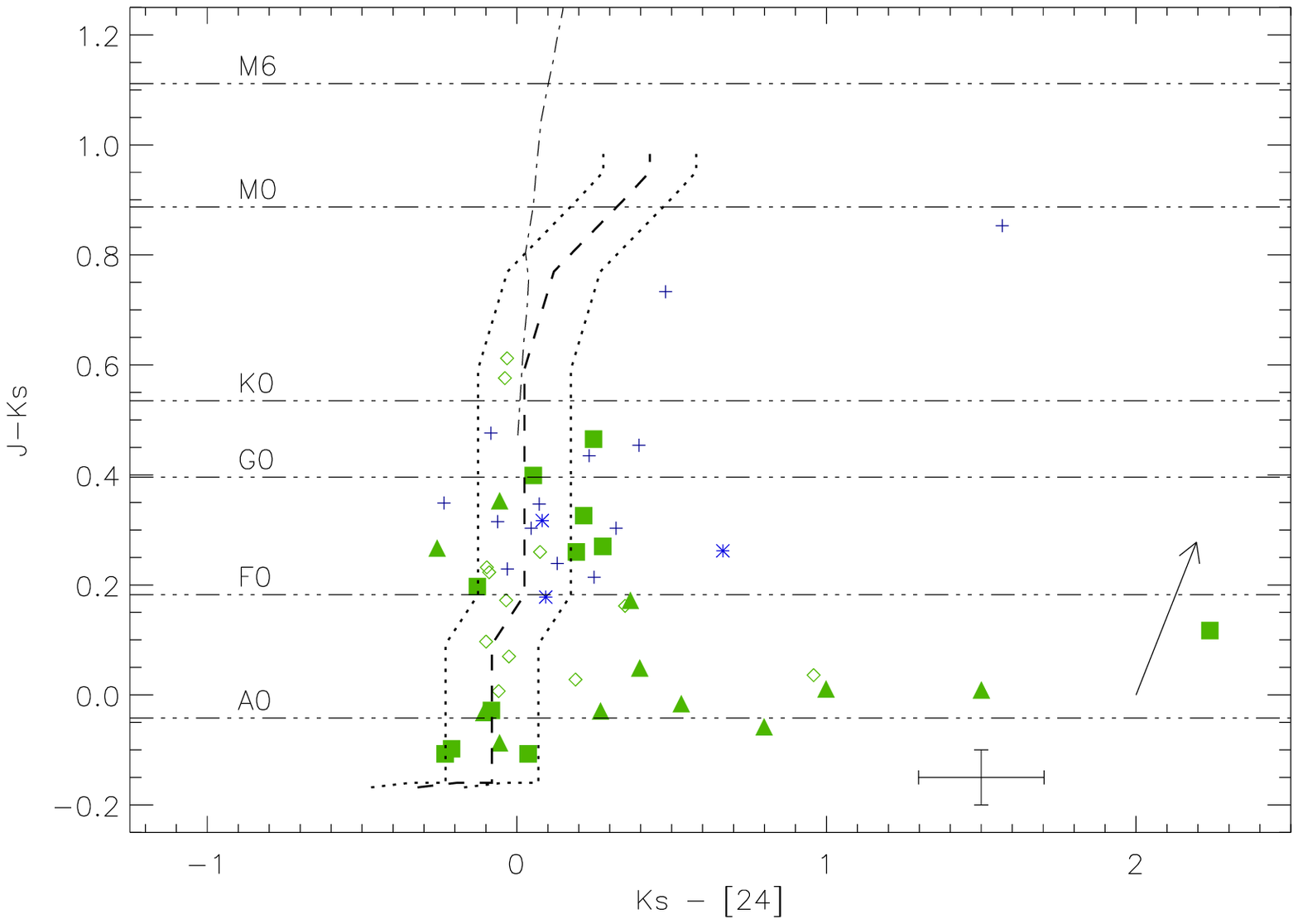}{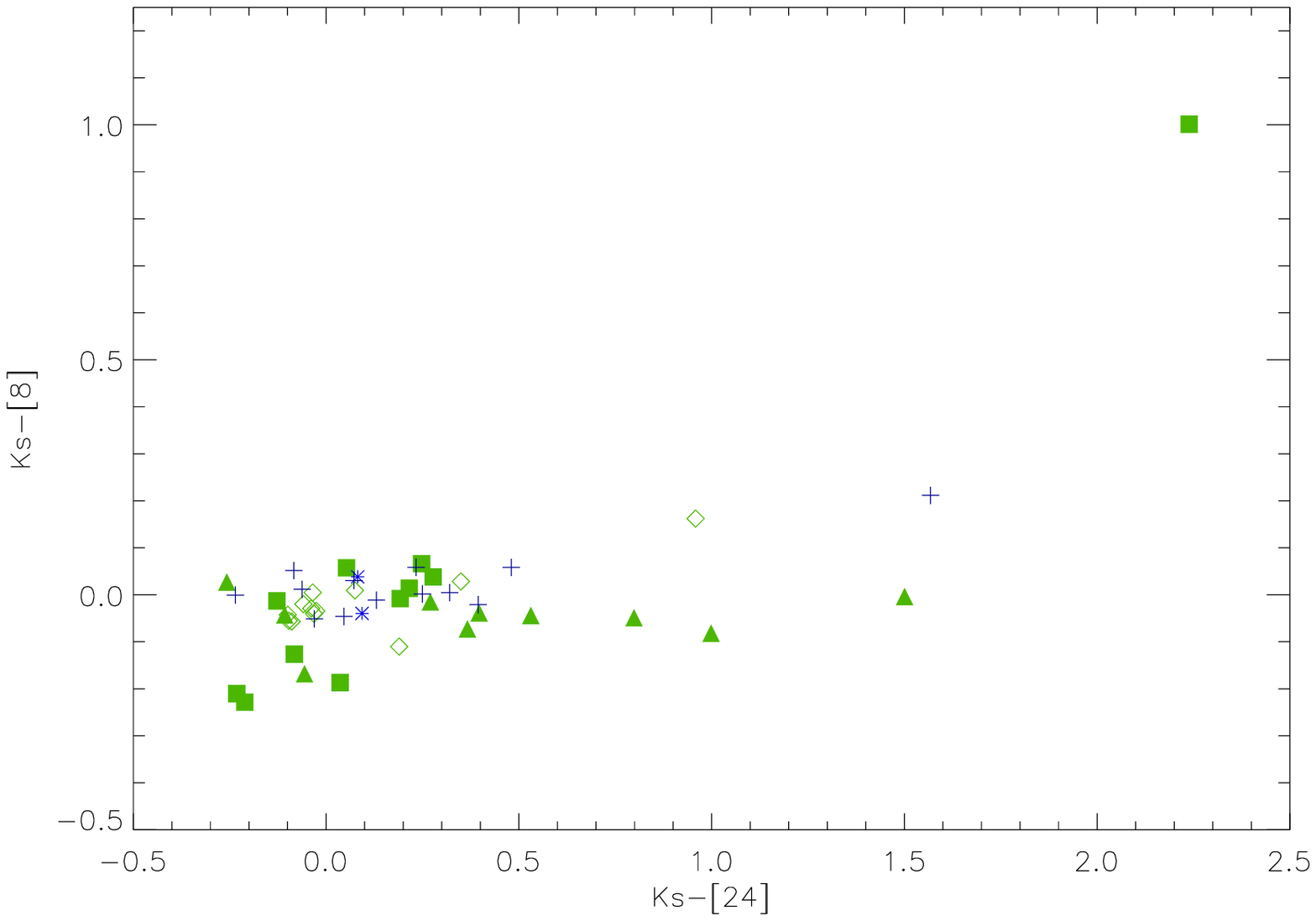}
%\plottwo{k2k4t.ps}{k3k4t.ps}
\plottwo{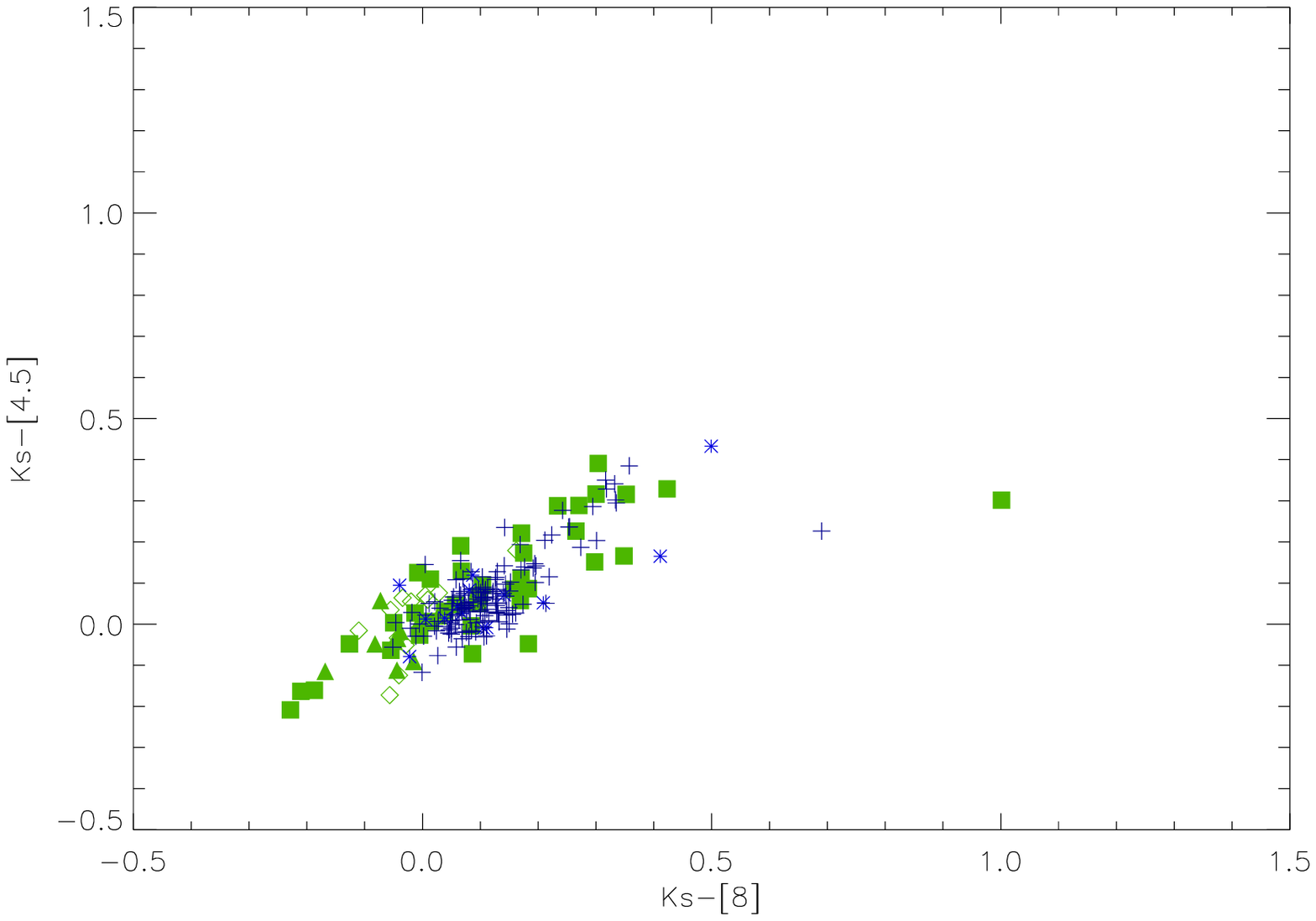}{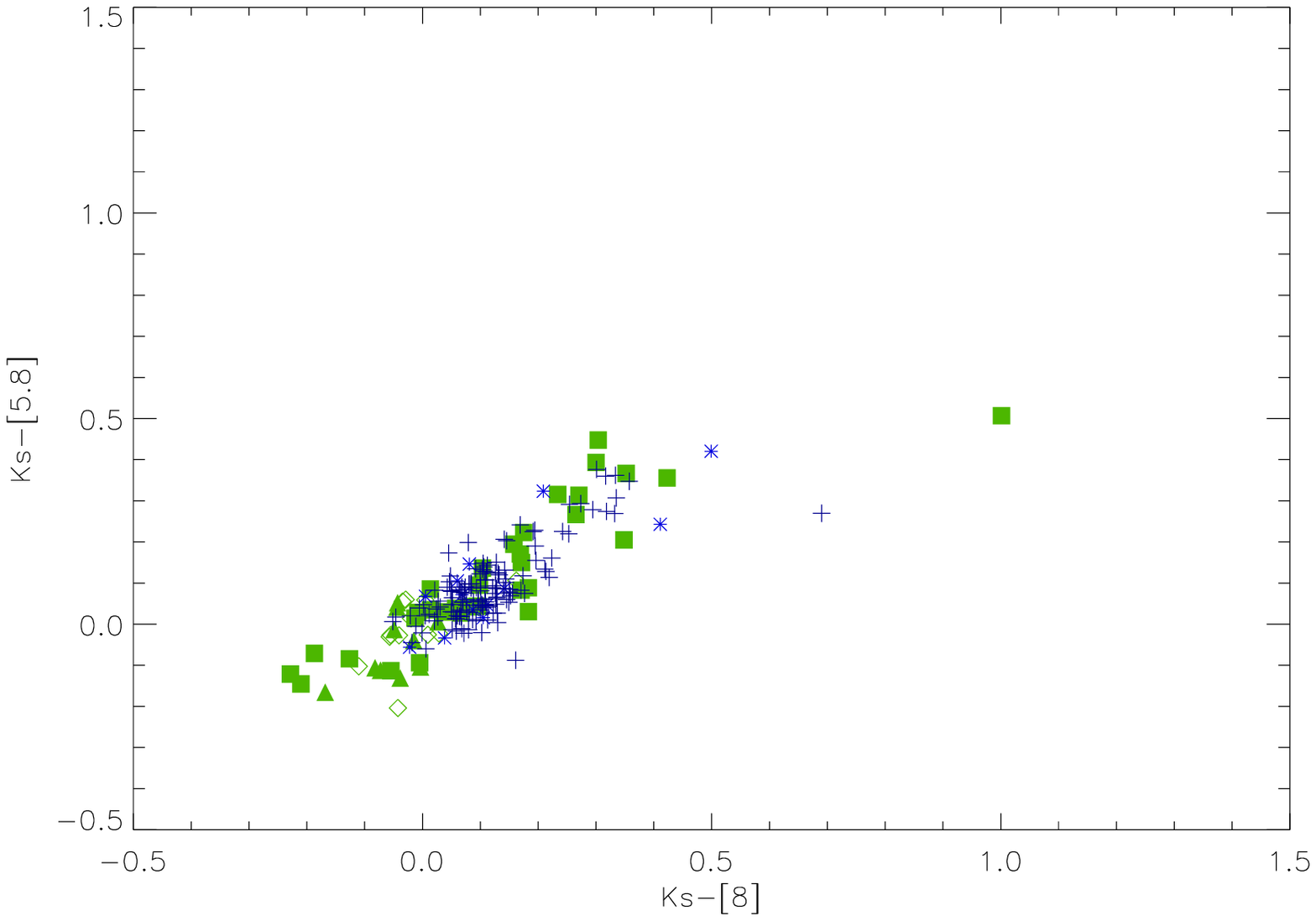}
\caption{Color-color diagrams for cluster members and 
\citet{Cl72} field stars.  
  The symbols are the same as in previous figures.   
(top left) J-K$_{s}$ vs. K$_{s}$-[24] colors: many cluster members 
have 24 $\mu m$ excesses.  (top right) K$_{s}$-[8] vs. K$_{s}$-[24] colors for members and 
field stars.  With one exception, the sources with clear MIPS excesses have 
photospheric IRAC colors.  (bottom) K$_{s}$-[4.5] and K$_{s}$-[5.8] vs. K$_{s}$-[8] color-color 
diagrams.  Almost all sources lie in a locus consistent with photospheric colors from early to 
late-type stars.  The spectral types shown in the top-left panel (dash-three dots) 
are derived from \citet{Si00} using the color conversions from \citet{Kh95}.}
\label{jkmipsmem}
\end{figure}

%Armed with a list of confirmed members and candidate members from x-ray observations, optical/IR colors, 
%proper motion, and spectroscopy, 
We now analyze the IRAC and MIPS colors of confirmed and candidate NGC 2232 members.  Figures \ref{jkiracmem} and 
\ref{jkmipsmem} show the K$_{s}$-[4.5], [5.8], [8], and [24] colors for these stars.  
Overplotted are the loci for photospheric colors of (pre) main sequence stars (vertical dashed line) and the 
giant locus (vertical dot dashed line); J-K$_{s}$ colors for A0-M5 stars are identified by horizontal dotted lines.  
The loci for photospheric colors were derived from the STAR-PET interactive tool on the 
\textit{Spitzer Science Center} website, which computes 2MASS-Spitzer colors based on the 
Kurucz-Lejeune stellar atmosphere models \citep[e.g][]{Ku93, Le97} convolved with the IRAC and MIPS filter 
responses (Table \ref{photcolors}).  We also overplot the reddening vector assuming the 2MASS/IRAC reddening 
laws from \citet{In05}; A$_{24}$ is estimated from \citet{Ma90}.  

For stars with photospheric near-IR emission and identical extinction,
  the J-K$_{s}$ color separates stars by spectral type.  
Cluster stars follow a well-defined locus in the near-IR/IRAC color-color diagrams 
(Figure \ref{jkiracmem}).  From early to late-type stars, the full dispersion about the 
photospheric locus ranges from $\sim$ 0.05 to 0.1 mag, 
roughly twice the maximum photometric error at [5.8, 8].
There are no cluster stars with red giant colors (J--K$_{s}$ $\gtrsim$ 0.9) or with the colors of the 
lowest-mass main sequence stars ($\gtrsim$ M4 spectral type).
The lack of red giant stars suggests our selection criteria are robust.
The lack of stars later than M4-M5 is consistent with our J-band magnitude 
cutoff for IRAC sources with 10 $\sigma$ detections.

\begin{figure}
%\plottwo{id18566.eps}{id18601ch4.eps}
\plottwo{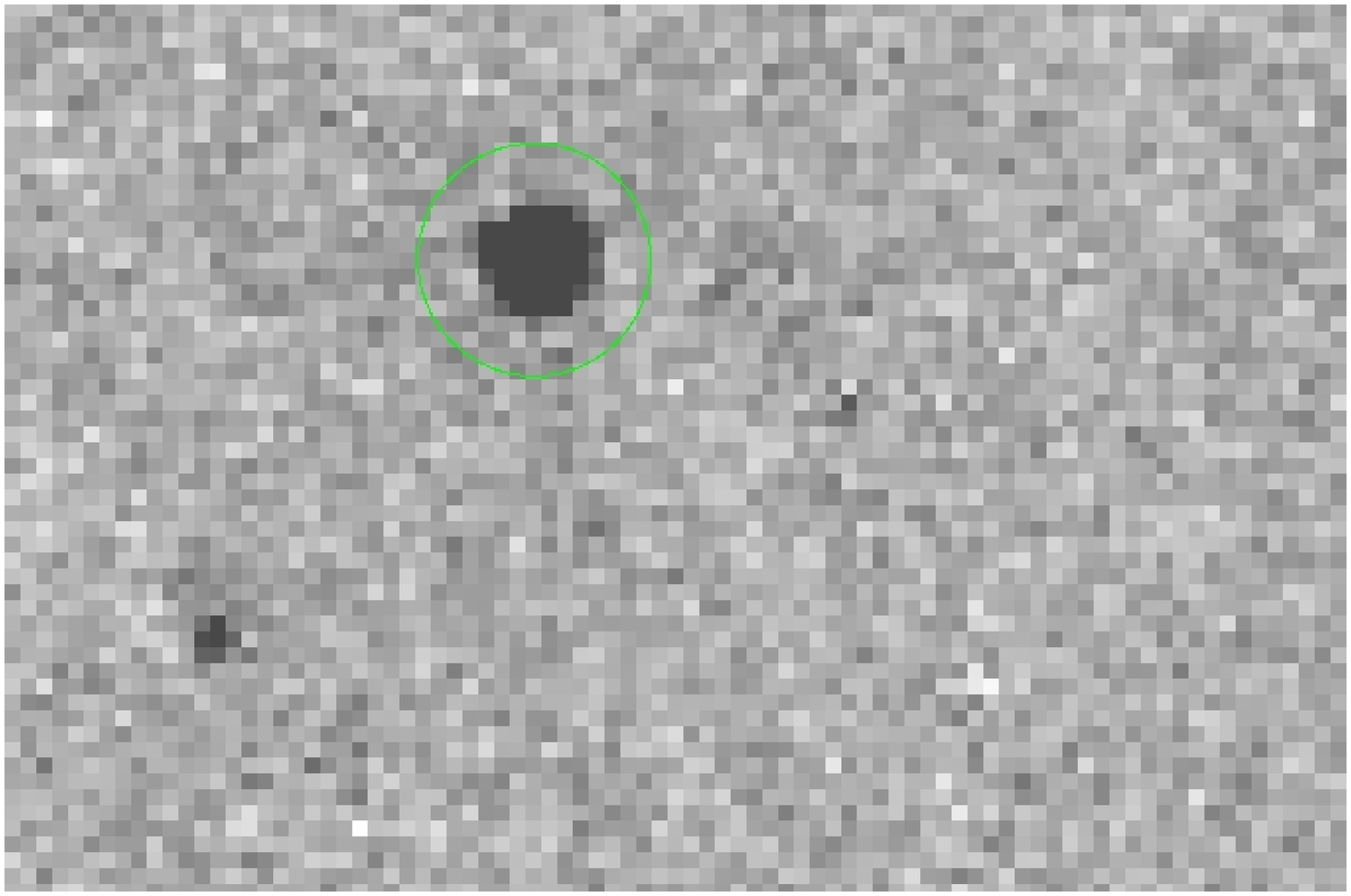}{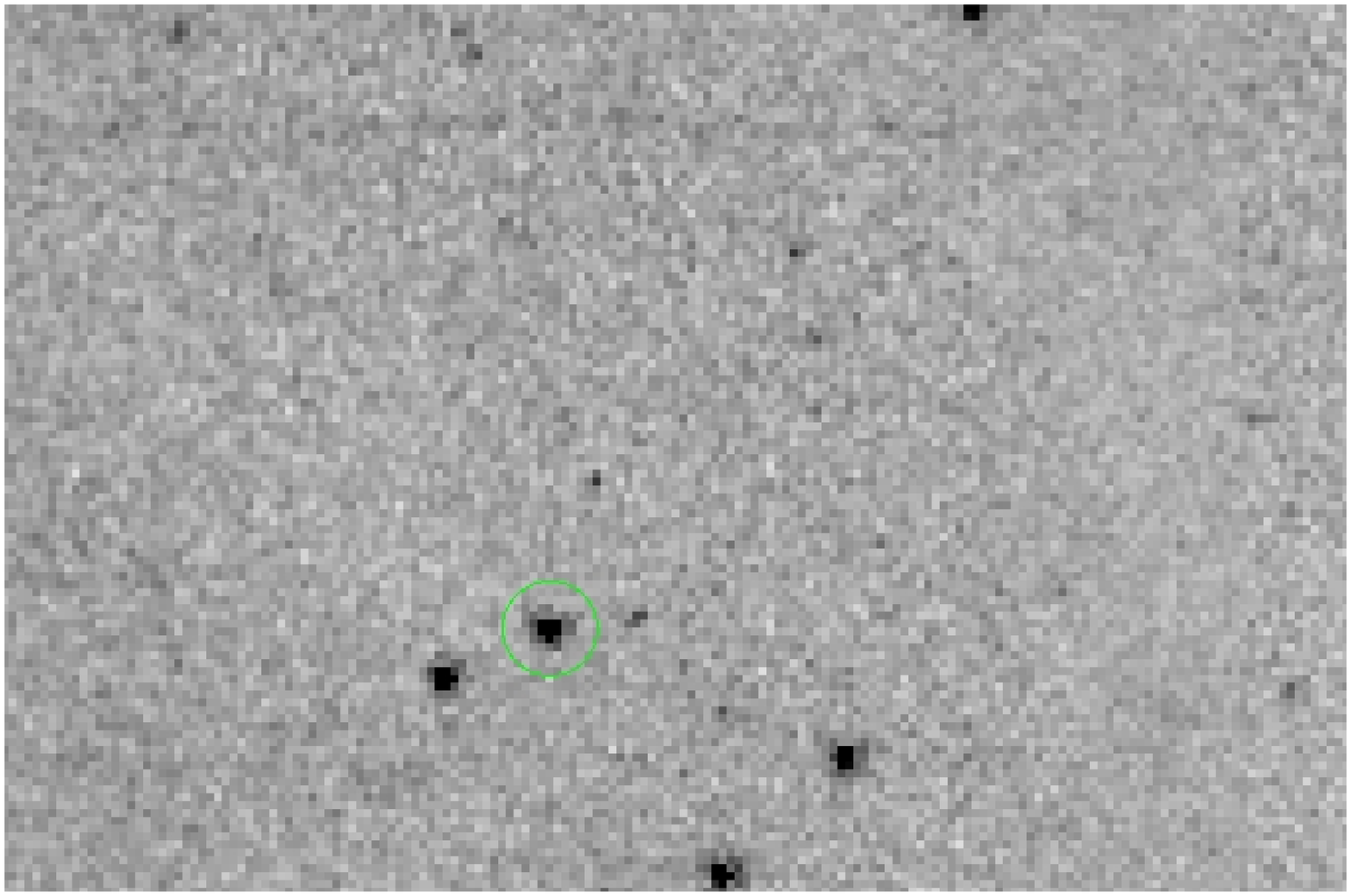}
\caption{IRAC images of ID 18566 (left) and ID 18601 (right).  Neither sources show evidence for 
source blending that could artifically brighten them in the IRAC bands.}
\label{ds9pic}
\end{figure}

NGC 2232 contains a small population of stars with 5.8 $\mu m$ and/or 8 $\mu m$ 
excesses.  One star (ID 18566) with a blue J-K$_{s}$ color has a K$_{s}$-[4.5, 5.8, 8] 
 excess ($\sim$ 0.2, 0.5, 1 mags) clearly larger than the 
zero K$_{s}$-[4.5, 5.8, 8] color expected for early-type stellar photospheres.
The stellar spectrum confirms the A spectral type 
suggested by the J--K$_{s}$ colors and shows no evidence for accretion (Figure \ref{fastspec}).
Visual inspection of the source on the IRAC and MIPS frames shows no
evidence for source blending (Figure \ref{ds9pic}, left panel).

While most late-type stars in the cluster have photospheric colors, a few may have 
K$_{s}$-[5.8, 8] colors $\gtrsim$ 0.2-0.3 mag redder than the photosphere.  
IDs 6540, 9220, 18601, and 18622 have 8 $\mu m$ excesses that are 
$\ge$ 3$\sigma_{[5.8],[8]}$ away from the right-hand vertical dotted line 
showing the 0.1 magnitude bound for photospheric sources.  Thus, these stars 
have red K$_{s}$-[IRAC] colors that may indicate the presence of warm, circumstellar dust.

One star (ID 6540) with red J-K$_{s}$ colors appears to have a $\sim$ 0.7 magnitude excess
at 8 $\mu m$.  ID 6540 may also have a borderline ($\sim$ 0.27 mag) 
excess at 5.8 $\mu m$.  Although the image 
mosaic at 5.8 $\mu m$ does not show contamination from another source, the 8 $\mu m$ 
image is slightly extended and may be blended.  Blending could arise from the 
superposition of ID 6540 with a background galaxy.  
Galaxies have much redder [5.8]--[8] colors than stars; thus a background galaxy 
can raise the 8 $\mu m$ flux of a stellar source and remain undetected at shorter 
wavelengths.  Thus, the excess for this source is questionable.

The other late-type stars with 8 $\mu m$ excess are not 
artificially brightened by source blending (e.g. ID 18601; see Figure \ref{ds9pic}, right panel).
Another source with very red K$_{s}$-[5.8, 8] colors, ID 18622, also lacks evidence for blending.
The IRAC mosaics show that these stars have no neighbors whose light partially 
falls within their aperture radii.
These sources are also $\sim$ 12th-13th magnitude, corresponding to an IRAC flux 
$\sim$ 4-15 times brighter than the 10$\sigma$ limits in these filters.
Thus, NGC 2232 may contain a late-type population of stars with warm 
dust emission.  In contrast, all sources with J-K$_{s}$ $\sim$ 
0.2-0.7 (plausibly F0-K5 stars) have photospheric colors.

The population of potential 24 $\mu m$ excess sources is larger (Figure \ref{jkmipsmem}).  Many 
cluster stars with blue J-K$_{s}$ colors have 24 $\mu m$ fluxes up to $\approx$ 3--10$\times$ 
their predicted photospheric fluxes (K$_{s}$-[24] $\sim$ 1- 2.5).  Other 
members with slightly redder colors (J-K$_{s}$ $\sim$ 0.2-0.5; plausibly F0-K0 stars) may also have 
weak excess emission ($\sim$ 0.25-1 mags).  Figure \ref{jkmipsmem} (left panel) reveals a third population of stars with 
K$_{s}$-[24] $\sim$ 0.5-1.5 that are plausibly K/M stars.  While the errors in [24] are 
 larger at a given magnitude than in the IRAC bands, the excesses are typically much larger 
than IRAC excesses.  In \S 2 we showed that the number counts of MIPS-detected stars peak at [24] $\sim$ 10.5  
and J $\sim$ 11.  Assuming a distance of 340 pc, an extinction of E(B-V) =0.07, and a cluster age of 25 Myr, 
11th magnitude cluster stars should have a G2 spectral type.  Thus, the 
MIPS survey is likely complete to early G stars.  While MIPS does not detect 
the photospheres of stars later than $\sim$ G2, it can detect late-type stars with 24 $\mu m$ excesses 
from circumstellar dust. 

All but one of the MIPS-excess sources appear to have photospheric IRAC colors (Figure \ref{jkmipsmem}, top-right panel).
The confirmed, early and intermediate-type (B-G) NGC 2232 members have a range of 24 $\mu m$ excesses ranging
from $\sim$ 0.3 to 2.2 mag.  However, all but one have a very narrow range of 
K$_{s}$-[8] colors ($\sim$ -0.1 to 0.1) consistent with stellar photospheres.  The later-type 
sources with $\sim$ 0.3-0.9 mag 24 $\mu m$ excesses from Figure \ref{jkmipsmem} (top-left panel and bottom-left panel) also have a narrow range of K$_{s}$-[8] 
colors ($\sim$ 0-0.15) consistent with late-type photospheres.  Sources lacking 8$\mu m$ excess also have a range 
of K$_{s}$-[5.8] and K$_{s}$-[4.5] colors consistent with photospheres.

\subsection{SED Modeling of HD 45435 (ID 18566): Evidence for 
Debris Emission Produced From Terrestrial Planet Formation}
%: Evidence for Debris Emission Produced 
%From Terrestrial Planet Formation}
NGC 2232 has many good candidates for debris disks.
In addition to a handful of stars with modest IRAC
excesses, there are many cluster members with 
24 $\mu$m excess emission. The level of this excess emission
($\sim$ 0.5--2.25 mags) is much smaller than excesses in primordial
disks ($\sim$ 5 mags; \citealt{Kh95, He07, Ck08}); however, the levels are similar
to the excesses observed in debris disks ($\sim$ 0.5--3 mags; \citealt{Ri05, 
Su06, Go07, Cu08a}).

To test whether the excess sources are primordial disks,
evolved primordial/transition disks, or debris disks, we rely on analyses
of spectral energy distributions (SEDs) and the ratio
of the observed fluxes to predicted photospheric fluxes
at 8 $\mu$m and at 24 $\mu$m. Only one star, ID 18566 with
an A3 spectral type, has clear excesses in the IRAC and
MIPS bands. These excesses allow us to
measure the temperature and luminosity of the dust. As
shown below, our analysis demonstrates that this star
harbors a debris disk.  Other sources with IRAC (MIPS)
excesses have no obvious MIPS (IRAC) excesses. Thus, we
cannot analyze the SED in detail. Because the excesses
for these sources are smaller than those for ID 18566,
these stars probably also harbor debris disks.  
In \S4.3, we show that the observed excesses and upper limits at
other wavelengths are consistent with this conclusion.

%Our goal is to determine if the SEDs of all 24 $\mu m$ excess sources identify debris disks or 
 %\textit{evolved primordial disks} (e.g. transition disks).  Specifically, we model the 
%spectral energy distribution (SED) of ID 18566, an A3 star with clear IRAC and MIPS excesses.  
To analyze the SED for HD 45435, we follow the procedure from \citet{Au99} and \citet{Cu07b, Cu08a}.
The star+disk emission is first fit to a sum of the stellar blackbody plus two scaled blackbodies from 
the disk to derive a) the 
amount of disk emission relative to the star and b) the temperature, and thus location, of the 
disk emission.  We assume uncertainties of 5\% (10\%) in the IRAC (MIPS) 
bands due to possible systematic effects (e.g. uncertainty in the zero-point 
flux, phase-dependent response, and photometric errors).  Evolved primordial disks such as transition 
disks in IC 348 and Taurus have fractional luminosities, L$_{d}$/L$_{\star}$, $\approx$ 0.1 
\citep{Cal05, Lo05, Ck08}.  Debris disks have disk luminosities 
$\gtrsim$ 100 times smaller than evolved primordial disks \citep[$\approx$ 
10$^{-5}$--10$^{-3}$, e.g.][]{Pa06, Su06, Rh07, Ck08}.
Thus, the disk luminosity discriminates between disks of different evolutionary states.

Second, the source SED is compared to more rigorous disk models.  We consider an evolved 
primordial disk model from \citet{Kh87}, a terrestrial zone debris disk model from \citet{Kb04}, and 
a cold debris disk model from \citet{Kb08}.  For the evolved primordial disk model, 
the disk temperature scales as T $\propto$ r$^{-3/4}$.  We vary the inner hole size to match the 
level of IRAC emission.   Debris from terrestrial planet formation is modeled at 1.5--7.5 AU from 
a 2.0 M$_{\odot}$ star \citep{Kb04}.  For the cold debris disk model, debris from planet formation is modeled at
30--150 AU from a 2.0 M$_{\odot}$ star \citep{Kb08}. 

\begin{figure}
\epsscale{0.7}
\centering
%\plotone{SED1.ps}
\plotone{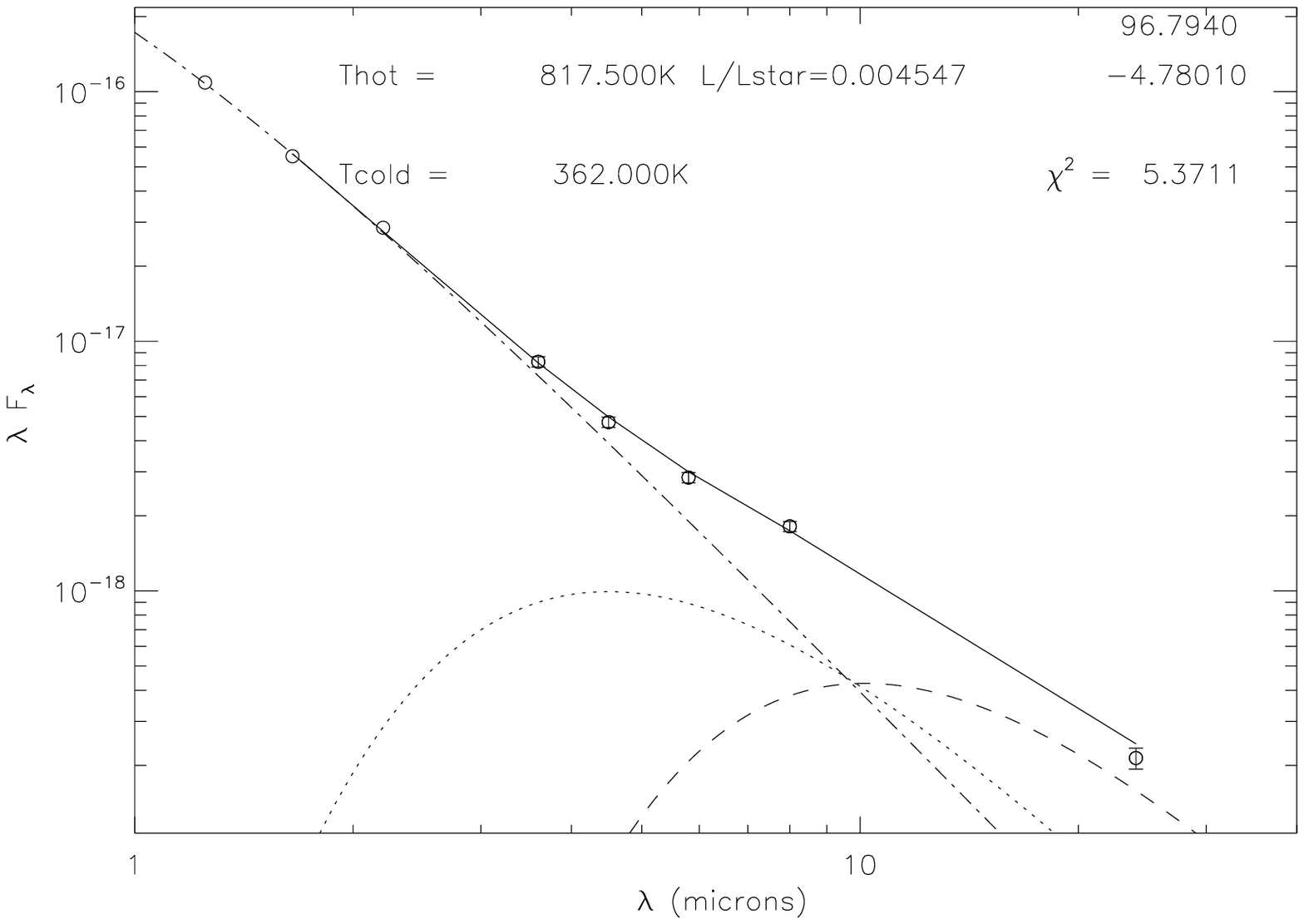}
%\plotone{id185661.ps}
\plotone{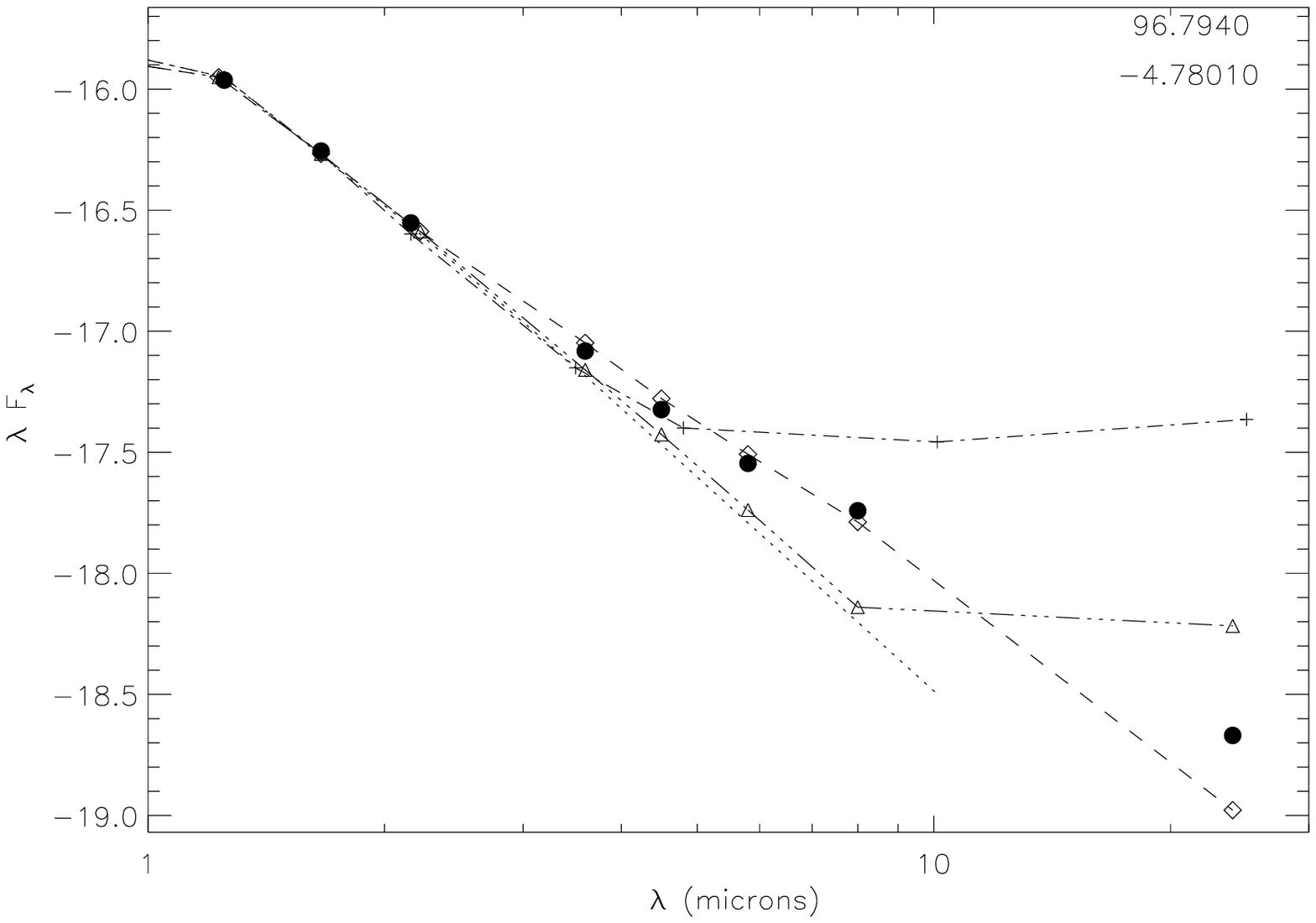}
\caption{SED models for ID 18566.  
(top)Blackbody fits to the SED of ID 18566, using a stellar blackbody (dot-dashed line), a 
hot dust component (dotted line), and a warm dust component (dashed line).
  The blackbody fits reveal a population of hot/warm 
dust.  The disk has a fractional luminosity characteristic of luminous debris disks.  (bottom) 
SED modeling of the star compared to a transition disk model (dashed-dotted line/crosses), 
a terrestrial zone debris disk model (dashed line/diamonds), and a cold debris disk model 
(dashed-three dots/triangles).  The stellar photosphere is shown for reference (dotted line).} 
\label{sedfit}
\end{figure}

The min($\chi^{2}$) scaled blackbody fit for HD 45435 shown in Figure \ref{sedfit} demonstrates 
 that the amplitude of disk emission (L$_{d}$/L$_{\star}$ $\sim$ 5$\times$10$^{-3}$) 
is comparable to the most luminous debris disks ( e.g. HD 113766A and h and $\chi$ Per-S5, see \citealt{Li08, Cu08a})
 that the disk has very hot dust.  The fits require dust populations at two different temperatures, 
 $\sim$ 360 K and 815 K.  This emission is consistent with warm dust in regions comparable 
to the orbits of Mercury and Earth in the Solar System.  Disk emission clearly emerges above 
the stellar photosphere by 3.6 $\mu m$ and dominates over the photosphere longwards of 8 $\mu m$.  

%Unique to this system is the possible presence of K-band emission ($\sim$ 0.05-0.1 mag excess).  This excess 
%is not consistent with photometric errors which are small ($<<$ 0.01 mags).  The excess gradually becomes larger 
%at longer wavelengths.  If confirmed by additional observations, this system is a rare 
%debris disk with a K-band excess from hot dust \citep[e.g. Vega][]{Abs06}.

Comparing the source SED to predictions from transition and debris disk models confirms that the HD 45435 
likely harbors a terrestrial zone debris disk.  The transition disk model drastically overpredicts 
the mid-IR flux longwards of $\sim$ 4-5 $\mu m$.  The cold debris disk fares better in reproducing the 
MIPS excess.  However, it predicts only marginal 5.8 $\mu m$ and 8 $\mu m$ excesses in contrast to 
the strong excesses observed.  The terrestrial zone debris disk models accurately reproduces the IRAC 
and MIPS fluxes. 

In summary, HD 45435 (ID 18566) has the strongest 24 $\mu m$ excess of any NGC 2232 member and is the only 
source with a strong IRAC excess.  Compared to other cluster 
members, its disk is the most likely to be at an early evolutionary state, 
perhaps an evolved primordial disk/transition disk.  However, 
SED modeling suggests that this source is instead a warm debris disk.  
The SEDs of sources with weaker 24 $\mu m$ emission are even more 
optically thin and more likely to be debris disks.
Thus, optical spectroscopy and SED modeling suggest that the 
mid-IR excesses from early-type stars in NGC 2232 are due to 
dust in circumstellar debris disks.

\subsection{Frequency of 8 $\mu m$ and 24 $\mu m$ Emission from Debris Disks}
We now consider the frequency and nature of the stars with
excess IRAC (MIPS) emission but no statistically significant
MIPS (IRAC) excess. After defining robust criteria for these
sources, we measure the frequency of disk emission in both
bands. To place initial constraints on the disk properties
of these sources, we use IRAC and MIPS upper limits to
constrain the dust temperature and to derive a lower limit
to the dust luminosity. These results demonstrate that 
NGC 2232 members include at least two stars with warm debris disks and 
many stars with cold debris disks.
%Finally, to investigate any connection between 
%star/primordial disk interactions and the subsequent debris disk phase, we 
%compare the 24 $\mu m$ excess emission to stellar rotation.

%\subsection{Disk Modeling: Evidence for Terrestrial Planet Formation and Gas/Ice Giant Planet Formation}
%\subsection{Frequency of 8 $\mu m$ and 24 $\mu m$ Excess Emission from Debris Disks}
 \subsubsection{Frequency of 8 $\mu m$ Emission}
With the exception of HD 45435, NGC 2232 lacks any stars with 
very luminous 8 $\mu m$ excess emission (e.g. greater than twice 
the photospheric flux).  Warm dust around other stars 
must be far weaker.  The preceding section (\S 4.1) identifies 
several stars with K$_{s}$-[8] colors redder than 3$\sigma$ 
away from theoretical photospheric colors.  However, more robust 
measures for establishing excess sources are needed.
Specifically, the Kurucz-Lejeune synthetic K$_{s}$-[8] colors used to determine 
photospheric colors (and thus excess sources) are prone to 
systematic uncertainties in the model atmospheres.  
Young M stars, which comprise the majority of the 8 $\mu m$ excess candidates,
 are chromospherically active and can have large 
fractions of their surfaces covered with starspots, causing changes in the 
brightness of colors of $\sim$ 0.1 magnitudes \citep{Bou93}.  Thus, sources with near IR 
and mid-IR data taken at different epochs may have an intrinsically larger dispersion 
in colors at a given spectral type.  This dispersion results in a larger uncertainty 
in measuring the amount of IR excess emission.
Because NGC 2232 was not observed in K$_{s}$ and 
[8] simultaneously, starspots may also contribute small uncertainties in identifying 
excess sources based on the K$_{s}$-[8] color.  
%Errors in determining the true photospheric colors have led 
%to claims of weak mid-IR excesses around late-type stars \citep[e.g.,][]{Rj99} which were 
%later cast into doubt \citep{Met04}.

To minimize the impact of systematic errors in identifying stars with 8 $\mu m$ excess 
emission, we determine the locus of 
photospheric colors empirically from IRAC fluxes.
Following \citet{Ca06}, we use the [4.5] filter to define the stellar photosphere
and compare its flux to the 8$\mu m$ flux to identify excess sources:
log(F[8]/F[4.5]).  Using the 4.5 $\mu m$ flux has several advantages over using 
K$_{s}$.  The scatter in the [4.5]-[8] color as a function of [8] for all IRAC-detected 
stars (not shown) is slightly smaller than that in K$_{s}$-[8].  All sources 
with 8 $\mu m$ detections should also have detections at 4.5 $\mu m$.  Furthermore, the 
theoretical K$_{s}$-[8] colors from the Kurucz-Lejeune stellar atmosphere models 
 suggest that the [8] to K$_{s}$ flux ratios as a function of 
spectral type cannot be fit by a linear function, whereas the [8] to [4.5] flux ratios 
can \citep{Ca06}.  Finally, the 8 $\mu m$ to 4.5 $\mu m$ flux 
ratio is insensitive to variations in reddening because stars are 
reddened by the same amount in either filter \citep{In05}.
%Therefore, 
%like \citet{Ca06} we use the [4.5] filter to define the stellar photosphere 
%and compare its flux to the 8$\mu m$ flux to identify excess sources:
%log(F[8]/F[4.5]).  
%Because stars with excesses at 8 $\mu m$ may also 
%have excesses at 5.8 $\mu m$, we define the photospheric locus for 
%log(F[5.8]/F[4.5]) as well.

Using 4$\sigma$ iterative clipping, we calculate a best-fit line using a standard linear least-squares fit 
to the locus of cluster stars in J-K$_{s}$ vs. log(F[8]/F[4.5]) 
space.  The final linear fit has a 
slope of 0.046 and a y-intercept of -0.465 with rms residuals of $\sim$ 3.92\%.  
To determine an emprical excess criterion, we identify the threshold, in $\sigma_{rms}$, 
beyond which there are no sources with 'negative excesses' (deficient flux ratios): 
$\sim$ 3.42$\sigma_{rms}$.  
Assuming a dispersion in IRAC colors dominated by photon noise, we expect less than 
1 star out of 200 cluster members detected at 8 $\mu m$ to be more 
than 3--4$\sigma_{rms}$ away from the photospheric locus.  We adopt 
a 4$\sigma_{rms}$ threshold to identify cluster stars with 8 $\mu m$ excesses. 
Similar thresholds have been adopted by other authors to identify stars with weak excesses 
in both IRAC and MIPS \citep[e.g.][]{Ca06, Hi08}.
 
\begin{figure}
%\plotone{fluxslope.ps}
\plotone{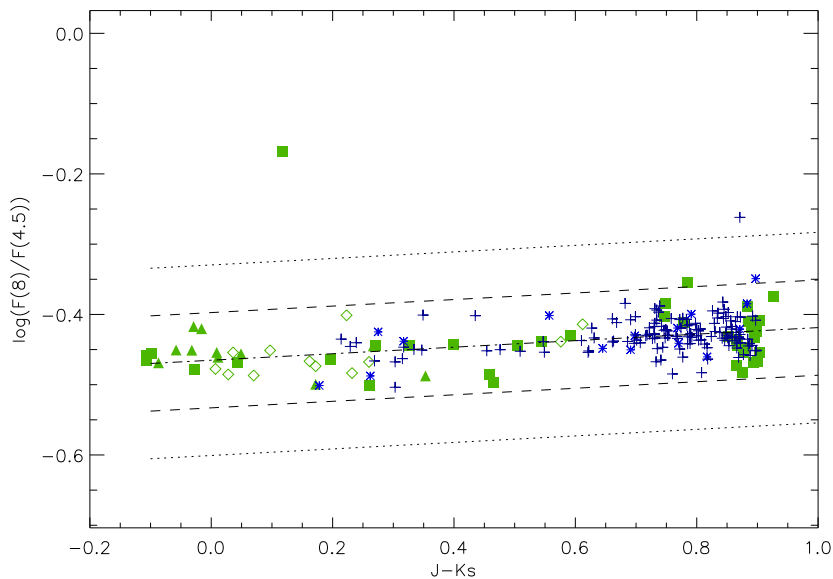}
\caption{The 8 $\mu m$ to 4.5 $\mu m$ flux ratio as a function of J-K$_{s}$ used to 
identify NGC 2232 cluster stars with IRAC excesses.  The dash-dotted line shows 
the linear least-squares fit to the flux ratios (rms residuals are $\sim$ 3.92\%).  
Dashed lines identify the 4$\sigma$ limit beyond which we identify excess sources.  
The dotted lines show the 8$\sigma$ limits.  IR excess sources are 
identified as those with flux ratios greater than 4$\sigma$ from the photosphere.  
Four stars in NGC 2232 have flux ratios consistent with IRAC excesses from warm dust.  
From left to right, these sources are ID 18566 (J--K$_{s}$ $\approx$ 0.1), ID 18613 (J--K$_{s}$ $\approx$ 0.8), 
ID 6540 (J--K$_{s}$ $\approx$ 0.85), and ID 9220 (J--K$_{s}$ $\approx$ 0.9).}
\label{fluxslope}
\end{figure}

NGC 2232 includes four stars that are $\ge$ 4$\sigma_{rms}$ away from the 
photospheric locus in log(F[8]/F[4.5]) (Figure \ref{fluxslope}).  For confirmation 
of these excesses, we compare their K$_{s}$-[8] colors to photospheric predictions.  ID 18613, 
a cluster member identified by x-ray activity, has 
a 4.35$\sigma$ excess in the 8$\mu m$ to 4.5$\mu m$ flux ratio but has a rather 
blue K$_{s}$-[8] color ($\sim$ 0.183) consistent with photospheric predictions.  
However, three stars -- IDs 6540, 9220, and 18566 -- have flux ratio excesses and
 red K$_{s}$-[8] colors lying outside the locus of predicted 
photospheric colors.  The apparent excesses for these stars are more 
than 3$\sigma$ larger than their photometric errors. 
  Even though ID 6540 has a $>$ 8$\sigma$ excess, 
the previous section cites source blending 
as a possible source for its excess emission.  In contrast, ID 18566 (HD 45435)
is not blended and clearly has excess emission from 4.5 $\mu m$ onwards.  ID 9220 also shows no evidence 
for blending and has red K$_{s}$-[8] \textit{and} K$_{s}$-[5.8] colors suggestive of mid-IR excess in 
addition to its excess in log(F[8]/F[4.5]).  ID 6540 is identified as a candidate member 
based solely on near-IR color-magnitude diagrams; ID 9220 is identified as a member from both
optical/near-IR and near-IR color-magnitude diagrams.  Therefore, at least early-type star and likely 
 one late-type star (ID 9220) in NGC 2232 have bona fide IRAC excesses indicative of warm dust; two other late-type 
stars may also have warm dust emission.  Properties of the four 8 $\mu m$ 
excess sources are listed in Table \ref{8exclist}.

While our method can identify IR-excess sources without some systematic uncertainties,  it 
may conceal some bona fide excess sources with very warm dust (T$_{d}$ $\sim$ 350--600 K).  Using the 
8 $\mu m$ to 4.5 $\mu m$ flux ratios to define excess sources best 
identifies warm dust whose flux peaks at $\lambda$ $\gtrsim$ 8 $\mu m$.  
In terms of their photometric errors, IDs 18601 and 18622 
have both K$_{s}$-[8] and K$_{s}$-[5.8] colors ($\sim$ 0.4 mags) greater than 3$\sigma$  
 away from the locus of predicted photospheric colors.  However, their 
8 $\mu m$ to 4.5 $\mu m$ flux ratios are less than 3$\sigma$ away from the 
empirically-defined photospheric locus.  If these systems have weakly-emitting 
warm dust that peaks in the IRAC bands, then their 8 $\mu m$ to 4.5 $\mu m$ 
flux ratios will be bluer than that for dust whose 8 $\mu m$ emission is 
on the Wien tail of a slightly colder blackbody.

We adopt a simple model to show that the weak 8 $\mu m$ excesses around 
late-type stars yield disk luminosities consistent with 
debris disks.  To estimate the disk luminosity at 8 $\mu m$, 
we calculate the flux of the 8 $\mu m$ excess, assume that 
the disk emission peaks at 8 $\mu m$ and 
originates from a 360 K blackbody, and 
assume the stellar luminosity of a 25 Myr-old M0 star 
from \citet{Si00}.  We derive 
 L$_{d}$/L$_{\star}$ $\approx$ 10$^{-4}$--10$^{-3}$, 
consistent with debris disk luminosities.  The 
maximum disk luminosity from a 120 K blackbody consistent with the 24 $\mu m$ 
upper limits is also L$_{d}$/L$_{\star}$ $\approx$ 10$^{-3}$.

To estimate the frequency of 8 $\mu m$ emission as a function of stellar mass/spectral 
type, we divide the sample into 'early' (BA), 'intermediate' (FG), and late (KM) stars 
based on their J-K$_{s}$ colors.
%Because most of the stars here lack spectra, we assume that the reddening is 
%low and use the photospheric J-K$_{s}$ colors reddened E(B-V) $\sim$ 0.07 
%to divide stars into these groups.
The frequency of 8 $\mu m$ emission in these groups is 6.3\% (1/16), 
$<$ 3.8\% (0/26), and 0--1.9\% (0-3/158).  With the exception of the warm debris disk source 
HD 45435 (ID 18566), no star plausibly earlier than K0 has 8 $\mu m$ excess.  
At least one and as many as three late-type stars have large 
8 $\mu m$ to 4.5 $\mu m$ flux ratios indicative of warm dust.  If two stars (IDs 18601 and 18622) with 
small log(F[8]/F[4.5]) but red K$_{s}$-[5.8, 8] colors are also classified as 
excess sources, the frequency of warm dust around late-type stars could be 
as high as 3.2\%.  Optical spectra is required to confirm the membership status 
of late-type candidate cluster members and thus better constrain the frequency 
of late-type stars with warm dust.

These results suggest that warm dust surrounding NGC 2232 cluster stars 
is rare.  However, the derived frequencies of warm dust for early-type members are based 
on a small sample, which precludes a robust determination of the excess frequency 
for 25 Myr-old B and A stars.  Lack of deep x-ray data and incomplete 
optical spectroscopy prevents robust estimates of membership probabilities 
for intermediate and late-type stars.  These data would allows us to separate 
true cluster members from foreground M dwarfs and background red giants.

%Membership for intermediate and late-type stars is incomplete because of a 
%lack of deep x-ray data.  Furthermore, the census of members identified  
%from J/J-K$_{s}$ and J/J-H color-magnitude diagrams is uncertain due to contamination 
%from foreground M stars and background M giants.  Defining members based on these colors 
%is especially problematic for stars with J-H, K$_{s}$ = 0.6-0.7, 0.75-0.85, where 
%the locus of pre-main sequence colors clearly intersects the giant locus (see Figure \ref{jjmhk}).  
%Identifying candidate members by a narrow cluster locus in both J/J-K$_{s}$ and J/J-H color-magnitude diagrams 
%reduces but does not eliminate contamination.  
%
%Therefore, the frequencies of warm dust around intermediate and late-type cluster 
%members is also highly uncertain.  Follow-up 
%x-ray observations and spectroscopy are needed to identify true cluster members and 
%thus better constrain the frequency of warm dust.  Regardless of the uncertainties in 
%defining a reliable census of cluster members, it is clear that 
%the frequency of warm dust around all stars must be low ($<<$ 10\%).
\subsubsection{Frequency of 24 $\mu m$ Emission}
To identify sources with 24 $\mu m$ excess, 
we consider the sensitivity of the MIPS data and the photospheric 
colors of likely cluster members.  For 5$\sigma$ detections with 
[24] $\approx$ 10.5, MIPS cannot detect the photospheres of 
cluster stars with spectral types later than $\sim$ G2.  
For early G stars, the Kurucz-Lejeune stellar atmosphere models 
predict K$_{s}$-[8] $\sim$ -0.020 and K$_{s}$-[24] $\sim$ 0.005.  
Observations of $\sim$ 30 solar-type stars from \citet{Rie08} 
suggest K$_{s}$-[8] $\sim$ 0.046 and K$_{s}$-[24] $\sim$ 0.045.  
Uncertainties in absolute calibration are small, $\sim$ 1.5\% \citep{Rie08}.
Older calibrations suggest K$_{s}$-L'/M $\lesssim$ 0.02 for solar-type 
stars \citep[e.g. ][]{Kh95}.  Thus, we expect small photospheric 
colors -- K$_{s}$-[24] $\lesssim$ 0 -- for all cluster stars 
detected by MIPS.

Small number statistics and different areal coverage prevent developing an empirical 
locus of photospheric MIPS colors for all spectral types.
There are only 38 candidate/confirmed cluster members 
detected by MIPS compared to 200 detected at 8 $\mu m$.
Therefore, it is difficult to define the empirical locus of 
photospheric MIPS colors from early to late spectral types.  
Because the MIPS coverage is slightly larger than the IRAC coverage at 4.5 $\mu m$, 
not all MIPS-detected stars have measured 4.5$\mu m$ fluxes.  Therefore,  
using the [4.5] filter as a photospheric baseline is not ideal.

%The K$_{s}$-[24] colors predicted by the Kurucz-Lejuene stellar atmosphere models
 %are about zero and are nearly independent of spectral type from B5 to G0.  
%Absolute calibration errors for early type and solar-type stellar photospheres at 2.2 $\mu m$ 
%and 24 $\mu m$ are small ($\sim$ 1.5\%, \citealt{Rie08}), making comparisons between theoretical and 
%observed K$_{s}$-[24] colors for $\gtrsim$ solar-mass stars reliable.
%Because of the lack of MIPS sensitivity, all late-type cluster stars 
%detected by MIPS should have clear excesses.  Therefore, 
%to identify 24 $\mu m$ debris emission around NGC 2232 stars, 
Because the expected photospheric colors of B to G stars are small and vary little with 
spectral type, we develop an excess criterion based on the K$_{s}$-[24] color and 
the MIPS photometric error:
\begin{equation}
(K_{s}-[24])_{obs} - (K_{s}-[24])_{phot} \ge 3\sigma_{24}.
\end{equation}
To account for systematic errors (e.g. uncertainities in the zero-point flux, 
phase-dependent response errors, and uncertainties in the model atmospheres) 
and the observed dispersion of colors, we also require that 
the excess contribute greater than $\sim$ 15\% of the total flux at 24 $\mu m$:
\begin{equation} 
(K_{s}-[24])_{obs} - (K_{s}-[24])_{phot} \gtrsim 0.15. 
\end{equation}

From Figure \ref{jkmipsmem}, it is clear that the frequency of 24 $\mu m$ excess emission is 
much higher than for 8 $\mu m$ excess emission.  As previously noted, the poorer sensitivity 
of MIPS results in completeness-related 
bias for cluster members later than $\sim$ G2;
%\footnote{This follows from comparing the 
%typical J-K$_{s}$ and K$_{s}$-[24] colors for G0 stars along with their expected 
%brightnesses in J for NGC 2232 members.}.  
F and G stars are also typically too faint 
to have proper motion data and robust x-ray detections.  In contrast with the 
A star population, the census of F and G cluster members is likely incomplete.  Thus, 
we identify excess sources for FGK stars but do not attempt to derive a frequency of 
24 $\mu m$ excess emission for these stars.  

Based on the criteria for excess, 8/15 B and A cluster stars have 24 $\mu m$ emission 
from circumstellar dust\footnote{We count ID 18648 as a late A star from its J-K$_{s}$ colors, though 
removing this source from the list of B and A stars does not qualitatively change our analysis.}.  We identify two additional excess sources (IDs 
4144 and 5494) with J-K$_{s}$ colors suggestive of F0-K0 stars.  
  Other sources with spectral types between F0 and 
K0 have red K$_{s}$-[24] colors suggestive of IR excess.
Because their colors are less than 
3$\sigma$ away from the locus of photospheric sources, we do not classify these 
sources (IDs 11320, 18591, and 18605) as excess sources.   
Among the later-type stars, we identify ID 18012 an excess source.
  The spectral types estimated from photometry ($\sim$ K7-M0) and spectroscopy (M0) 
are consistent.  Thus, ID 18012 is a rare 
example of a 10--30 Myr-old M star with substantial debris emission.
Table \ref{24exclist} lists properties of the early-type members of NGC 2232 and late-type 
members with 24 $\mu m$ excess.
\subsection{Evidence for a Correlation between Rotation and Debris Disk Emission for 
High-Mass Stars: Massive Primordial Disks May Evolve into Massive Debris Disks}
%In addition to comparing the frequency and luminosity of disk emission with stellar 
%mass/spectral type, we can also investigate the connection between disk luminosity/frequency 
%and stellar rotation.  
Stellar rotation is another property that can impact the frequency and luminosity of disk 
emission.  With studies traditionally focused on near-IR ground-based observations 
of young clusters with primordial disks, the connection between disk emission and 
stellar rotation has been controversial.  Many authors \citep{Stau99, Re04, Mak04} 
fail to find a correlation. Others, most notably \citet{Herb02}, 
claim that slower rotators are more likely to have disk emission.  Because  
 the level of disk emission compared to the stellar photosphere is larger 
in the mid infrared than in the near-IR, \textit{Spitzer Space Telescope} 
observations of young stars provide a more definitive 
test of the connection between stellar rotation and disk emission.

Recently, \citet{Cb07} investigated the stellar rotation-disk emission connection in two young 
clusters, NGC 2264 ($\sim$ 2--3 Myr) and the Orion Nebula Cluster ($\sim$ 1 Myr).  By 
comparing the rotation periods and Spitzer/IRAC excesses of high-mass cluster stars 
(see also \citealt{Re06, Ir08}),
they identify a clear increase in the disk fraction with rotational period.  
\citeauthor{Cb07} interpret this correlation as evidence in favor of the 
"disk locking" model for rotational evolution of pre-main sequence stars \citep[e.g.][]{Sh94, Bo97}, 
where the star loses angular momentum from magnetic field interactions with the disk.  
Thus, the more massive disks (presumably identified by their stronger 
mid-IR emission) are able to spin down the star more substantially.

If massive primordial disks evolve into massive debris disks, then stellar rotation rates 
 should correlate with the level of debris emission.  
While MIPS data for the $\beta$ Pic Moving Group hint at a possible connection between
the rotation rate and IR excess in debris disks \citep{Re08}, the authors 
caution that such a correlation is preliminary.  The sample is 
dominated by GKM stars that can drive strong winds from their convective atmospheres, thus affecting 
stellar rotation.  A more conclusive plan is to compare rotation and MIPS excess in B/A star cluster members, which 
lack convective atmospheres and generally have stronger 24 $\mu m$ excesses.  

To test this hypothesis, we combine the MIPS data for B/A star cluster members with the 
rotational velocity study of \citet{Le74}.  The study derives v\textit{sini} from the He I (4471 \AA) and 
Mg II (4481 \AA) lines with an uncertainty of $\approx$ 10-20\%.  
We match all but three of the members studied by \citeauthor{Le74}.
All of these stars are photometrically, spectroscopically, and astrometrically confirmed as members. 
None have been identified as Be stars, though the rotation rate of one star, 
with v\textit{sini} $\sim$ 400 km s$^{-1}$, is close to the breakup velocity 
for 2-2.5 M$_{\odot}$ stars.  

\begin{figure}
%\plotone{excrot.ps}
\plotone{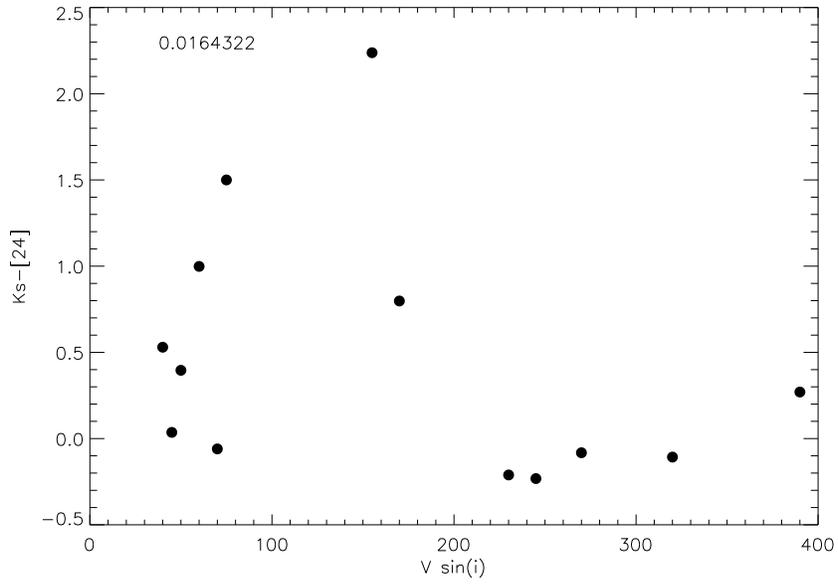}
\caption{V\textit{sini} (km s$^{-1}$) vs. K$_{s}$-[24] for B1--A3 stars in 
NGC 2232.  The probability that K$_{s}$-[24] is uncorrelated with V\textit{sini} is 
$\sim$ 1.6\%.  Thus, the data show that slower rotators typically have stronger IR excesses.}
\label{excrot}
\end{figure}

Figure \ref{excrot} shows that rotation and 24 $\mu m$ excess appear correlated.  
Only one star with v\textit{sini} $\ge$ 200 km s$^{-1}$ has a (marginal) excess of $\sim$ 0.27 
mag.  In contrast, 75\% of stars with v\textit{sini} $\le$ 200 km s$^{-1}$ 
have 24 $\mu m$ excesses.  
The MIPS excesses of the slower rotators also extend to much 
larger amplitudes, up to $\sim$ 10$\times$ photospheric levels, 
compared to those of the faster rotators ($\approx$ 25\% of the photospheric levels).
%The amplitude of excesses are also much larger than that 
%of the MIPS-excess star with v\textit{sini} = 390 km s$^{-1}$: up to $\approx$ 10$\times$ 
%photospheric levels compared to a 25\% excess.  
The Kolmogorov-Smirnov test yields a 
probability of $\sim$ 1.6\% that there is no correlation between 
rotation and 24 $\mu m$ excess (d=0.8).  Both B stars and early A stars have 
excesses and photospheric emission.  The mean and median spectral types 
of stars with v\textit{sini} $\le$ ($>$) 200 km s$^{-1}$ are B8.7 and A0 (B7.6 and B6.7).

While a larger sample of B and A stars would 
provide better statistics, these data support the picture that 
\textit{massive primordial disks evolve into massive debris disks}.
However, several caveats about rotational velocities for early-type stars complicate 
our interpretation of Figure \ref{excrot}.  For example, we only know the component 
of the rotational velocity along our line of sight, so the true rotation rate is 
not known.  While the effects of \textit{sini} should wash out in very large samples, 
they may be important for the NGC 2232 sample alone.  To better investigate any 
connection between rotation and debris emission, these data should be combined with 
data for other similarly-aged A stars to provide a much stronger statistical 
significance to any correlation.
\section{The Evolution of Debris Emission from Planet Formation: NGC 2232 in Context}
\subsection{Evolution of 8 $\mu m$ Emission}
The low frequency of 8 $\mu m$ emission from disks in NGC 2232 is 
consistent with recent Spitzer surveys of other young clusters  
with substantial populations of warm, terrestrial zone debris disks (Table \ref{8evo}).
  In h and $\chi$ Persei (13 Myr), the frequency of 8 $\mu m$ circumstellar 
dust emission rises from $\sim$ 1.3--2.2\% (14/1023; 42/1878) for B and A stars 
to $\gtrsim$ 6\% (34/523; 50/618) for FG stars \citep{Cu07a, Cu08b}.  
\citet{Cu07a} cite a lower limit of $\approx$ 4\% for the frequency of 8 $\mu m$ emission 
from intermediate-type stars.  This lower limit is still 
larger than the upper limit for the early-type excess frequency of B and A stars 
for h and $\chi$ Per ($\approx$ 3.2\%) 
and much larger than that for B stars alone ($\lesssim$ 1\%).  
Thus, the frequency of 8 $\mu m$ emission is spectral-type dependent 
at a given age and environment.  
%SED modeling 
%of 8 $\mu m$ excess sources suggest that the emission may result from terrestrial 
%planet formation \citet{Cu07b, Cu08a}.  

For other young clusters, small number statistics prevent a
clear measure of the variation of excess 8 $\mu$m emission
with spectral type. In NGC 2547 (38 Myr), \citet{Go07} derive
excess frequencies of $<$5\% (0/19) for B and A stars, $\sim$
1.5--5\% (1--3/62) for F and G stars, and $\sim$ 1\% (3/347)
for K and M stars.  Several stars studied in \citeauthor{Go07} 
have 8 $\mu m$ excesses but were not classified as excess sources 
because they lacked detectable excesses at 5.8 $\mu m$.  
This criterion may preclude detection of colder dust 
where the 8 $\mu m$ excess is on the Wien side of the blackbody. 
Relaxing the requirement of a clear
5.8 $\mu$m excess in NGC 2547 stars yields a larger frequency of 
warm dust emission, up to $\sim$ 2\% (8/347) for
M-type stars in this cluster. In NGC 2232, our estimates for
the frequency of 8 $\mu$m excess -- $\sim$ 6\% (BA), $<$ 4\% (FG), and
$\sim$ 0--1.9\% (KM)-- are nearly identical to the results for NGC 2547.
Within the errors, the low frequency of 8 $\mu$m emission among
AB and FG stars in NGC 2232 and NGC 2547 is qualitatively consistent with
results for h and $\chi$ Per.

 Combining results for all four clusters, the frequency of warm
dust among 10--30 Myr cluster stars at a given age and 
environment is small.  Across all
spectral types, the frequency of 8 $\mu$m emission is $\lesssim$
5--10\%. Although NGC 2232 and NGC 2547 have relatively sparse
populations of stars with G or earlier spectral types ($\sim$
150), the large population in h and $\chi$ Per ($\gtrsim$ 2000)
yields a robust estimate for the frequency of warm dust among
early and intermediate-type stars \citep{Cu07a}. Current data preclude
8 $\mu$m measurements for K and M stars in h and $\chi$ Per.
However, the large populations in NGC 2232 (158) and NGC 2547
(347) demonstrate that warm dust at levels L$_{d}$/L$_{\star}$ 
$\gtrsim$ 10$^{-3}$ is rare among 10--30 Myr old
K and M stars.
\subsection{Evolution of 24 $\mu m$ Emission}
Previous Spitzer studies of 5--100 Myr-old clusters examine the
time evolution of 24 $\mu$m excess emission around A stars.
Using somewhat different samples of nearby A stars, \citet{Ri05}
and \citet{Su06} show that the maximum amount of excess
emission at a given stellar age declines inversely with age from
$\sim$ 30--50 Myr to $\sim$ 1 Gyr. \citet{Cu08a} add results
for several young clusters to this sample and demonstrate that
the maximum level rises for stars with ages of 5--10 Myr, peaks
for stars with ages of 10--20 Myr, and then declines inversely
with time. With new results for NGC 2232 and NGC 2547, we can
reinvestigate `the rise and fall of debris disks' and consider, for
the first time, the relative frequency of debris disks for A stars
in young clusters.
%\subsubsection{Frequency of 24 $\mu m$ Emission with Time}

To investigate the evolution of 24 $\mu$m emission among A stars
in young clusters, we consider clusters with good statistics and
ages of 5--100 Myr.  In addition to our results for NGC 2232, 
we include data from Orion OB1a and 
b (10 and 5 Myr), Upper Centaurus Lupus (16 Myr), h and $\chi$ Persei (13 Myr), 
NGC 2547 (38 Myr), IC 2602 (50 Myr), IC 2391 (50 Myr), and the Pleiades 
\citep[100 Myr][]{He06, Cu08a, Go07, Sie07, Go06}. 

\begin{figure}
%\plotone{24freq.ps}
\plotone{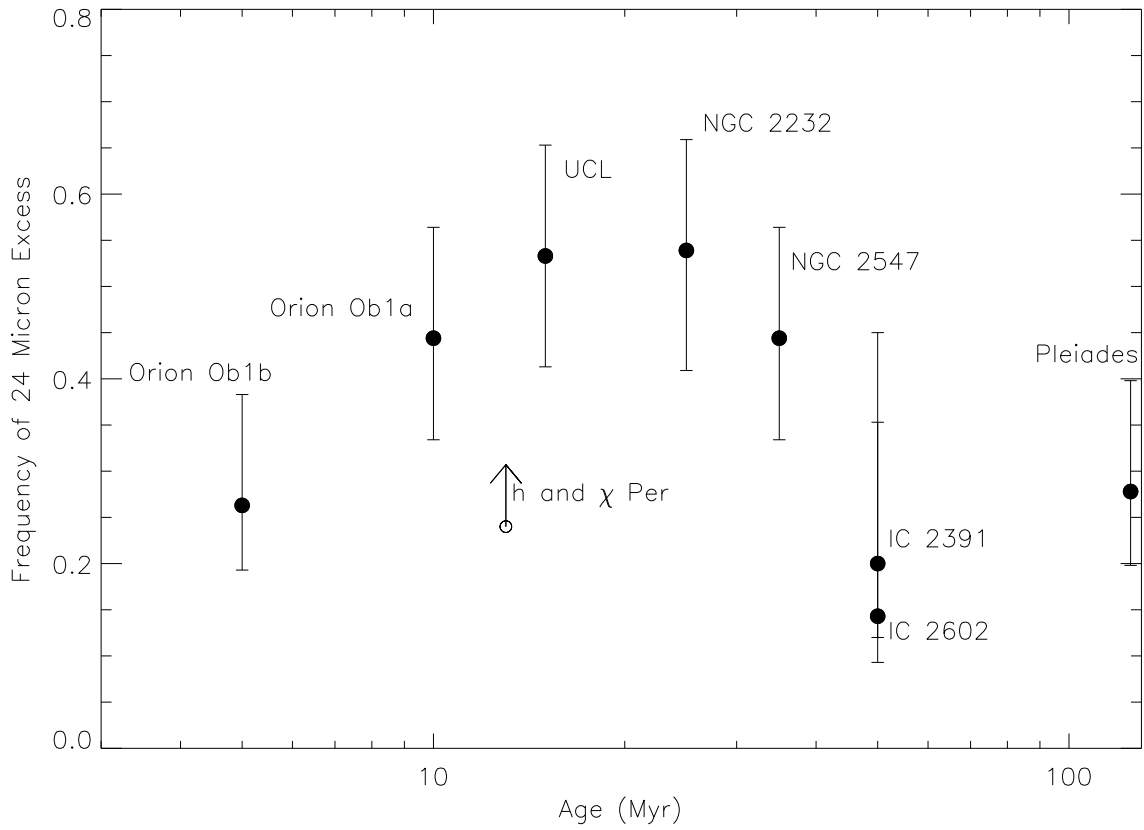}
\caption{Frequency of 24 $\mu m$ excess vs. age for B and A stars in 
several open clusters studied by Spitzer.  The frequency of debris emission apparently 
\textit{increases} from 5 Myr to $\approx$ 10-25 Myr before declining after 
$\approx$ 40 Myr.}
\label{freq24time}
\end{figure}

These data demonstrate a time-dependent frequency of 24 $\mu$m
emission from dust among young A stars (Figure \ref{freq24time}). For stars
with ages 5--25 Myr, the frequency appears to rise from $\lesssim$
40\% for 5--10 Myr old stars to $\gtrsim$ 50\% for 15--25 Myr old
stars. For older stars (50--100 Myr), the 24 $\mu$m excess
frequency declines to $\sim$ 10--20\%.
%Data from other 5--25 Myr old clusters (e.g. $\beta$ Pic Moving Group, NGC 2362) 
%are not shown here because of 
 %$support this trend \citep{Clp08}.  
%The evolution of 24 $\mu m$ excess 
%is in stark contrast to the evolution of 8$\mu m$ 
%excess, which rapidly declines from $\sim$ 3 Myr through 20 Myr \citep[e.g.][]{Cu07a,
%He07}.  

The increasing 24 $\mu m$ excess frequency from $\sim$ 5 Myr to $\sim$ 10-25 Myr 
can be explained within the context of standard models of planet formation \citep[e.g.][]{Kb08}.  
During the early stages of planet formation, much of the 
primordial dust mass may be incorporated into $\gtrsim$ km-sized planetesimals.  This 
growth to larger sizes reduces the disk opacity and thus level of emission.
Some disks may have undetectable levels of emission.  Once forming planets reach $\gtrsim$ 100-1000 km 
sizes, they stir the leftover planetesimals they accrete to high velocities.  When planetesimals 
collide with each other at these velocities, they produce copious amounts of dust which 
is then detectable in the mid infrared.  

%\subsubsection{Luminosity of 24 $\mu m$ Debris Emission with Time}

Next, we examine the level of 24 $\mu m$ excess emission with respect to time 
(see also \citealt{Ri05} and \citealt{Cu08a}), focusing on 5--40 Myr-old 
early-type stars.  We include the complete sample of 
B and A stars in NGC 2232.
In addition to our data, we include stars from clusters with debris disks:
 Orion OB1a and b (10 and 5 Myr; \citealt{He06}), $\eta$ Cha (8 Myr; \citealt{Gau08}),
the $\beta$ Pic Moving Group (12 Myr; \citealt{Re08}), 
h and $\chi$ Persei ($\sim$ 13 Myr; \citealt{Cu08a}), 
Sco-Cen, and NGC 2547 (38 Myr; \citealt{Ri05}). 
For the Sco-Cen, we include data for Upper Scorpius (5 Myr), 
Upper Centaurus Lupus (16 Myr), and Lower Centaurus Crux (17 Myr) 
from \citet{Ch05}.  We limit our literature sample to 
BAF stars.

\begin{figure}
%\plotone{exc_v_age2.ps}
\plotone{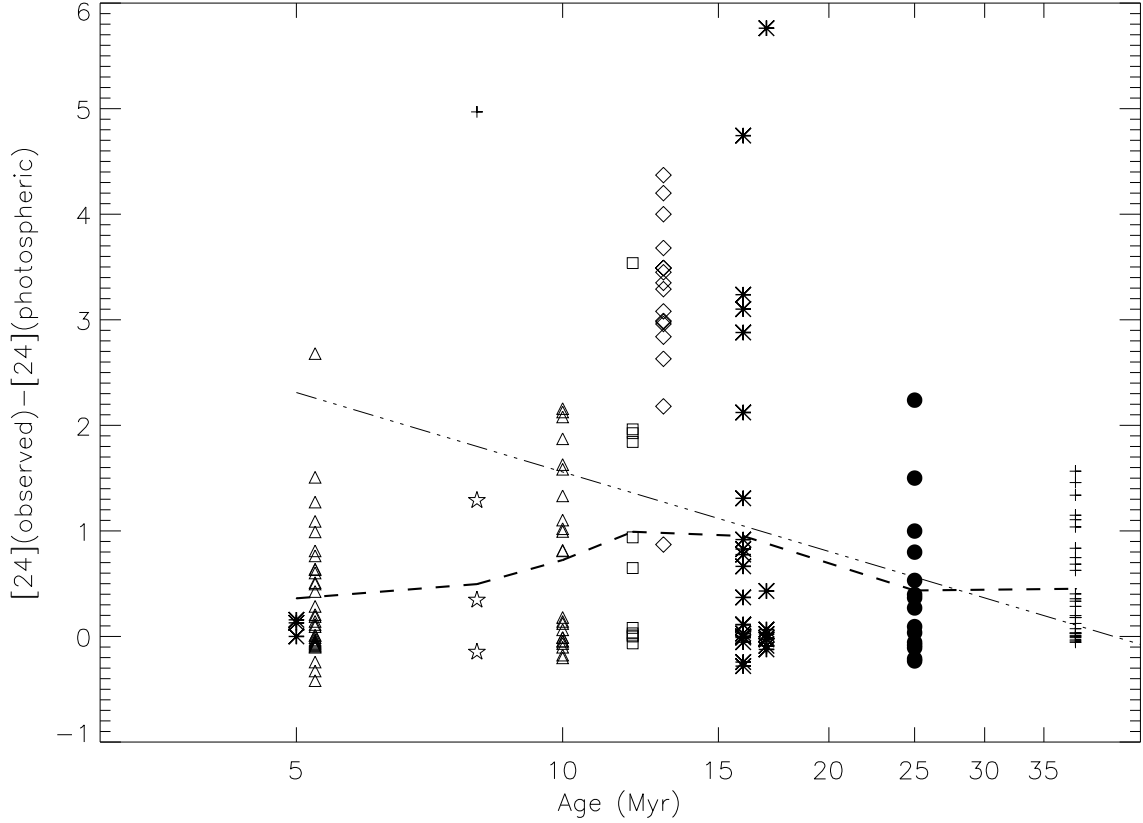}
\caption{Level of 24 $\mu m$ excess around BAF stars 
vs. time for 5--40 Myr old clusters.  We include 
Orion Ob1a and b (10 and 5 Myr; triangles), $\eta$ Cha (8 Myr; stars), the $\beta$ Pic Moving Group (12 Myr; 
squares), h and $\chi$ Persei (13 Myr; diamonds), Sco-Cen (5, 16, and 17 Myr; 
asterisks), NGC 2232 (25 Myr filled circles), and NGC 2547 (38 Myr; small crosses).  
HR 4796A, from \citet{Lo05} is also shown as a cross at t$\sim$ 8 Myr and 
[24]-[24]$_{phot}$ $\sim$ 5.  The data 
are consistent with a rise in debris emission from 5 Myr to 10 Myr, a peak from 
$\sim$ 10 Myr to 20 Myr.  For stars older than $\approx$ 40 Myr, the distribution 
of 24 $\mu m$ luminosity defines an envelope which is consistent with a t$^{-1}$ decline 
\citep{Ri05}.  The dashed line shows the mean excess. Predictions for the decline in emission from 
a steady-state collisional evolution model (dash-three dots) follow a t$^{-1}$ decline.} 
\label{excvage}
\end{figure}

Figure \ref{excvage} shows the amplitude of 24 $\mu m$ excess vs. time.
In support of \citet{Cu08a}, there is a rise in the level of debris emission 
from $\sim$ 5 Myr to 10 Myr and a peak at $\sim$ 10--20 Myr 
\footnote{Unpublished data for other 5-40 Myr old clusters -- NGC 2362 and IC 4665 -- 
also support this view (\citealt{Clp08}; Currie 2008b, unpublished).}.   Adding $\gtrsim$ 
30--50 Myr-old clusters \citep{Cu08a, Ri05} shows that the emission is consistent with a 1/t decline. 
%Connecting the mean excess level for each cluster, 
The mean level of excess (dashed line) for $\lesssim$ 20 Myr-old clusters 
clearly deviates from a t$^{-1}$ decline (dot-dashed line)
observed for $\gtrsim$ 30--50 Myr-old stars. 
%This is shown by connecting the mean excess level 
%for each cluster (dashed line).  

These new data improve our understanding of the evolution 
of debris disk emission at $\sim$ 20--40 Myr.  
Data for NGC 2232 suggest a slow 
 decline in debris emission from 10-20 Myr to 25 Myr.
The mean level of excess emission from the $\beta$ Pic Moving Group (12 Myr)
and Sco-Cen (16--17 Myr) to NGC 2232 (25 Myr) clearly drops.  However, 
the excess levels for NGC 2547 are comparable to those for NGC 2232.
 The trend in the decline of 24 $\mu m$ excess emission is then slightly 
shallower than a 1/t (dash-three dots) through $\sim$ 30--40 Myr.  
As shown in \citet{Kb08}, debris disk evolution models for 
A stars predict a high level of emission through $\sim$ 30 Myr.
Debris emission declines more rapidly for sources older than $\sim$ 40-50 Myr 
\citep{Ri05, Su06}.

These results demonstrate that an epoch of detectable, strong mid-IR emission  
follows an epoch with weak or undetectable emission.  When observed 
in an ensemble of stars at a range of ages, this sequence manifests itself as an 
increase of the 24 $\mu m$ excess frequency and luminosity with respect to time.  
This trend is not apparent at 8$\mu m$ because the time for 
protoplanets to form in the terrestrial region
probed by 8$\mu m$ dust emission is far shorter than in the ice giant/Kuiper belt 
regions which are probed by 24 $\mu m$ dust emission \citep{Kb04}.  If 
8 $\mu m$ excess emission is systematically weaker than 24 $\mu m$ excess, scatter 
in the photometry for intermediate and late-type members may conceal bona fide 
excess sources.

The high frequencies of 24 $\mu m$ dust emission from 
multiple 10--40 Myr-old clusters imply that at least the majority of primordial disks 
pass through the debris disk stage at later times.  Dust with emission at 24 $\mu m$ 
but not at 8 $\mu m$ probes active planet formation in regions with equilibrium 
temperatures of $\approx$ 120 K \citep{Cu08a}, which is colder than the water-ice condensation 
temperature in the solar nebula ($\approx$ 170 K).

Within the context of the \citeauthor{Kb04} debris disk evolutionary 
models, 24 $\mu m$ dust probes 
active planet formation beyond the ice line, so the frequency of 
icy planets around A stars ($\approx$ 1.5-3 M$_{\odot}$ at this age) 
is $\eta_{i}$ $\approx$ 50--60\%.  
Thus, \textbf{\textit{most $\approx$ 1.5--3 M$_{\odot}$ stars should form icy planets}}.

%In contrast, the frequency of warm dust emission is far lower. 
%NGC 2232, NGC 2547, and h and $\chi$ Persei show that most disks show 
%no evidence of active terrestrial planet formation when disks are forming icy planets.
%This is likely because the observable stages of terrestrial planet formation last far shorter 
%than the observable stages of icy planet formation, as suggested by planet formation models 
%\citep[e.g.][]{Kb04, Kb08}.
\section{Discussion}
%\subsection{Summary of Results}
We have described the first search for debris disks 
 in the 25 Myr-old nearby open cluster, NGC 2232. 
Using ROSAT x-ray observations and optical/IR colors and 
spectroscopy, we identified probable cluster members and 
cross correlated this list with Spitzer IRAC and MIPS data.
  By comparing 
these data with results for other 5-40 Myr old clusters, we 
reconstruct the time evolution of 8 $\mu m$ and 24 $\mu m$ emission from debris 
disks and probe the connection between debris emission and 
stellar rotation.

Our analysis identifies one A-type cluster member, HD 45435, with 
strong warm dust emission and 1--3 late-type stars 
with weakly-emitting warm dust.  
HD 45435 has dust with a temperature of T $\sim$
350--800 K and a luminosity $L_d/L_{\star} \sim 5 \times 10^{-3}$. For
the late-type stars, the lack of 24 $\mu$m excesses yields good
limits for the dust luminosity, $L_d/L_{\star} \sim 1-10 \times 10^{-4}$,
but poor limits on the dust temperature, T $\lesssim$ 200-300 K. 
%It is not 
%clear whether the 'excess' emission from later-type stars is due to 
%dust or due larger photometric errors.  
The vast majority of 
cluster stars lack evidence for marginal warm dust emission.  Combined
with results for h and $\chi$ Per and NGC 2547, our analysis suggests
that warm dust emission from debris disks around 10--30 Myr old stars
is rare, with a frequency $\lesssim$ 10\% (see also \citealt{Ma04, Cu07a, Go07}).

 In contrast, cold dust emission around 10--30 Myr old A-type stars is
common. Roughly half of all A stars with ages between 10 Myr and 30 Myr
have excess emission at 24 $\mu$m. Upper limits to the excess emission at
8 $\mu$m yield good constraints on the dust temperature,
$T \sim$ $\lesssim$ 200-300 K, and luminosity, $L_d/L_{\star} \lesssim$ 10$^{-3}$.

These results place interesting limits on the frequency of planet around
young stars. For most A-type stars with 24 $\mu$m excesses, dust emission
probes active planet formation beyond the ice line, according to 
debris disk evolutionary models \citep[e.g.][]{Kb08}.  Thus, at least 50\%
of A-type ($\sim$ 1.5--3 $M_{\odot}$) stars likely have icy planets. 
Warm dust emission probes terrestrial planet formation
inside the ice line.  Thus, $\lesssim$ 10\% of all stars show evidence
for rocky planet formation. Despite fairly large samples, it is not yet
clear whether terrestrial planet formation (a) is rare, (b) produces only 
weak emission concealed by intrinsic photometric scatter, or (c) concludes
much more rapidly than icy planet formation, as predicted by planet
formation models \citep{Kb04}.

For well-sampled clusters with ages 5--50 Myr, the frequency and magnitude
of cold dust emission among early-type stars depends on stellar age. The
frequency of 24 $\mu$m emission increases from 5--10 Myr, peaks at
10--30 Myr, and then declines. The typical amount of cold dust emission
correlates with the frequency of dust emission.
  Both of these trends are consistent with predictions from debris
disk models \citep{Kb04}.  Comparing the rotation rates of B and A stars with their 
24 $\mu m$ excesses shows that slower rotators have stronger excesses.  
Because primordial disk emission is also stronger for slower rotators \citep{Cb07}, 
there may be an evolutionary link between massive primordial disks 
and massive debris disks.

Despite our conclusion that icy planets are common
around higher-mass stars, we cannot infer the masses
of planets from the luminosity of the debris disk emission.  The 
objects responsible for debris-producing planetesimal collisions 
have radii $\gtrsim$ 500 km and masses $\ge$ 10$^{-4}$ M$_{\oplus}$.
While the debris luminosity increases as planets grow to hundreds of 
kilometers in size during runaway growth, the luminosity is independent 
of the planet size during oligarchy\footnote{Even though planetesimals 
collide at higher velocities around more massive planets, the 
scale height of the planetesimals also increases.  Thus, the collision rate, which 
controls the debris luminosity, does not increase dramatically (S. Kenyon, unpublished; \citealt{Kb08})}.
The observed L$_{d}$/L$_{\star}$ depends on the initial disk mass -- and thus the total mass 
in larger objects -- more than the mass of the largest object \citep{Kb08}.

Radial velocity surveys also constrain the 
frequency of planets around high-mass stars\footnote{However, constraining the 
frequency of \textit{icy} planets around high-mass stars requires an observing 
baseline that is much longer than that which is currently available.}.
Current surveys show that the frequency of planets is higher 
around $\ge$ 1.5 M$_{\odot}$ stars than around solar and subsolar-mass 
stars \citep{jjohn07}, consistent with theoretical expectations \citep{gk08}. 
From the standpoint of debris disk studies, however, it is not clear whether a) high-mass stars 
should make more icy planets than solar-mass stars or b)  
 high-mass stars produce more debris than they do around solar-mass stars 
simply because their disks are likely more massive. 

This work highlights two problems that impede our 
understanding of terrestrial planet formation from surveys of 
young, open clusters with Spitzer observations.  The lack of massive young clusters prevents 
robust constraints on terrestrial planet formation
around high and intermediate-mass stars.
The low luminosity of warm debris around low-mass stars limits 
our ability to construct large samples of stars with warm dust.
High-precision photometry of nearby clusters like NGC 2232 
may probe terrestrial planet formation around low-mass stars.
However, NGC 2232 and most other nearby clusters are not 
massive enough to provide 
constraints on terrestrial planet formation around 
intermediate and high-mass stars.
The only cluster studied so far that is clearly massive enough to provide 
robust constraints is h and $\chi$ Persei \citep{Cu07a, Cu07b}.  

To investigate the evolution of warm debris emission around high and 
intermediate-mass stars more conclusively, observations of clusters more populous than 
 NGC 2232 are necessary.  Spitzer Cycle 4 and 5 observations of 
h and $\chi$ Persei and other 
8--20 Myr old clusters will fill this role.  Preliminary analysis 
of MIPS h and $\chi$ Per data taken during March 2008 show a huge increase 
in the number of cluster stars detected, including several 
terrestrial zone debris disk candidates from \citet{Cu07b} that 
previously had MIPS upper limits.  Observations for h and $\chi$ Persei 
will be bracketed by observations of 
massive 8 and 20 Myr old clusters, NGC 6871 and 
NGC 1960, to provide better observational constraints on 
terrestrial planet formation during the critical 10--30 Myr epoch.

This work justifies the need for follow-up observations of 
NGC 2232.  New optical photometry, spectroscopy, and x-ray surveys 
will provide a far better constraint on cluster membership and 
member properties, which will improve the utility of NGC 2232 as a 
laboratory for studying debris disk evolution and planet formation.  
\citet{Ly06} showed that optical colors provide a useful diagnostic 
of cluster membership; a deep spectroscopic survey to identify 
late-type members is also feasible with multi-object spectrographs 
like Hectospec on the MMT and Hydra at Kitt Peak.
Identifying late-type members with a deep Chandra survey of NGC 2232 
would better constrain the evolution of x-ray activity from 10 Myr 
to 50 Myr in addition to identifying late-type, chromospherically active 
cluster stars.  All these observations should provide a much more 
complete census of cluster members, which in turn will yield a 
better analysis of the disk population for late-type stars.

Finally, the proximity of NGC 2232 to the Sun allows for 
follow-up ground-based/space-based observations of disks, 
some of which are not possible with distant, populous 
clusters such as h and $\chi$ Persei, NGC 6871, and NGC 1960.  The 
hot dust emission from ID 18566 may be resolvable by ground-based 
interferometers such as the Large Binocular Telescope Interferometer.  
The ESO space telescope \textit{Herschel} should be able to survey 
NGC 2232 for evidence of cold debris disks emitting longwards of 
$\sim$ 40 $\mu m$.  With a significant disk population established, 
NGC 2232 should be an excellent target for mid-IR observations 
with the \textit{James Webb Space Telescope}.

Although NGC 2232 has been virtually ignored 
by the star/planet formation community for the past 30 years, this work shows it to 
be a potentially important laboratory for understanding debris disk evolution and planet formation 
which deserves further study.
\begin{acknowledgements}
We thank the anonymous referee for a thorough review and helpful suggestions which 
improved the quality of this paper.
We also thank Nancy Remage Evans for advice on ROSAT archival data and Jesus Hernandez
 for use of the SPTCLASS spectral-typing code.  Finally, we thank Ken Rines 
for taking a spectrum of ID 18012 on short notice.  This work is supported 
by Spitzer GO grant 1320379, NASA Astrophysics Theory grant NAG5-13278, and 
NASA grant NNG06GH25G.  This paper makes use 
of the WEBDA open cluster database.  
\end{acknowledgements}

\begin{deluxetable}{lllllllllllllllll}
 \tiny
%\rotate
%\documentstyle[10pt]
%SPMquot"(0pt)
%\setlength{\tabcolstep}{0.02in}
%\linewidth{0.1 in}
\tabletypesize{\tiny}
%\tabletypesize{\scriptsize}
\tablecolumns{17}
\tablecaption{Photometry Catalogue for Sources with Detected on NGC 2232 field}
\tiny
\tablehead{{ID}&{RA}&{DEC}&{V}&{J}&{H}&{K$_{s}$}&{[3.6]}&{$\sigma$([3.6])}&{[4.5]}&{$\sigma$([4.5])}&
{[5.8]}&{$\sigma$([5.8])}&{[8]}&{$\sigma$([8])}&{[24]}&{$\sigma$([24])}}
\startdata
 1& 96.4706& -5.3994&0.00&13.51&13.05&12.86& 0.00  & -99.00& 0.00  & -99.00& 0.00  & -99.00& 0.00  & -99.00&10.25& 0.15\\
 2& 96.4708& -5.4639&0.00&11.19&10.46&10.19& 0.00  & -99.00& 0.00  & -99.00& 0.00  & -99.00& 0.00  & -99.00& 9.78& 0.14\\
 3& 96.4790& -5.1864&0.00&15.82&15.12&14.90& 0.00  & -99.00& 0.00  & -99.00& 0.00  & -99.00& 0.00  & -99.00& 9.22& 0.06\\
 4& 96.4816& -5.3179&0.00&10.52& 9.91& 9.71& 0.00  & -99.00& 0.00  & -99.00& 0.00  & -99.00& 0.00  & -99.00& 9.66& 0.09\\
 5& 96.4836& -5.0643&0.00& 9.90& 9.67& 9.56& 0.00  & -99.00& 0.00  & -99.00& 0.00  & -99.00& 0.00  & -99.00& 9.68& 0.09
\enddata
\tablecomments{The table includes sources with at least one detection in a Spitzer IRAC or MIPS band.  
Values of 0.00 in the photometry column and -99.00 in the photometric uncertainty columns denote 
sources that were not observed in a given filter.  Sources observed in the MIPS band but not in the 
IRAC bands fall outside the IRAC coverage.} 
\label{fullphot}
\end{deluxetable}

\begin{deluxetable}{llllllllll}
 \tiny
%\rotate
%\documentstyle[10pt]
%SPMquot"(0pt)
%\setlength{\tabcolstep}{0.02in}
%\linewidth{0.1 in}
\tabletypesize{\tiny}
%\tabletypesize{\scriptsize}
\tablecolumns{9}
\tablecaption{Spectroscopy of Selected NGC 2232 sources}
\tiny
\tablehead{{ID}&{RA}&{DEC}&{Spectral Type} & {$\sigma$(Spectral Type)}& 
{EW(H$_{\alpha}$)}&{FWHM(H$_{\alpha}$)}&{ROSAT counts ks$^{-1}$}&{Member?}}
%\tablehead{{ID}&{RA}&{DEC}&{J}&{H}&{K$_{s}$}&{[3.6]}&{$\sigma$([3.6])}&{[4.5]}&{$\sigma$([4.5])}&
%{[5.8]}&{$\sigma$([5.8])}&{[8]}&{$\sigma$([8])}{[24]}&{$\sigma$([24])}}
\startdata
  1050  &96.5720   & -4.5989   &  G0.7   &   3.4   &   2.70   &  99 & --&no \\
  6041  &96.8377   & -4.9424   &  G6.3   &   2.4   &   2.40   &  99 & --&no \\
  6639  &96.8691   & -4.8748   &  G3.7   &   2.0   &   0.00   &  99 & --&no \\
 10258  &97.0566   & -4.5612   &  K6.0   &   1.2   &   1.60   &  99 & --&yes\\
 10763  &97.0836   & -4.6977   &  G2.5   &   2.6   &   2.00   &  99 & --&yes \\
 16505  &97.3680   & -4.6234   &  F4.0   &   1.4   &   4.50   &  99 & --&no \\
 18012  &97.4388   & -5.0088   &  M0.0   &   1.5   &   -0.7   &  99 & --&yes\\
 18553  &96.8783   & -4.6588   &  F9.8   &   1.9   &   2.30   &  99 & 6.1&yes \\
 18554  &96.6436   & -4.5974   &  B3.1   &   1.3   &   5.90   &  99 & 1.4&yes \\
 18555  &96.8426   & -4.5690   &  K1.8   &   1.2   &  -0.15   &  99 & 2.1&yes \\
 18558  &96.7537   & -4.3556   &  B3.9   &   1.4   &   5.90   &  99 & 0.9&yes \\
 18559  &96.6810   & -4.7450   &  F5.0   &   3.0   &   5.00   &  99 & 1.5&yes \\
 18560  &96.5763   & -4.5993   &  M3.2   &   0.9   &  -7.40   &   5.20 &0.9& yes \\
 18562  &96.6217   & -4.3049   &  M3.6   &   1.0   &  -6.90   &   5.00 & 0.7&yes \\
 18566  &96.7940   & -4.7801   &  A2.8   &   2.5   &   9.90   &  99  & 1.5&yes\\
 18568  &97.0254   & -4.6412   &  M0.8   &   0.7   &  -1.80   &   4.30 & 0.9&yes \\
 18570  &96.8788   & -4.5815   &  M3.1   &   0.8   &  -5.10   &   5.40 & 1.2&yes \\
 18572  &96.8402   & -4.9407   &  F7.3   &   2.3   &   3.90   &  99  & 2.6&no\\
% 18573  &97.1983   & -4.7466   &  99  &   99   &   3.20   &  99  \\
 18577  &97.0173   & -4.8984   &  K3.9   &   1.2   &  -0.76   &   5.90 & 3.6&no \\
 18578  &96.8713   & -4.8730   &  M3.6   &   0.9   &  -6.30   &   6.00 & 1.3&yes \\
 18579  &96.9484   & -4.8247   &  B7.6   &   1.2   &   7.20   &  99 & 2.1&yes \\
 18580  &97.0004   & -4.7930   &  M2.9   &   1.2   &  -3.70   &   4.50 & 0.8&yes \\
 18581  &96.9899   & -4.7621   &  B0.8   &   1.4   &   4.60   &  99  & 5.9&yes\\
 18582  &96.8650   & -4.7564   &  M0.8   &   0.8   &  -1.80   &   6.00 & 1.9&yes \\
 18587  &97.1874   & -4.7139   &  M0.5   &   0.8   &  -2.28   &   4.70 & 2.2&yes\\
 18590  &96.9365   & -4.7342   &  M4.2   &   1.1   &  -8.80   &   5.70 & 0.6&yes \\
 18599  &96.9009   & -4.9428   &  M3.5   &   1.0   &  -5.10   &   5.30 & 0.8&yes \\
 18605  &97.1404   & -4.7829   &  G1.4   &   2.5   &   2.80   &  99 & 1.0&yes \\
 18606  &96.9688   & -4.7669   &  B9.5   &   1.3   &  11.10   &  99 & 0.4&yes \\
 18607  &97.0515   & -4.7617   &  M3.9   &   1.2   &  -7.00   &   5.00 & 0.5&yes \\
 18612  &97.0250   & -4.6057   &  M2.4   &   0.7   &  -5.05   &   5.20 & 1.2&yes \\
 18613  &97.0584   & -4.5622   &  K7.5   &   0.8   &  -1.50   &   6.30 & 1.5&yes \\
 18617  &97.0399   & -4.7984   &  M2.0   &   0.8   &  -3.80   &   5.10 & 1.1&yes \\
 18620  &97.0354   & -4.7400   &  M2.2   &   0.8   &  -9.25   &   5.50 & 0.6&yes \\
 18621  &96.9177   & -4.7127   &  M3.7   &   1.2   &  -6.06   &   4.90 & 2.1&yes \\
 18622  &97.0234   & -4.6411   &  M3.4   &   1.1   &  -8.20   &   5.80 & 2.6&yes \\
% 18623  &96.8445   & -4.5721   &  99     &    99   &   3.50   &  99  \\
 18633  &96.5333   & -4.6282   &  B8.8   &   1.2   &   8.80   &  99  & --&yes\\
\enddata
\label{fullspectra}
\tablecomments{FAST spectra of selected stars in the NGC 2232 field.  ID 10258 has a 
near-IR colors lying just outside the range for candidate members but a spectral 
type possibly consistent with membership.  The uncertainties in spectral type ($\sigma$(ST)) 
are determined from the dispersion in spectral types computed from each of the spectral indices 
\citep[see][for more information]{He04}.  Entries in the FWHM(H$_{\alpha}$) column with a 
value of '99' did not have H$_{\alpha}$ clearly in emission and/or a high enough signal to noise 
to calculate a reliable full-width half-maximum.  }
\end{deluxetable}

%\begin{deluxetable}{llllllllllllllll}
%\begin{minipage}{40mm}
\begin{deluxetable}{cccc}
%\begin{minipage}{40mm}
\centering
% \tiny
%\rotate
%\documentstyle[10pt]
%SPMquot"(6pt)
%\setlength{\tabcolstep}{0.02in}
%\linewidth{0.1 in}
%\tabletypesize{\tiny}
\tabletypesize{\scriptsize}
\tablecolumns{4}
\tablecaption{Confirmed and Candidate Cluster Members with Spitzer Data}
\tablehead{{ID}&{RA}&{DEC}&{Mem. Type}} 
\startdata
    150 & 96.5238 & -4.7164& 4\\
    845 & 96.5620 & -4.6887& 4\\
    964 & 96.5676 & -4.4612& 4\\
    985 & 96.5687 & -4.4479& 4\\
   1084 & 96.5738 & -4.8108& 4\\
\enddata
\tablecomments{The table includes sources with at least one detection in a Spitzer IRAC or MIPS band.  
Membership type 1 denotes confirmed x-ray active cluster members; type 2 identifies members 
from \citet{Cl72}.  Type 3 lists 'candidate' members as determined from V/V-J colors, spectroscopy, 
and proper motion.  Type 4 lists 'candidate' members determined from J/J-K$_{s}$ and J/J-H colors.}
\label{fullmember}
%\end{minipage}
\end{deluxetable}
%\end{minipage}

\begin{deluxetable}{lllllll}
 \tiny
%\rotate
%\documentstyle[10pt]
%SPMquot"(0pt)
%\setlength{\tabcolstep}{0.02in}
%\linewidth{0.1 in}
\tabletypesize{\tiny}
%\tabletypesize{\scriptsize}
\tablecolumns{4}
\tablecaption{Photospheric 2MASS-IRAC/MIPS Colors}
\tiny
\tablehead{{Spectral Type}&{K$_{s}$-[5.8]}&{K$_{s}$-[8]}&{K$_{s}$-[24]}}
\startdata
B0 & -0.17 & -0.24 & -0.34\\
B3 & -0.11 & -0.16 & -0.21\\
B5 & -0.07 & -0.12 & -0.1\\
B8 & -0.05 & -0.1 & -0.1\\
A0 & -0.02 & -0.05 & -0.1\\
A5 & -0.01 & -0.05 & -0.1\\
F0 & 0.02 & -0.02 & 0.005\\
F5 & 0.02 & -0.02 & 0.005\\
G0 & 0.01 & -0.02 & 0.005\\
G5 & 0.01 & -0.02 & 0.005\\
K0 & 0.01 & 0 & 0.005\\
K5 & 0.08 & 0.1 & 0.1\\
M0 & 0.21 & 0.24 & 0.34\\
M2 & 0.25 & 0.30 & 0.41\\
M5 & 0.28 & 0.33 & 0.41\\
\enddata
%\tablecomments{
\label{photcolors}
\end{deluxetable}

\begin{deluxetable}{llllllll}
 \tiny
%\rotate
%\documentstyle[10pt]
%SPMquot"(0pt)
%\setlength{\tabcolstep}{0.02in}
%\linewidth{0.1 in}
\tabletypesize{\tiny}
%\tabletypesize{\scriptsize}
\tablecolumns{8}
\tablecaption{8$\mu m$ Excess Sources in NGC 2232} 
\tiny
%\tablehead{{ID}&{Cross-ID}&{Spectral Type}&{K$_{s}$-[8]}&{$\sigma$[8]}&{K$_{s}$-[8]$_{phot}$+3$\sigma$[8]}&{K$_{s}$-[8]$_{excess}$}&{Excess/$\sigma$[8]}}
\tablehead{{ID}&{Cross-ID}&{Spectral Type}&{log(F$_{[8]}$/F$_{[4.5]}$)}&{log(F$_{[8]}$/F$_{[4.5]}$)--Photosphere}&{Excess/$\sigma_{rms}$}}
\startdata
 6540 & -- &K5-M0 & -0.262 & 0.163&9.61\\
 9220 & --&K7-M2 & -0.349 & 0.074&4.39\\
% 10559 & --& K5-K7 & 0.335 & 0.029 & 0.182 & 0.269 & 5.3\\
 18566 & HD 45435&A2.8 & -0.168 & 0.292& 17.24\\
18613 &--& K2-K7&-0.355&0.074&4.35\\
 %18601 & --&M0 & 0.499 & 0.031 & 0.244 & 0.337 & 8.18\\
% 18622 & --&M3.4 & 0.423 & 0.036 & 0.261 & 0.370 & 4.44\\
\enddata
\tablecomments{Sources with warm, 8 $\mu m$ excess emission from circumstellar dust 
as determined by their 8 $\mu m$ to 4.5 $\mu m$ flux ratios.  
Sources with a range of spectral types have this range estimated photometrically.  
Two other sources, IDs 18601 and 18622, have K$_{s}$-[8] colors that are greater than 
3$\sigma_{[8]}$ away from photospheric predictions.  However, these sources do not have 
[8]/[4.5] flux ratios that are more than 3$\sigma$ away from the empirically-determined 
locus of photospheric ratios.}
\label{8exclist}
\end{deluxetable}

\begin{deluxetable}{lllllllll}
 \tiny
%\rotate
%\documentstyle[10pt]
%SPMquot"(0pt)
%\setlength{\tabcolstep}{0.02in}
%\linewidth{0.1 in}
\tabletypesize{\tiny}
%\tabletypesize{\scriptsize}
\tablecolumns{9}
\tablecaption{Early-Type Members and Intermediate/Late-Type Members with 24 $\mu m$ Excess}
\tiny
\tablehead{{ID}&{Cross-ID}&{Spectral Type}&{K$_{s}$-[8]}&{K$_{s}$-[24]}&{24 $\mu m$ Excess?}&{V\textit{sin i} (km s$^{-1}$)}}
%\tablehead{{ID}&{Cross-ID}&{Spectral Type}&{K$_{s}$-[8]}}
%\tablehead{{ID}}
\startdata
 18554&HD 45321 & B3.1 & -0.21 & -0.23&no&245\\
 18558&HD 45418, 9 Mon & B3.9 & -0.23 & -0.21&no&230\\
 18566&HD 45435 & A2.8 & 1.00 & 2.24&yes&155\\
 18579&HD 45516 & B7.6 & -0.13 & -0.08&no&270\\
 18581&HD 45546 & B0.8 & -0.19 & 0.04&no&45\\
 18606&BD-04 1526B&B9.5& -0.05 & --&no data&--\\
 18616&--& --&-0.13 & -0.13&no&--\\
 18630&HD 45583, V682 Mon& B8&-0.17 & -0.06&no&70\\
 18631&HD 45532 & A0 &-0.04 & -0.11&no&320\\
 18633&HD 45238 & B8.8&-0.02 &0.27&yes&390\\
 18634&HD 45399 & A0&-0.05 &0.80&yes&170\\
 18635&HD 295102& B8&-0.04 & 0.53&yes&$\le$40\\
 18636&HD 45627 & A0&-0.04&0.40&yes&50\\
 18638&HD 45565 & A0&-0.08&1.00&yes&60\\
 18639&HD 45691 & A0&0.0 &1.50&yes&75\\
 18648&HD 295100& A7-F2&-0.07&0.37&yes&--\\
 \\
 4144 & -- & G2-G7 & -0.07 & 0.39 &yes&--\\
 5494 & -- & F2-F7 & --&0.75 &yes&--\\
 18012 & -- & M0 & 0.21 & 1.57 &yes&--\\
\enddata
%\tablecomments{Sources with warm, 8 $\mu m$ excess emission from circumstellar dust.
%Sources with a range of spectral types have this range estimated photometrically.}
\tablecomments{Spectral types come from the VIZIER database except those determined
in this work.  Sources with a range of spectral types have this range estimated 
from their J-K$_{s}$ colors. ID 4144 is identified as a candidate member from its position on the 
 J/J-H and J/J-K$_{s}$ color-magnitude diagrams.  ID 5494 is identified as a candidate member 
from its position on the V/V-J, J/J-H, and J/J-K$_{s}$ color-magnitude diagrams.  ID 18012 is 
identified as a candidate member from the J/J-H and J/J-K$_{s}$ diagrams and then confirmed 
as a member spectroscopically.}
\label{24exclist}

\end{deluxetable}

\begin{deluxetable}{lllllll}
 \tiny
%\rotate
%\documentstyle[10pt]
%SPMquot"(0pt)
%\setlength{\tabcolstep}{0.02in}
%\linewidth{0.1 in}
\tabletypesize{\tiny}
%\tabletypesize{\scriptsize}
\tablecolumns{7}
\tablecaption{Frequency of 8$\mu m$ Excess from $\gtrsim$ 10 Myr-old Clusters}
\tiny
%\tablehead{{Spectral Type}&{K$_{s}$-[5.8]}&{K$_{s}$-[8]}&{K$_{s}$-[24]}}
\tablehead{{Cluster}&{Age}&{Parent Sample Size}&{Freq. BA}&{Freq. FG}&{Freq. KM}&{References}}
\startdata
%NGC 2362 & 5 & 335 & 0. & 0. & xxx (xxx/xxx) & \citet{Dh07}\\
h and $\chi$ Persei & 13 & 4,700 & 1.3\% (14/1023) & 6.5\%(34/523)& --& 1\\
   & & & 2.2\% (42/1878) & 8\% (50/618)& -- &\\
NGC 2232 & 25& 238 & 6.7\% (1/15) & 0 (0/33) & 0-1.9\% (0-3/158) & 2\\
NGC 2547 & 38 & 450 & 0 (0/19)& 1.6-4.9\% (1-3/62)& 1-2.3\% (3-8/347)& 3\\
\enddata
\tablecomments{References are 1) \citet{Cu07a, Cu08b}, 2) this work, and 
3) \citet{Go07}.  Parent sample size refers to the initial photometric sample.  The sizes of individual 
subsets are listed in parentheses in columns 4-6.  Values for the fraction of excess sources 
in NGC 2547 include an estimate on the fraction of interloping intermediate/late-type sources 
that are not likely to be proper motion members.  Values for h and $\chi$ Persei include the 
sample of sources that are photometrically consistent with cluster membership as well as a subset 
with spectra.  The frequency of 8 $\mu m$ excess around late-type stars for NGC 2232 is highly uncertain 
due to the incomplete membership list.}
\label{8evo}
\end{deluxetable}

%\begin{figure}
%\centering
%\plottwo{ch1edist.ps}{ch2edist.ps}
%\plottwo{ch3edist.ps}{ch4edist.ps}
%\caption{Magnitude vs. error for IRAC-detected sources with 2MASS counterparts.  The [5.8] and [8] channels 
%do not reach a $\le$ 10$\sigma$ photometric uncertainty until m([5.8])=15 and m([8])=14.5.  The errors in the 
%first two channels are much lower, though residual image artifacts caused by bright sources may make the
%actual photometric errors for nearby faint sources larger than quoted here. } 
%\label{errordist}
%\end{figure}
%\begin{figure}
%\centering
%\plotone{jjmk.ps}
%\caption{J/J-K$_{s}$ color-magnitude diagram for sources on the NGC 2232 field.  Sources 
%with possible 8 $\mu m$ excess are identified with blue squares; red triangles 
%denote 24 $\mu m$ excess candidates.}
%\label{whatcolorexc}
%\end{figure}


\clearpage

\epsscale{1}
\clearpage
\clearpage

\begin{thebibliography}{}
%\bibitem[Absil et al.(2006)]{Abs06}Absil, O,. et al., 2006, A\&A, 452, 237
\bibitem[Augereau et al.(1999)]{Au99}Augereau, J. C., Lagrange, A. M., Mouillet, D., Papaloizou, J. C. B.,
 Grorod, P. A., 1999, A\&A, 348, 557
\bibitem[Backman and Paresce(1993)]{Bp93}Backman, D., Paresce, F., 1993, in Protostars and Planets III (E.H. Levy and J.I. Lunine, eds.)
(Univ. of Arizona, Tucson), 1253, (1993)
\bibitem[Baraffe et al.(1998)]{Ba98}Baraffe, I., et al., 1998, A\&A, 337, 403
\bibitem[Bertin and Arnouts(1996)]{Be96}Bertin, E., \& Arnouts, S., 1996, A\&AS, 117, 393
\bibitem[Bouvier et al.(1993)]{Bou93}Bouvier, J., Cabrit, S., Fernandez, M., Martin, E. L., \& Matthews, J. M, 1993, A\&A, 26, 272
\bibitem[Bouvier et al.(1997)]{Bo97}Bouvier, J., Forestini, M., Allain, S., 1997, A\&A\, 326, 1023
\bibitem[Bouwman et al.(2008)]{Bou08}Bouwman, J., et al., 2008, \apj\ accepted, arXiv:0802.3033
\bibitem[Calvet et al.(2005)]{Cal05}Calvet, N., et al., 2005, \apj, 630, 185L
\bibitem[Carpenter et al.(2006)]{Ca06} Carpenter, J., Mamajek, E., Hillenbrand, L., Meyer, M., 2006, \apj, 651, 49L
\bibitem[Chen et al.(2005a)]{Ch05} Chen, C., Jura, M., Gordon, K. D., Blaylock, M., 2005, \apj, 623, 493
\bibitem[Chen et al.(2006)]{Ch06}Chen, C., et al., 2006, \apjs, 166, 351
\bibitem[Cieza and Baliber(2007)]{Cb07}Cieza, L., Baliber, N., 2007, \apj, 671, 605
\bibitem[Claria(1972)]{Cl72}Claria, J., 1972, A\&A, 19, 309
\bibitem[Currie et al.(2007a)]{Cu07a}Currie, T., et al., 2007, \apj, 659, 599
\bibitem[Currie et al.(2008c)]{Ces08}Currie, T., Evans, N., et al., 2008, \aj in prep.
\bibitem[Currie et al.(2007b)]{Cu07b}Currie, T., Kenyon, S., Rieke, G., Balog, Z., Bromley, B., 2007, \apj, 663, 105L
\bibitem[Currie et al.(2007c)]{Cu07c}Currie, T., Kenyon, S., Balog, Z., Bragg, A., Tokarz, S., 2007, \apj, 669, 33L
\bibitem[Currie et al.(2008a)]{Cu08a}Currie, T., Kenyon, S., Balog, Z., Rieke, G., Bragg, A., Bromley, B., 2008, \apj, 672, 558
\bibitem[Currie et al.(2008d)]{Clp08}Currie, T., et al., 2008, in prep.
\bibitem[Currie(2008)]{Cu08b}Currie, T., 2008, Ph.D. thesis, University of California-Los Angeles
\bibitem[Currie and Kenyon(2008)]{Ck08} Currie, T., Kenyon, S. J., 2008, submitted, arXiv:0801.1116
\bibitem[Currie et al.(2008b)]{Cu08c}Currie, T., Irwin, J., Hernandez, J., et al., 2008, in prep.
%\bibitem[Currie et al.(2008c)]{Clp08}Currie, T., Lada, C. J., Plavchan, P., et al., 2008, in prep.
\bibitem[Fabricant et al.(1998)]{Fa98} Fabricant, D., et al., 1998, \pasp, 110, 79
\bibitem[Fazio et al.(2004)]{Faz04}Fazio, G., et al., 2004, \apjs, 154, 10
\bibitem[Gautier et al.(2007)]{Ga07}Gautier, T., et al., 2007, \apj, 667, 527
\bibitem[Gautier et al.(2008)]{Gau08}Gautier, T., et al., 2008, \apj\ accepted, arXiv:0804.3113
\bibitem[Gorlova et al.(2006)]{Go06}Gorlova, N., et al., 2006, \apj, 649, 1028
\bibitem[Gorlova et al.(2007)]{Go07}Gorlova, N., Balog, Z., Rieke, G. H., Muzerolle, J., Su, K. Y. L.,
Ivanov, V. D., Young, E. T.,  2007, \apj, 670, 516
\bibitem[Herbst et al.(2002)]{Herb02}Herbst, W., Bailer-Jones, C. A. L., \& Mundt, R. 2001, \apj, 554, 197L
\bibitem[Hernandez et al.(2004)]{He04}Hernandez, J., Calvet, N., Briceno, C., Hartmann, L., Berlind, P., 2004, \aj, 127, 1682
\bibitem[Hernandez et al.(2006)]{He06}Hernandez, J., Briceno, C., Calvet, N., Hartmann, L., 
Muzerolle, J., \& Quintero, A., 2006, \apj, 652, 472
\bibitem[Hernandez et al.(2007)]{He07}Hernandez, J., et al., 2007, \apj, 662, 1067
\bibitem[Hillenbrand(2005)]{Hi05} Hillenbrand, L., 2005, astro-ph/0511083
\bibitem[Hillenbrand et al.(2008)]{Hi08}Hillenbrand, L., et al., 2008, \apj, 677, 630
\bibitem[Indebetouw et al.(2005)]{In05}Indebetouw, R., et al., 2005, \apj, 619, 9311
\bibitem[Irwin et al.(2008)]{Ir08}Irwin, J., et al., 2008, \mnras, 384, 675
\bibitem[Jacoby et al. (1984)]{Ja84} Jacoby, G., et al., 1984, \apjs, 56, 257
%\bibitem[Jayawardhana et al.(1999)]{Rj99}Jayawardhana, R., Hartmann, L., Fazio, G., Fisher, R. S., 
%Telesco, C., and Pina, R., 1999, \apj, 521, 129L
\bibitem[Johnson et al.(2007)]{jjohn07}Johnson, J., et al., 2007, \apj, 670, 833
\bibitem[Kennedy and Kenyon(2008)]{gk08}Kennedy, G., Kenyon, S. J., 2008, \apj, 673, 502
\bibitem[Kenyon and Bromley(2004)]{Kb04}Kenyon, S., Bromley, B., 2004, \apj, 602, 133L
\bibitem[Kenyon and Bromley(2006)]{Kb06}Kenyon, S., Bromley, B., 2006, \aj, 131, 1837
\bibitem[Kenyon and Bromley(2008)]{Kb08}Kenyon, S., Bromley, B., 2008, \apjs, submitted
\bibitem[Kenyon and Hartmann(1987)]{Kh87}Kenyon, S., Hartmann, L., 1987, \apj, 323, 714
\bibitem[Kenyon and Hartmann(1995)]{Kh95}Kenyon, S., Hartmann, L., 1995, \apjs, 101, 117
\bibitem[Kurucz(1993)]{Ku93}Kurucz, R. L., 1993, SYNTHE Synthesis Programs and Line Data Kurucz CD-ROM No. 18, Cambridge, MA: 
Smithsonian Astrophysical Observatory, 1993, 18
\bibitem[Lada et al.(2006)]{La06}Lada, C. J., et al., 2006, \aj, 131, 1574
\bibitem[Lejeune et al.(1997)]{Le97}Lejeune, Th., Cuisinier, F., and Buser, R., 1997, A\&AS, 125, 229
\bibitem[Levato and Malaroda(1974)]{Lm74}Levato, H., Malaroda, S., 1974, \aj, 79, 890
\bibitem[Levato(1974)]{Le74}Levato, H., 1974, \aj, 79, 1269
\bibitem[Lisse et al.(2008)]{Li08}Lisse, C., Chen, C., Wyatt, M., Morlock, A., 2008, \apj, 673, 1106
\bibitem[Low et al.(2005)]{Lo05} Low, F., Smith, P., Werner, M., Chen, C., Krause, V., Jura, M., 
Hines, D., 2005, \apj, 631, 1170
\bibitem[Lyra et al.(2006)]{Ly06}Lyra, W., et al., 2006, A\&A, 453, 101
\bibitem[Makidon et al.(2004)]{Mak04}Makidon, R. B., Rebull, L. M., Strom, S. E., Adams, M. T., \& Patten, B. M., 
2004, \aj, 127, 2228
\bibitem[Mamajek et al.(2004)]{Ma04} Mamajek, E., et al., 2004, \apj, 612, 496
\bibitem[Mathis (1990)]{Ma90}Mathis, J., 1990, \araa, 28, 37
\bibitem[Mayne and Naylor(2008)]{Mn08}Mayne, N. J., Naylor, T., 2008, \mnras\ accepted, arXiv:0801.4085
%\bibitem[Metchev et al.(2004)]{Met04}Metchev, S., Hillenbrand, L., Meyer, M., 2004, \apj, 600, 435
\bibitem[Meyer et al.(2006)]{Me06}Meyer, M., et al., 2006, \pasp, 118, 1690
\bibitem[Naylor and Jeffries(2006)]{Nj06} Naylor, T., Jeffries, R. D., 2006, \mnras, 373, 1251
\bibitem[Padgett et al.(2006)]{Pa06} Padgett, D., et al., 2006, \apj, 645, 1283
\bibitem[Panzera et al.(2003)]{Pa03}Panzera, M. R., Campana, S., Covino, S., Lazzati, D., Mignani, R. P., 
Moretti, A., Tagliaferri, G., 2003, A\&A, 399, 351
\bibitem[Papovich et al.(2004)]{Pa04} Papovich, C., et al., 2004, ApJS, 154, 70
\bibitem[Plavchan et al.(2005)]{Pl05}Plavchan, P., et al., 2005, \apj, 631, 1161
\bibitem[Plavchan et al.(2008)]{Pl08}Plavchan, P., et al., 2008, in prep.
\bibitem[Preibisch and Feigelson(2005)]{Pr05}Preibisch, T., Feigelson, E., 2005, \apjs, 160, 390
\bibitem[Quijada et al.(2004)]{Qu04}Quijada, M., et al., 2004, SPIE, 5487, 244
\bibitem[Rebull et al.(2004)]{Re04}Rebull, L., Wolff, S. C., \& Strom, S. E. 2004, \aj, 127, 1029
\bibitem[Rebull et al.(2006)]{Re06}Rebull, L., et al., 2006, \apj, 646, 297
\bibitem[Rebull et al.(2008)]{Re08}Rebull, L., et al., 2008, \apj\ in press, arXiv:0803.1674
\bibitem[Rhee et al.(2007a)]{Rh07}Rhee, J., Song, I., Zuckerman, B., McElwain, M., 2007, \apj, 660, 1556
\bibitem[Rhee et al.(2007b)]{Rh07a}Rhee, J., Song, I., Zuckerman, B., 2007, \apj, 671, 166
\bibitem[Rhee et al.(2008)]{Rh07b}Rhee, J., Song, I., Zuckerman, B., 2008, 675, 777
\bibitem[Rieke et al.(2004)]{Rie04}Rieke, G., et al., 2004, \apjs, 154, 25
\bibitem[Rieke et al.(2005)]{Ri05}Rieke, G., 2005, \apj, 620, 1010
\bibitem[Rieke et al.(2008)]{Rie08}Rieke, G., et al., 2008, \aj, 135, 2245
\bibitem[Sargent et al.(2006)]{Sa06}Sargent, B., et al., 2006, \apj, 645, 395
\bibitem[Shu et al.(1994)]{Sh94}Shu, F., et al., 1994, \apj, 429, 781
\bibitem[Siegler et al.(2007)]{Sie07}Siegler, N., et al., 2007, \apj, 654, 580
\bibitem[Siess et al.(2000)]{Si00} Siess, L., Dufour, E., Forestini, M., 2000, A\&A, 358, 593
\bibitem[Song et al.(2005)]{Sz05} Song, I., Zuckerman, B., Weinberger, A., Becklin, E., 2005, Nature, 436, 363
\bibitem[Stauffer et al.(2005)]{St05}Stauffer, J., et al., 2005, \aj, 130, 1834
\bibitem[Staussun et al.(1999)]{Stau99}Staussun, K. G., Mathieu, R. D., Mazeh, T., \& Vrba, F. J. 1999, \aj, 117, 2941
\bibitem[Su et al.(2006)]{Su06} Su, K., et al., 2006, \apj, 653, 675
\bibitem[Voges et al.(1999)]{Vog99}Voges, W., et al., 1999, A\&A, 349, 389
\bibitem[Werner et al.(2004)]{We04}Werner, M., et al., 2004, \apjs, 154, 1
\bibitem[Wetherill and Stewart(1993)]{Ws93}Wetherill, G., Stewart, G., 1993, Icarus, 106, 190
\bibitem[White and Basri(2003)]{Wb03}White, R., Basri, G., 2003, \apj, 582, 1109
\bibitem[Wyatt et al.(2007)]{Wy07}Wyatt, M., et al., 2007, \apj, 658, 569
\bibitem[Yin et al.(2002)]{Yi02}Yin, Q., et al., 2002, Nature, 418, 949
\bibitem[Zuckerman and Song(2004)]{Zs04} Zuckerman, B., Song, I., 2004, \araa, 42, 685
\end{thebibliography}
\end{document}